\documentclass[oldversion]{aa}
\usepackage{graphicx}
\def\H0 {$H_{\rm o}$}

\def\CH3C2H {\hbox{${\rm CH}_3{\rm C}_2{\rm H}$}} %CH3C2H
\def\ffas {\hbox{$\,.\!\!^{\prime\prime}$}}
\def\ffcirc {\hbox{$\,.\!\!^{\circ}$}}

\def\ffs {\hbox{$\,.\!\!^{\rm s}$}}
\def \la{\mathrel{\mathchoice   {\vcenter{\offinterlineskip\halign{\hfil
$\displaystyle##$\hfil\cr<\cr\sim\cr}}}
{\vcenter{\offinterlineskip\halign{\hfil$\textstyle##$\hfil\cr
<\cr\sim\cr}}}
{\vcenter{\offinterlineskip\halign{\hfil$\scriptstyle##$\hfil\cr
<\cr\sim\cr}}}
{\vcenter{\offinterlineskip\halign{\hfil$\scriptscriptstyle##$\hfil\cr
<\cr\sim\cr}}}}}
\def \ga{\mathrel{\mathchoice   {\vcenter{\offinterlineskip\halign{\hfil
$\displaystyle##$\hfil\cr>\cr\sim\cr}}}
{\vcenter{\offinterlineskip\halign{\hfil$\textstyle##$\hfil\cr
>\cr\sim\cr}}}
{\vcenter{\offinterlineskip\halign{\hfil$\scriptstyle##$\hfil\cr
>\cr\sim\cr}}}
{\vcenter{\offinterlineskip\halign{\hfil$\scriptscriptstyle##$\hfil\cr
>\cr\sim\cr}}}}}

\begin{document}

\title{Molecular line emission in NGC~4945, imaged with ALMA}

\author{C. Henkel \inst{1,2,3}, 
        S. M{\"u}hle\inst{4}, 
        G. Bendo\inst{5,6},
        G.~I.~G. J{\'o}zsa\inst{4,7,8},
        Y. Gong\inst{1,9},
        S. Viti\inst{10},
        S. Aalto\inst{11},
        F. Combes\inst{12},
        S. Garc\'{\i}a-Burillo\inst{13},
        L.~K. Hunt\inst{14},
        J. Mangum\inst{15},
        S. Mart\'{\i}n\inst{16,17},
        S. Muller\inst{11},
        J. Ott\inst{18},
        P. van der Werf\inst{18},
        A.~A. Malawi\inst{2},
        H. Ismail\inst{2}, 
        E. Alkhuja\inst{2},
        H.~M. Asiri\inst{2},
        R. Aladro\inst{1},
        F. Alves\inst{4,20},
        Y. Ao\inst{21},
        W. A. Baan\inst{22},
        F. Costagliola\inst{11},
        G. Fuller\inst{5},
        J. Greene\inst{23},
        C.~M.~V. Impellizzeri\inst{16},
        F. Kamali\inst{1},
        R.~S. Klessen\inst{24},
        R. Mauersberger\inst{1},
        X.~D. Tang\inst{1,3,25},
        K. Tristram\inst{16},
        M. Wang\inst{9},
        J.~S. Zhang\inst{26}
       }

\offprints{C. Henkel, \email{chenkel@mpifr-bonn.mpg.de}}

\institute{
  Max-Planck-Institut f{\"u}r Radioastronomie, Auf dem H{\"u}gel 69, 
  53121 Bonn, Germany
 \and 
  Astronomy Department, Faculty of Science, King Abdulaziz University, P.O. Box 80203, 
  Jeddah 21589, Saudi Arabia
 \and
  Xinjiang Astronomical Observatory, Chinese Academy of Sciences, 830011 Urumqi, China
 \and
  Argelander Institut f{\"u}r Astronomie, Universit{\"a}t Bonn, Auf dem H{\"u}gel 71, 
  53121 Bonn, Germany
 \and 
  Jodrell Bank Centre for Astrophysics, University of Manchester, Oxford Road, 
  Manchester M13\,9PL, UK
 \and
  UK ALMA Regional Centre Node, University of Manchester, Oxford Road, 
  Manchester M13\,9PL, UK
 \and
  SARAO, SKA South Africa, The Park, Park Road, Pinelands 7405, South Africa
 \and
  Rhodes University, RARG, RATT, P.O.Box 94, Grahamstown 6140, South Africa
 \and
  Purple Mountain Observatory \& Key Laboratory for Radio Astronomy, 
  Chinese Academy of Sciences, 2 West Beijing Road, 210008 Nanjing, PR China
 \and
  Department of Physics and Astronomy, UCL, Gower St., London, WC1E 6BT, UK
 \and
  Dept. of Earth and Space Sciences, Chalmers University of Technology, 
  Onsala Observatory, 43992 Onsala, Sweden 
 \and
  LERMA, Observatoire de Paris, College de France, CNRS, PSL Univ., UPMC, Sorbonne Univ., Paris, France
 \and
  Observatorio de Madrid, OAN-IGN, Alfonso XII, 3, 28014-Madrid, Spain
 \and
  INAF-Osservatorio Astrofisico di Arcetri, Largo E. Fermi, 5, 50125, Firenze, Italy 
 \and
  National Radio Astronomy Observatory, 520 Edgemont Road, Charlottesville, VA 22903, 
  USA 
 \and
  European Southern Observatory, Alonso de C{\'o}rdova 3107, Vitacura Casilla 763 0355, 
  Santiago, Chile   
 \and
  Joint ALMA Observatory, Alonso de C{\'o}rdova 3107, Vitacura Casilla 763 0355, 
  Santiago, Chile   
 \and
  National Radio Astronomy Observatory, P.O. Box O, 1003, Lopezville Road, Socorro, 
  NM 87801-0387. USA
 \and
  Leiden Observatory, Leiden University, P.O. Box 9513, 2300 RA Leiden, The Netherlands 
 \and
  Max-Planck-Institut f{\"u}r Extraterrestrische Physik, Giessenbachstra{\ss}e 1, 
  85748 Garching, Germany
 \and
  National Astronomical Observatory of Japan, 2-21-1 Osawa, Mitaka, 181-8588, Tokyo, Japan
 \and
  Netherlands Institue for Radioastronomy ASTRON, 7991 PD Dwingeloo, The Netherlands
 \and
  Dept. of Astrophysical Sciences, Princeton University, Princeton, NJ 08544, USA
 \and
  Universit{\"a}t Heidelberg, Zentrum f{\"u}r Astronomie, Inst. f{\"u}r Theoretische 
  Astrophysik, Albert-Ueberle-Str. 2 {\it and} Interdisziplin{\"a}res Zentrum f{\"u}r 
  Wissenschaftliches Rechnen, Im Neuenheimer Feld 205, 69120 Heidelberg, Germany
 \and
  Key Laboratory of Radio Astronomy, Chinese Academy of Sciences, 830011 Urumqi, China
 \and 
  Center for Astrophysics, Guangzhou University, Guangzhou 510006, China
}

\date{Received date ; accepted date}
 
\abstract
{
NGC~4945 is one of the nearest ($D$$\approx$3.8\,Mpc; 1$''$ $\approx$ 19\,pc) starburst 
galaxies. To investigate structure, dynamics, and composition of its dense nuclear gas, 
ALMA band 3 ($\lambda$$\approx$3--4\,mm) observations were carried out with $\approx$2$''$ 
resolution. Measured were three HCN and two HCO$^+$ isotopologues, CS, C$_3$H$_2$, SiO, 
HCO, and CH$_3$C$_2$H. Spectral line imaging demonstrates the presence of a rotating 
nuclear disk of projected size 10$''$$\times$2$''$ reaching out to a galactocentric 
radius of $r$$\approx$100\,pc with  position angle PA = 45$^{\circ}$$\pm$2$^{\circ}$, 
inclination $i$ = 75$^{\circ}$$\pm$2$^{\circ}$ and an unresolved bright central core 
of size $\la$2$''$. The continuum source, representing mostly free-free radiation from 
star forming regions, is more compact than the nuclear disk by a linear factor of two but 
shows the same position angle and is centered 0\ffas39 $\pm$ 0\ffas14 northeast of the nuclear 
accretion disk defined by H$_2$O maser emission. Near the systemic velocity but outside the 
nuclear disk, both HCN $J$=1$\rightarrow$0 and CS $J$=2$\rightarrow$1 delineate molecular arms of 
length $\ga$15$''$ ($\ga$285\,pc) on opposite sides of the dynamical center. These are connected 
by a (deprojected) $\approx$0.6\,kpc sized molecular bridge, likely a dense gaseous bar seen 
almost ends-on, shifting gas from the front and back side into the nuclear disk. Modeling 
this nuclear disk located farther inside ($r$$\la$100\,pc) with tilted rings 
provides a good fit by inferring a coplanar outflow reaching a characteristic deprojectd 
velocity of $\approx$50\,km\,s$^{-1}$. All our molecular lines, with the notable exception 
of CH$_3$C$_2$H, show significant absorption near the systemic velocity ($\approx$571\,km\,s$^{-1}$), 
within a range of $\approx$500 -- 660\,km\,s$^{-1}$. Apparently, only molecular transitions 
with low critical H$_2$-density ($n_{\rm crit}$ $\la$ 10$^{4}$\,cm$^{-3}$) do not 
show absorption. The velocity field of the nuclear disk, derived from CH$_3$C$_2$H, 
provides evidence for rigid rotation in the inner few arcseconds and a dynamical mass of 
$M_{\rm tot}$ = (2.1$\pm$0.2) $\times$ 10$^8$\,M$_\odot$ inside a galactocentric radius of 
2\ffas45 ($\approx$45\,pc), with a significantly flattened rotation curve farther out. Velocity 
integrated line intensity maps with most pronounced absorption show molecular peak positions 
up to $\approx$1\ffas5 ($\approx$30\,pc) southwest of the continuum peak, presumably due to 
absorption, which appears to be most severe slightly northeast of the nuclear maser disk. 
A nitrogen isotope ratio of $^{14}$N/$^{15}$N $\approx$ 200--450 is estimated. This range 
of values is much higher then previously reported on a tentative basis. Therefore, with 
$^{15}$N being less abundant than expected, the question for strong $^{15}$N enrichment 
by massive star ejecta in starbursts still remains to be settled. }


\titlerunning{ALMA observations of NGC~4945}

\authorrunning{Henkel, C. et al.}

\maketitle

\section{Introduction}

Observing the optically highly obscured central regions of galaxies hosting 
an active supermassive nuclear engine is fundamental for our understanding of 
galaxy evolution. The transfer of mass into sub-parsec scale accretion disks, 
the feeding of the nuclear engine and its feedback, potentially affecting the 
large scale appearance of the parent galaxy, are basic phenomena to be studied.
The nearby, almost edge-on H$_2$O megamaser galaxy NGC~4945 is such an active 
galaxy. Its central region is known to show a rich molecular spectrum hosting 
not only a nuclear starburst but also an active galactic nucleus (AGN; see, 
e.g., Marconi et al. 2000; Yaqoob 2012). Past molecular single-dish studies 
include Henkel et al. (1990, 1994), Dahlem et al. (1993), Mauersberger et 
al. (1996), Curran et al. (2001), Wang et al. (2004), Hitschfeld et al. (2008), 
P{\'e}rez-Beaupuits et al. (2011), and Monje et al. (2014). As an outstanding 
galaxy in the Cen~A/M~83 group at a distance of only $D$ $\approx$ 3.8\,Mpc (e.g., 
Karachentsev et al. 2007; Mould \& Sakai 2008; 1$^{\prime\prime}$ corresponds to 
$\approx$19\,pc), NGC~4945 hosts one of the three brightest IRAS (Infrared Astronomical 
Satellite) point sources beyond the Magellanic Clouds (Wang et al. 2004). 

Although this Seyfert 2 galaxy has an infrared luminosity a few times that of the 
Milky Way, it nevertheless suffers from a lack of interferometric studies of molecular 
high density tracers, a consequence of its southern location. Greenhill et al. (1997) 
used the 22\,GHz H$_2$O maser lines to map the circumnuclear accretion disk at a 
galactocentric radius of $r_{\rm GC}$ $\approx$ 0.45\,pc (0\ffas025), which shows 
a similar position angle as the large scale disk, revealing a binding mass of order 
$M$ = (1--2) $\times$ 10$^6$\,M$_{\odot}$. More recently, 183 and 321\,GHz H$_2$O 
and even Class~I methanol maser emission has also been detected (Hagiwara et al. 
2016; Humphreys et al. 2016; Pesce et al. 2016; McCarthy et al. 2017). Cunningham 
\& Whiteoak (2005) mapped the $\lambda$ $\approx$ 3\,mm lines of HCN (partially), 
HNC, and HCO$^+$ with an angular resolution of 5\ffas6 $\times$ 3\ffas5 and found 
an elongated morphology, with a position angle consistent with the inner H$_2$O 
and the outer large scale disk. If this gas were to trace a rotating disk, its radius 
would be $r_{\rm GC}$ $\approx$ 57\,pc, with a rotational velocity of $V_{\rm rot}$ $\approx$ 
135\,km\,s$^{-1}$, enclosing a mass of $M$ $\approx$ 2.5$\times$10$^8$\,M$_{\odot}$. 
Also based on ATCA (Australia Telescope Compact Array) data, more recently Green et 
al. (2016) published HCN, HCO$^+$, and HNC spectra, maps of integrated intensity, 
and the velocity field with $\approx$7$''$ resolution. 

A basic concern for the present study is the morphology of the dense molecular gas.
In the following, we try to improve our understanding of the interplay between the
active galactic nucleus, the surrounding dense gas environment, and its connection 
to clouds at larger galactocentric radii. After providing a comprehensive view 
on measured spatial and kinematical components we will look (1) for the presence 
of a molecular hole or the opposite, a molecular hotspot, at the very center,
(2) for the presence of a rotating molecular disk, (3) for the existence of a 
molecular bar and (4) for the development of spiral patterns. Radial motions,
frequently observed in other galaxies (e.g., Turner 1985; Feruglio et al. 2010; 
Bolatto et al. 2013; Garc\'{\i}a-Burillo et al. 2014) are another phenomenon
to be adressed.

\begin{figure}[t]
\vspace{0.0cm}
%\centering
\hspace{-0.2cm}
\resizebox{9.1cm}{!}{\rotatebox[origin=br]{0.0}{\includegraphics{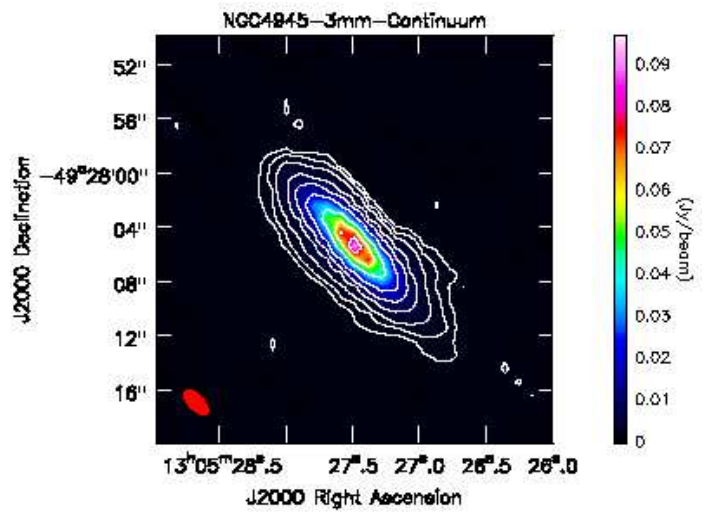}}}
\vspace{-0.7cm}
\caption{An image of the $\lambda$ $\approx$ 3\,mm  continuum emission from the central 
region of NGC~4945 (1$''$ coresponds to a projected linear scale of $\approx$19\,pc). 
The map is based on the line-free sky frequencies of 88.27--88.38\,GHz and  
88.54--88.70\,GHz in the H$^{12}$C$^{14}$N window, 85.60--86.01\,GHz and 86.28--86.42\,GHz
in the H$^{12}$C$^{15}$N window, 97.35--97.72\,GHz and 98.01--98.24\,GHz in the CS
window, and 100.15--100.46\,GHz in the continuum window (see Sect.\,2). The beam size is 
2\ffas55 $\times$ 1\ffas34 (0.1\,Jy\,beam$^{-1}$ is equivalent to $\approx$6.8\,K) with a position angle 
of 44$^{\circ}\!\!$.2. Contour levels are 0.5, 1, 2, 5, 10, 20, 40, 60, and 80\% of the peak flux 
density of 96\,mJy\,beam$^{-1}$. The restored beam is shown in the lower left corner. In 
view of the limited extent of the emission, no primary beam correction has been applied.}
\label{continuum}
\end{figure}

\begin{table*}
\caption[]{Detected spectral features toward NGC~4945$^{a}$}
\begin{flushleft}
\begin{tabular}{lcccccl}
\hline
Line                                  & $\nu_{\rm rest}$ &SEST$^{b}$ & ALMA$^{b}$  &SEST$^{c}$&  ALMA$^{c}$ & Comments to the ALMA profiles$^{d}$\\
				      &        (GHz)     &\multicolumn{2}{c}{(mJy)}&\multicolumn{2}{c}{Jy\,km\,s$^{-1}$}                      \\
\hline 
                                      &                  &           &             &          &             &                                 \\
HC$^{18}$O$^+$ $J$=1$\rightarrow$0    &        85.16222  &  ---/---  &   20/20     &   ---    &  2.7$\pm$0.4& weak, with absorption           \\
c-C$_3$H$_2$ 2$_{12}$--1$_{01}$       &        85.33889  &  180/130  &  130/130    & 60$\pm$6 & 21.4$\pm$0.7& strong, absorption NE of nucleus\\
CH$_3$C$_2$H 5$_0$--4$_0$             &        85.45727  &   ---/--- &   40/40     &   ---    &  7.5$\pm$0.5& strong, only emission           \\
c-C$_3$H$_2$ 4$_{32}$--4$_{23}$       &        85.65642  &   ---/--- &    ---      &   ---    &      ---    & weak, blended, see Sect.\,4.3.1 \\
H42$\alpha$                           &        85.68840  &   ---/--- &   40/40     &   ---    & 12.9$\pm$0.8& strong, only emission           \\ 
HC$^{15}$N $J$=1$\rightarrow$0        &        86.05497  &    20?    &    20       &   10?    &  1.1$\pm$0.2& weak                            \\
H$^{13}$CN $J$=1$\rightarrow$0        &        86.33992  &   65/75   &   50/60     & 21$\pm$2 &  9.4$\pm$0.8& strong, with absorption         \\
HCO 1$_{01}$$\rightarrow$0$_{00}$     &        86.67076  &   ---/--- &   30/30     &   ---    &      ---    & weak, blended                   \\
H$^{13}$CO$^+$ $J$=1$\rightarrow$0    &        86.75429  &   ---/--- &   40/40     &   ---    &      ---    & strong, with absorption, blended\\   
SiO $J$=2$\rightarrow$1               &        86.84696  &   ---/--- &   40/40     &   ---    &      ---    & strong, with absorption, blended\\
HCN $J$=1$\rightarrow$0               &        88.63116  &  1300/1100& 1300/1000   &495$\pm$9 &249.3$\pm$3.0& strong, with absorption         \\
CS $J$=2$\rightarrow$1                &        97.98095  &   880/770 &  650/500    &375$\pm$20&112.4$\pm$1.2& strong, with absorption         \\
				      &                  &           &             &          &             &                                 \\
\hline
\end{tabular}
\end{flushleft}
a) For the integrated flux density and other information on the H42$\alpha$ line, see Bendo 
et al. (2016). \\
b) Approximate flux densities for the blue- and redshifted line peaks, obtained from the 
15-m Swedish-ESO Submillimeter Telescope (SEST; Wang et al. 2004) and ALMA. Uncertainties
in the ALMA data are $\approx$5--10\,mJy from the noise level and 10\% of the flux density 
from the calibration, while differences in the chosen area ($\gg$10\,arcsec) do not lead to 
significant changes. For the SEST, rms uncertainties are of order 40\,mJy (c-C$_3$H$_2$), 
12\,mJy (H$^{13}$CN/HC$^{15}$N), 150\,mJy (HCN), and 100\,mJy (CS). As indicated by a question 
mark, the SEST detected HC$^{15}$N only tentatively. HCO provides a quartet of hyperfine 
components among which the strongest feature is seen by ALMA. \\
c) From Gaussian fits (see Wang et al. 2004 for the SEST data). \\
d) For the weak lines, it is not clear whether absorption is not observed  because of a too low
signal-to-noise ratio or whether it is truly absent. In the case of HC$^{15}$N, with two 
more abundant isotopologues showing weak absorption in a specific velocity interval and with 
HC$^{15}$N not being seen within exactly this velocity range, absorption is likely present.
\label{tab-lines}
\end{table*}

While trying to bridge the gap between the well known large scale structure
(e.g., Dahlem et al. 1993; Ott et al. 2001) and the sub-parsec maser disk,
we will also compare observed morphologies with that of the simultaneously 
measured radio continuum, mainly representing free-free emission from star 
forming regions (Bendo et al. 2016). Furthermore, we will analyze our images in the 
light of data taken at other wavelengths and briefly compare encountered 
structures with those from other (carefully selected) prominent nearby 
starburst galaxies. A superficial first look into chemical properties is also 
included. Finally, a relatively simple kinematical model of a particularly well 
ordered component of the dense molecular gas is presented.

Initially, the origin of the rare nitrogen isotope $^{15}$N was the major 
motivation to carry out the present study. Traditionally being assigned to low mass stars 
including novae, more recent models of stellar nucleosynthesis, describing 
the evolution of massive rapidly rotating stars, also permit the production of 
significant amounts of $^{15}$N through mixing of protons into helium burning 
shells (e.g., Woosley \& Weaver 1995; Timmes et al. 1995). Direct confirmation 
of this scenario can be obtained by measuring a low ($\approx$100) $^{14}$N/$^{15}$N 
ratio in a well developed starburst galaxy, where enrichment of the interstellar 
medium (ISM) by ejecta from massive stars must have occurred. Chin et al. (1999) 
tentatively reported such a ratio for NGC~4945, based on observations with the 15-m 
SEST (Swedish-ESO Submillimeter Telescope). Making use of the higher sensitivity 
and resolution provided by ALMA (Atacama Large Millimeter/Submillimeter Array) 
this result can now be checked.

The focus of this article is the molecular gas. Describing the observations in 
Sect.\,2, overall distributions of molecular line emission (and absorption),
including the continuum, are presented in Sect.\,3. Sect.\,4 provides the 
analysis, while main results are summarized in Sect.\,5. The H42$\alpha$ radio
recombination line, which also lies within the covered frequency range, 
has already been discussed by Bendo et al. (2016) in an accompanying paper.

\begin{figure*}[t]
\vspace{0.0cm}
\hspace{-0.5cm}
\resizebox{19.5cm}{!}{\rotatebox[origin=br]{0.0}{\includegraphics{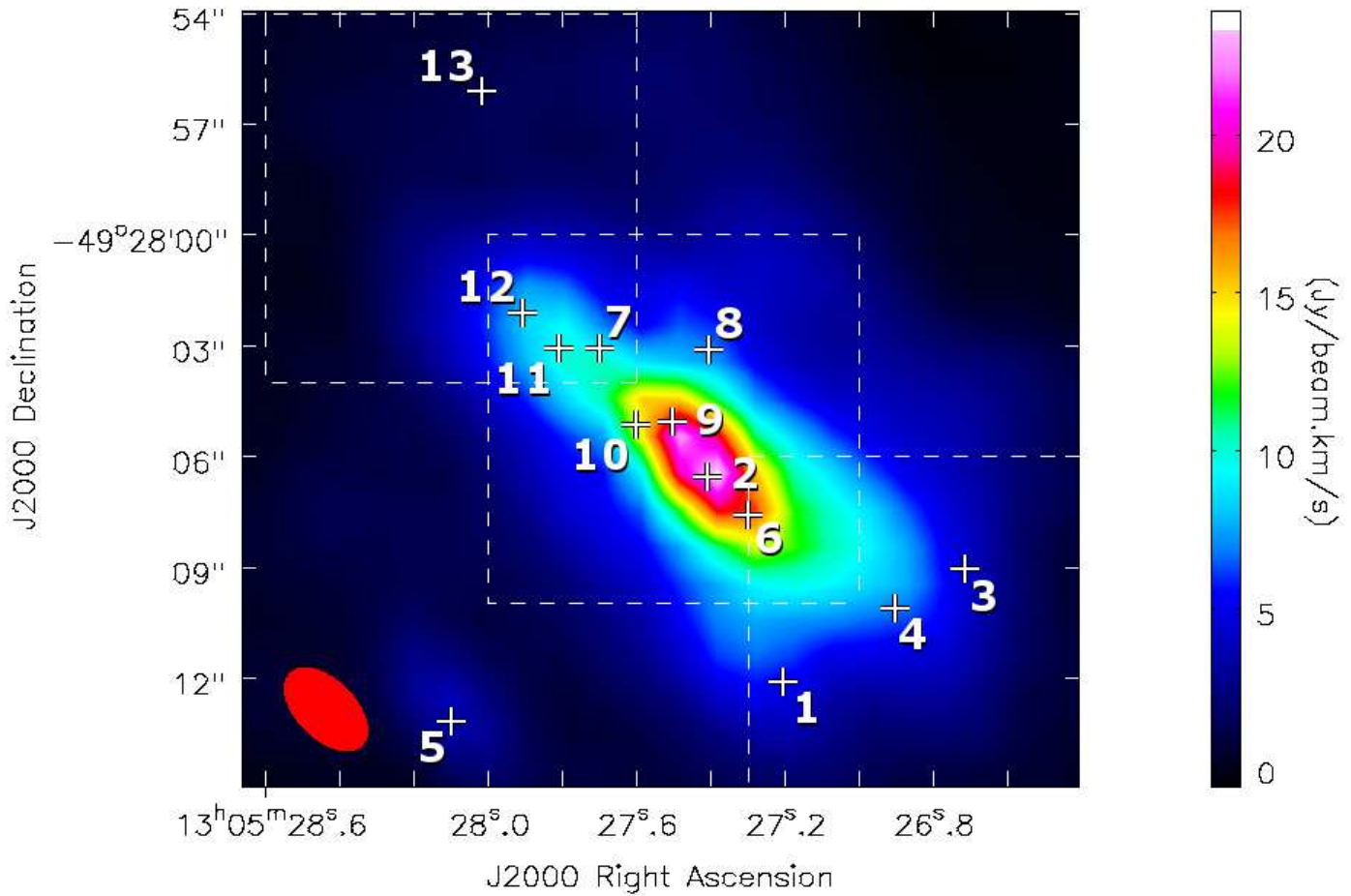}}}
\vspace{-1.4cm}
\caption{Distribution of the main spatial HCN $J$ = 1$\rightarrow$0 features
with integrated HCN intensity as background. Assigned numbers are also presented in Col.\,2 
of Table~\ref{tab-hcn}, together with additional information. This includes the offset from 
position 9, the peak of the $\lambda$$\approx$3\,mm continuum emission (see Fig.~\ref{continuum}),
which is responsible for the absorption features seen in most spectral lines. For the three
quadratic areas surrounded by white dashed lines, see Sect.\,3.3. 1$''$ corresponds 
to a projected linear scale of $\approx$19\,pc. The beam size (2\ffas71 $\times$ 1\ffas56, 
position angle 43$^{\circ}$) is given in the lower left.}
\label{hcn-hotspots}
\end{figure*}

\section{Observations and data reduction}

Band 3 ($\lambda$$\approx$3--4\,mm) images of NGC~4945 (project code: 2012.1.00912.S)
were acquired in Cycle 1 with ALMA in a compact 32 antennae configuration on 2013 January 24. 
The minimal and maximal baseline lengths were $\approx$20\,m and 420\,m. Among the 32 
antennas only one antenna, DV11, had to be flagged (bandpass). Four spectral windows were 
placed in Band 3, two in the lower sideband (LSB) and two in the upper sideband (USB) with 
the centers of the sidebands being separated by 12\,GHz. The data were obtained using 
spectral averaging from 
\begin{itemize}
\item (1) an H$^{12}$C$^{14}$N (hereafter HCN) window centered at a sky frequency of 
88.462948\,GHz (rest frequency: 88.6316\,GHz) with a bandwidth of 468.8\,MHz 
(1588\,km\,s$^{-1}$) and 480 channels with a spacing of 3.3\,km\,s$^{-1}$, 
\item (2) an H$^{12}$C$^{15}$N (hereafter HC$^{15}$N) window (this also includes 
H$^{13}$C$^{14}$N (H$^{13}$CN) and other lines) centered at 85.891221\,GHz (86.05497\,GHz) 
with a bandwidth of 1875\,MHz (6545\,km\,s$^{-1}$) and 1920 channels with a spacing of 
3.4\,km\,s$^{-1}$, 
\item (3) a $^{12}$C$^{32}$S (hereafter CS) window centered at 97.794508\,GHz 
(97.980950\,GHz) with a bandwidth of 937.5\,GHz (2874\,km\,s$^{-1}$) and 960 channels 
with a spacing of 3.0\,km\,s$^{-1}$, and 
\item (4) a continuum window (see also Sect.\,3.1) at 99.809726\,GHz (100\,GHz) with a bandwidth 
of 1875\,MHz. Here the spacing of each of the 128 channels is 44.0\,km\,s$^{-1}$. 
\end{itemize}

The systemic velocity is only known to an accuracy of about $\pm$25\,km\,s$^{-1}$ (e.g., 
Chou et al. 2007; Bendo et al. 2016). Here we adopt a barycentric velocity of $V_{\rm sys}$ 
= c$z$ = 571\,km\,s$^{-1}$ (optical convention; see, e.g., Dahlem et al. 1993). The 
differences between radio and optical velocities, $V_{\rm rad}$ -- $V_{\rm opt}$ 
= 1.1\,km\,s$^{-1}$, and between the barycentric and Local Standard of Rest reference 
systems, $V_{\rm barycentric}$ -- $V_{\rm LSR}$ = 4.6\,km\,s$^{-1}$, are too small to play 
an important role in the following analysis. 

Amplitude calibration was obtained by observing the quasar J1107--448. A flux 
density of 1.4\,Jy has been taken from the ALMA calibrator data base. Bandpass 
and phase calibrators were J1427--4206 and J1248--4559, respectively. While in 
the following relative, not absolute intensities are mainly addressed, we 
nevertheless estimate an absolute flux accuracy of $\pm$10\% (e.g., 
Fomalont et al. 2014; Wilson et al. 2014).

The phase tracking center was $\alpha_{\rm J2000}$ = 13$^{\rm h}$ 05$^{\rm m}$ 27\ffs28 
and $\delta_{\rm J2000}$ = --49$^{\circ}$ 28$'$ 04\ffas4 ($\alpha_{\rm B1950}$ = 
13$^{\rm h}$ 02$^{\rm m}$ 32\ffs07 and $\delta_{\rm B1950}$ = --49$^{\circ}$ 12$'$
01\ffas0; compare with Greenhill et al. 1997). The full width to half power (FWHP) field 
of view is $\approx$60$''$. To investigate basic mophological properties we used ``Briggs'' 
weighting with a robustness parameter of 0.0 (approximately half way between natural and 
uniform weighting) for a reasonable trade off between sensitivity and resolution. Full  
synthesized half power beam sizes are then $\approx$2\ffas6 $\times$ 1\ffas5, with position angles 
of $\approx$42$^{\circ}$. In those cases, where we intended to include emission from spatial 
scales as large as possible, i.e. for moment 1 and 2 as well as position-velocity plots 
(Sects.\,4.6.1 and 4.6.2), Briggs weighting with a robustness parameter of +2.0, to come
close to natural weighting, has been applied. This leads to almost circular full half power 
beam sizes of 2\ffas1 to 2\ffas3. According to the ``ALMA Cycle 1 Technical Handbook'' 
(their Fig.~6.1), the largest accessible angular scales are, at a 50\% intensity level, 
close to 20$''$. 

The observations lasted almost an hour, with an on-source integration time of 
about 30\,min. Five full cycles between phase calibrator and NGC~4945 were 
performed, integrating on the latter source for about $\approx$6\,min per cycle. 
Sensitivities in the HCN- and HC$^{15}$N-related windows are 0.6\,mJy\,beam$^{-1}$ in 
$\approx$25\,km\,s$^{-1}$ wide channels and 0.7\,mJy\,beam$^{-1}$ in $\approx$30\,km\,s$^{-1}$ 
wide channels of the CS window. For the continuum emission near 100\,GHz an rms noise 
level of $\approx$0.13\,mJy\,beam$^{-1}$ has been derived in a 310\,MHz wide window (100.15--100.46\,GHz,
see also Fig.~\ref{continuum}). Two versions of image cubes were created. One included the 
continuum emission. The other was continuum subtracted based on a linear fit to the line-free
regions given in the caption to Fig.~\ref{continuum}. The data were reduced, calibrated 
and imaged with the Common Astronomy Software Application (CASA)\footnote{http://casa.nrao.edu}, 
version 4.0 (see McMullin et al. 2007).

\begin{table*}
\caption[]{HCN $J$=$1\rightarrow$0 line features toward NGC~4945 (see also Fig.~\ref{hcn-hotspots})$^{a}$ }
\begin{flushleft}
\begin{tabular}{ccrrl}
\hline
Velocity        & Source & \multicolumn{2}{c}{$\Delta\alpha,\Delta\delta$} & Comments                                                            \\
(km\,s$^{-1}$)	&        & \multicolumn{2}{c}{(arcsec)}         &                                                                                 \\
\hline 
                &        &              &            &                                                                                           \\
$\approx$350    &    1   &      --3     &      --7   & Peak 1, becoming detectable at 315\,km\,s$^{-1}$; there strongest feature                 \\
                &    2   &      --1     &      --1   & Peak 2, at 350\,km\,s$^{-1}$ strongest feature, slightly SW from the center               \\
                &    3   &      --8     &      --4   & Peak 3, weak, southern extension at $V$$>$360\,km\,s$^{-1}$                               \\
390             & 1,2,3  &      --1     &      --1   & Peak 2 emission extends to the SW, Peaks 1 and 3 join the forming NE-SW ridge             \\         
420             &    2   &      --1     &      --1   & Peak 2 dominant, the molecular ridge expands toward the SW                                \\
460             &    2   &      --1     &      --2   & Peak 2 dominant, position angle of ridge changes from $\approx$40$^{\circ}$ to $\approx$60$^{\circ}$\\
                &    4   &      --6     &      --5   & Appearance of a secondary peak                                                            \\
480             &    2   &      --1     &      --2   & Peak 2 (inner SW ridge) gets weaker, appearance of weak extended arms                     \\ 
500             &    2   &      --1     &      --2   & Peak 2 not dominant, appearance of an extended arm in the NW                              \\
515             &        &      --1     &      --2   & Absorption, surrounded by emission in the SE, NW, and SW                                  \\ 
                &    5   &       +6     &      --8   & Weak extended arms in the NW; given offset marks a hotspot in the SE                  \\ 
530--570        &        &              &            & The morphology of the core emission resembles a $\Psi$ at position angle $\sim$40$^{\circ}$\\
575             &    6   &      --2     &      --3   & Absorption SW of the dynamical center, appearance of an extended arm in the SE            \\
600             &    6   &      --2     &      --3   & Absorption SW of the dynamical center, widespread weak line emission                      \\ 
                &    7   &       +2     &       +2   & A second region exhibiting absorption, located in the NE                                  \\
                &    8   &      --1     &       +2   & Prominent (at this velocity) but not a particularly strong emission peak                  \\
                &    5   &       +6     &       +8   & Prominent (at this velocity) but not a particularly strong emission peak                  \\
620--665        &    9   &        0     &        0   & Absorption dominates, morphology similar to that of the continuum emission                \\
                &        &              &            & (see Fig.~\ref{continuum})                                                                \\
                &        &              &            & Widespread weak line emission, strongest on the NW side of the absorbing region,          \\
                &        &              &            & from there extensions to the NE; also extended emission SE of the center                  \\
                &    5   &       +6     &       +8   & The SE hotspot (see 515\,km\,s$^{-1}$) fades                                              \\
650--675        &        &              &            & Absorption fades                                                                          \\
665             &   10   &       +1     &        0   & Appearance of an emission component slightly NE of the center                             \\
665--700        &   10   &       +1     &        0   & While the position of the line peak is kept, the emission spreads toward the NE           \\
                &        &              &            & The extended arms disappear                                                               \\
710             &   11   &       +3     &       +2   & At this position emission becomes most intense, at higher velocities                      \\
                &        &              &            & line intensities start to decrease, disappearance of the extended arms                    \\
                &        &              &            & in the NW and SE                                                                          \\
745             &        &        0     &        0   & SW edge of the ridge of line emission                                                     \\
                &   12   &       +4     &       +3   & NE edge of the ridge of line emission                                                     \\
770             &    9   &        0     &        0   & Weak line emission peak                                                                   \\
                &   13   &       +5     &       +9   & Weak line emission peak                                                                   \\
                &        &              &            &                                                                                           \\
\hline
\end{tabular}
\end{flushleft}
a) The table presents the positions of the main HCN $J$ = 1$\rightarrow$0 hotspots. The numbers
of the molecular hotspots given in Col.\,2 can be visualized in Fig.~\ref{hcn-hotspots} (white crosses).
In the table they follow, from top to bottom, the barycentic velocity axis of the data cube. Therefore
the table is best read from top to bottom.  Position offsets given in Cols.\,3 and 4 are relative to 
our $\lambda$ $\approx$ 3\,mm continuum peak at $\alpha_{2000}$ = 13$^{\rm h}$ 05$^{\rm m}$ 27\ffs5, 
$\delta_{2000}$ = --49$^{\circ}$ 28$'$ 05$''$ (Sect.\,3.1). 1$''$ corresponds to $\approx$19\,pc 
(Sect.\,1). 
\label{tab-hcn}
\end{table*}

\section{Results}

\subsection{Continuum emission and the position of the H$_2$O maser disk}

Figure~\ref{continuum} shows the 3\,mm nuclear continuum emission of NGC~4945 obtained from the line 
free parts of the four measured bands (see also Sect.\,2). Most of it (84\% $\pm$ 10\%) originates 
from free-free emission (see Bendo et al. 2016 for a detailed analysis, which also suggests that 
missing flux is not a problem here). This is the only detected source of continuum emission inside the area 
covered by the $\approx$1$^{\prime}$ sized primary beam. The integrated continuum flux density is 310$\pm$1\,mJy 
with a peak of 88.6$\pm$0.3\,mJy\,beam$^{-1}$. Note that the errors given are those of the fit and 
do not include the estimated calibration uncertainty ($\approx$10\%, see Sect.\,2). The beam deconvolved 
size (full width to half maximum) is well described by an ellipse, centered at $\alpha_{\rm J2000}$ = 
13$^{\rm h}$ 05$^{\rm m}$ 27\ffs4930 and $\delta_{\rm J2000}$ = --49$^{\circ}$ 28$'$ 05\ffas229. Because
of the high signal-to-noise ratio of several 100 (rms noise; $\approx$0.13\,mJy\,beam$^{-1}$, see Sect.\,2) 
the nominal position error is only $\approx$0\ffas003. In view of changing weather conditions, 
the angle between phase calibrator and the target ($\approx$5.$\!\!^{\circ}$4) and the level of errors 
in the antenna position measurements, we cautiously estimate that the absolute position uncertainty 
(compare, e.g., with Chou et al. 2007 and Lenc \& Tingay 2009) is of order 0\ffas1 (see also the ALMA 
knowledgebase in https://help.almascience.org/snoopi/index.php?, which provides (see the ``Recent Articles''
column, question 8), an absolute rms position uncertainty of 0\ffas04).

\begin{figure*}[t]
\vspace{0.0cm}
\centering
\hspace{0.4cm}
\resizebox{18.0cm}{!}{\rotatebox[origin=br]{0.0}{\includegraphics{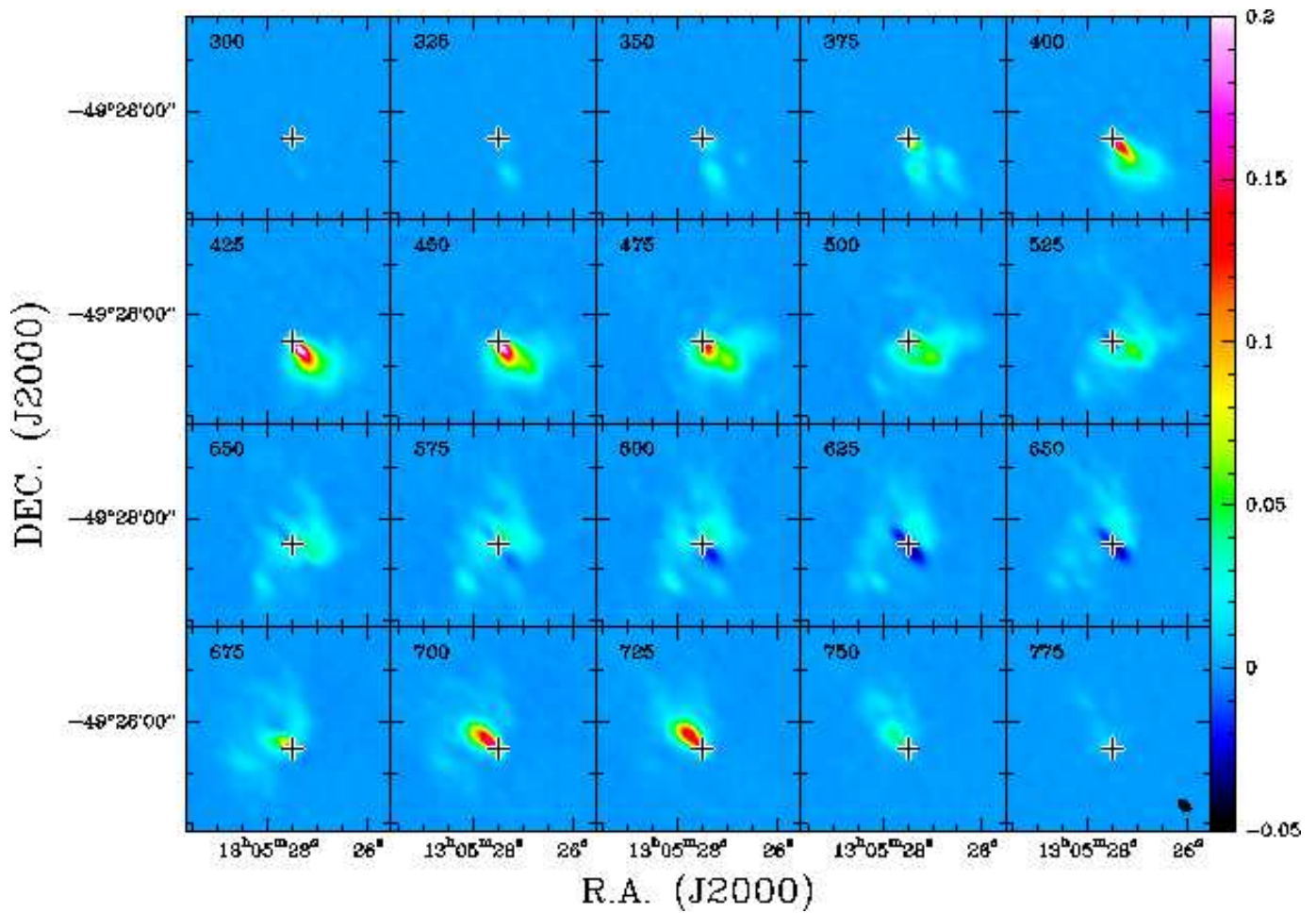}}}
\vspace{0.0cm}
\caption{H$^{12}$C$^{14}$N (HCN) $J$=1$\rightarrow$0 channel maps with the crosses 
indicating (as in case of Figs.~\ref{h13cn-channel}, \ref{ch3c2h-channel}, \ref{hc15n-channel}, 
\ref{cs-channel}, and \ref{c3h2-channel}) the position of the radio continuum peak. 
The spacing between tick-marks along the y-axis corresponds to 10$''$. Central barycentric 
velocities in km\,s$^{-1}$ are given in the upper left corners of each map. The systemic 
velocity is near 571\,km\,s$^{-1}$ (Sect.\,2). The wedge provides intensities in 
units of Jy\,beam$^{-1}$. Note the dark areas between 525 and 650\,km\,s$^{-1}$, where 
the line is seen in absorption. The large scale rotation of the highly inclined galaxy 
is such that emission at low velocities is mainly originating from the southwest of the 
center, while high velocity emission arises from the northeast. 0.1\,Jy correspond to 
$\approx$4.0\,K main beam brightness temperature. 1$''$ represents a projected 
linear scale of 19\,pc. The beam size is shown in the lower right corner of the figure.}
\label{hcn-channel}
\end{figure*}

Using the Very Long Baseline Array (VLBA) Greenhill et al. (1997) determined the position of the circumnuclear 
22\,GHz H$_2$O maser disk, obtaining $\alpha_{\rm B1950}$ = 13$^{\rm h}$ 02$^{\rm m}$ 32\ffs28 and $\delta_{\rm 
B1950}$ = --49$^{\circ}$ 12$'$ 01\ffas9 for the core of the curved line with a radius of $\approx$0\ffas025, 
formed by the maser hotspots. In the J2000 system this accretion disk is then centered at $\alpha_{\rm J2000}$ 
= 13$^{\rm h}$ 05$^{\rm m}$ 27\ffs48 and $\delta_{\rm J2000}$ = --49$^{\circ}$ 28$'$ 05\ffas6 and 
coincides within the limits of our resolution with the dynamical center (see Sect.\,4.6.1). In the 
following the position of the maser disk is therefore taken as synonymous for the dynamical center. The 
offset to our continuum peak is ($\Delta \alpha, \Delta \delta$) = (--0\ffas13, --0\ffas37), with estimated
uncertainties of 0\ffas1 in either measurement. This 2.8$\sigma$ discrepancy is right at the limit of our 
sensitivity, but supporting evidence for its reality is provided below, in Sects.\,4.1.2 and 4.6.1.

Beam deconvolved angular dimensions of the continuum source in NGC~4945 are 5\ffas83$\pm$0\ffas01 $\times$ 
1\ffas32$\pm$0\ffas01 (projected $\approx$110\,pc $\times$ 25\,pc) along the major and minor axes with a 
position angle of 41$^{\circ}\!\!$.4$\pm$0$^{\circ}\!\!$.1 (compare with Bendo et al. 2016 and, for 
$\lambda$ $\approx$ 13\,cm and 1.3\,mm, with Lenc \& Tingay 2009 and Chou et al. 2007, respectively).

\subsection{A brief summary of spectral line properties}

In this Sect.\,3, we provide information on spectral properties, referring exclusively 
to the ALMA data presented here. The basis for detected spatial features is provided 
by Fig.~\ref{hcn-hotspots}. Additional information related to this image and its positions
marked by white crosses is displayed in Table~\ref{tab-hcn}, which also provides relevant
radial velocities. For those mainly interested in the more general discussion of these 
data, also including results from other studies, we refer to Sect.\,4.

As expected, no spectral lines were found in the continuum 
spectral window and there are also no detected features close to those of HCN and CS
(cf. Sect.\,2). Table~\ref{tab-lines} summarizes the spectral features encountered in the 
three measured frequency bands containing molecular lines and coarsely characterizes 
their strength and line shape (i.e., exclusively emission or also absorption). 
The next subsections discuss estimates of the recovered flux for individual spectral 
features and provide a detailed description of the HCN $J$=1$\rightarrow$0 morphology. 
Finally, a short characterization of other lines is also presented. As a general 
rule, molecular lines from the nuclear region of NGC~4945 tend to show relatively 
strong emission at blueshifted velocities in the (barycentric) velocity interval $V$ $\approx$ 
400--475\,km\,s$^{-1}$ southwest of the dynamical center, less emission near the systemic 
velocity ($V_{\rm sys}$ $\approx$ 571\,km\,s$^{-1}$) and a second spectral peak 
of line emission near $V$ = 710\,km\,s$^{-1}$ northeast of the dynamical center. Also 
worth noting in this context is that the front side of the galactic plane is crossing 
the minor axis slightly southeast of the nucleus (e.g., Marconi et al. 2000).

\begin{figure*}[t]
\includegraphics[width=0.38\textwidth]{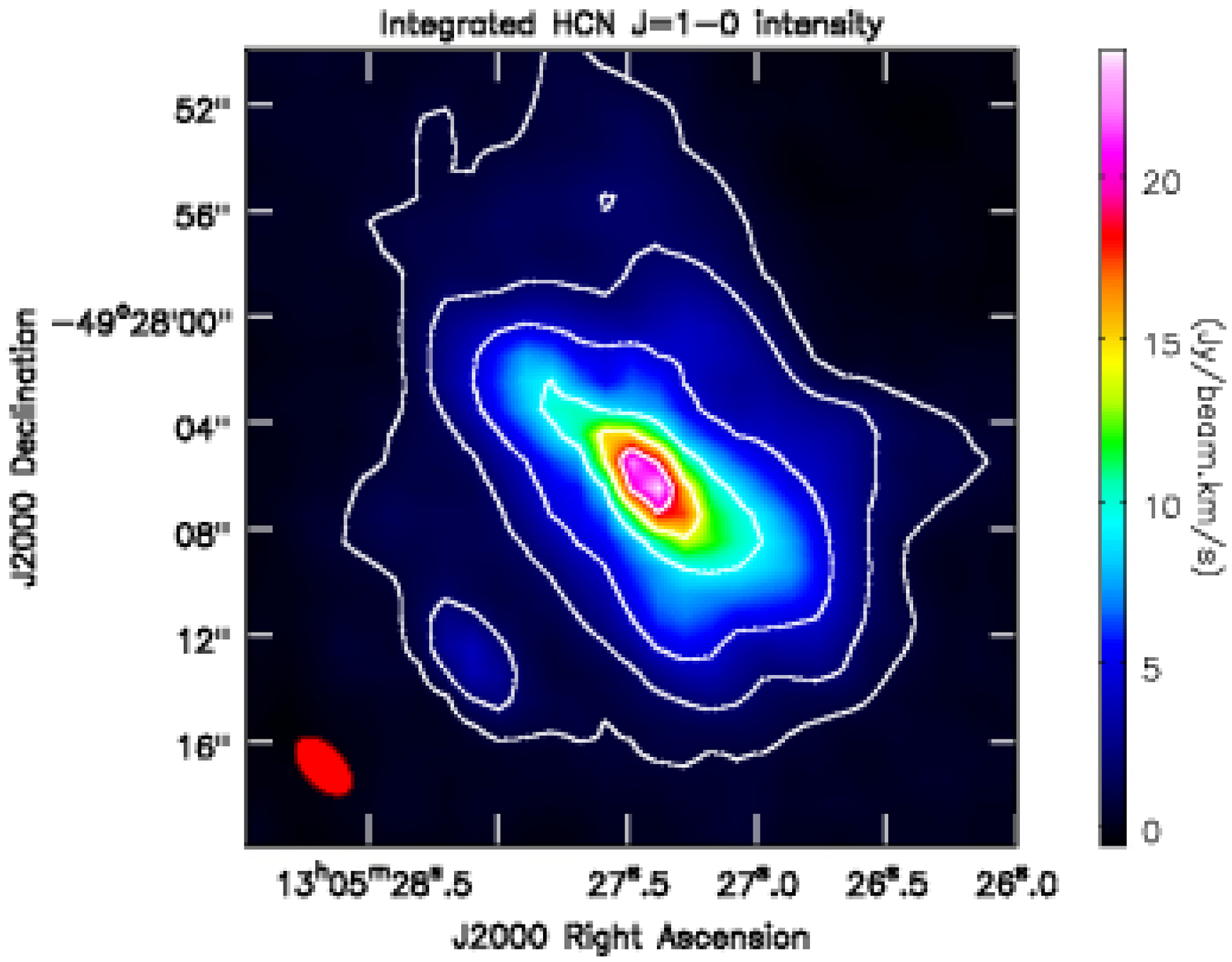}\hspace{1.5cm}\includegraphics[width=0.38\textwidth]{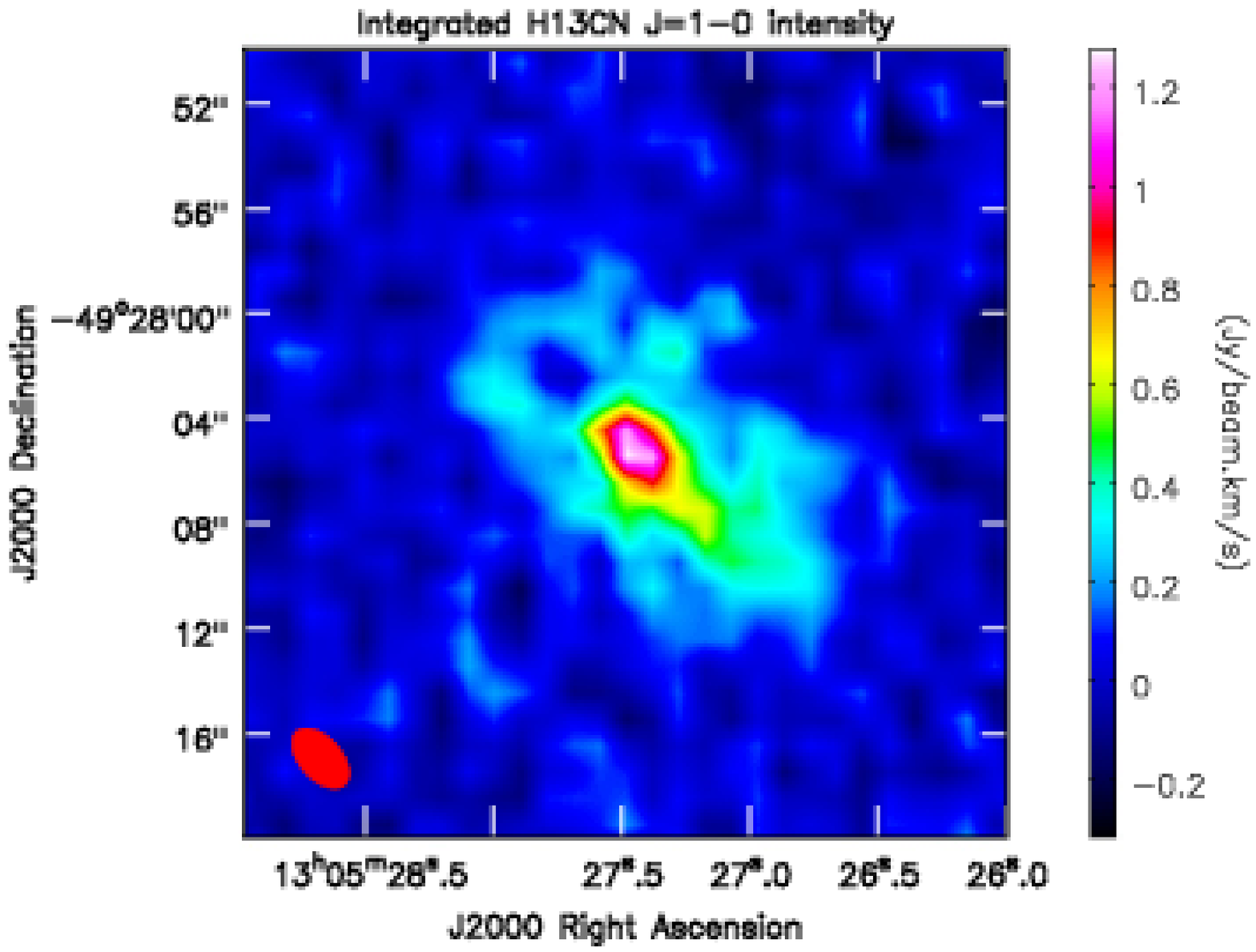}
\includegraphics[width=0.38\textwidth]{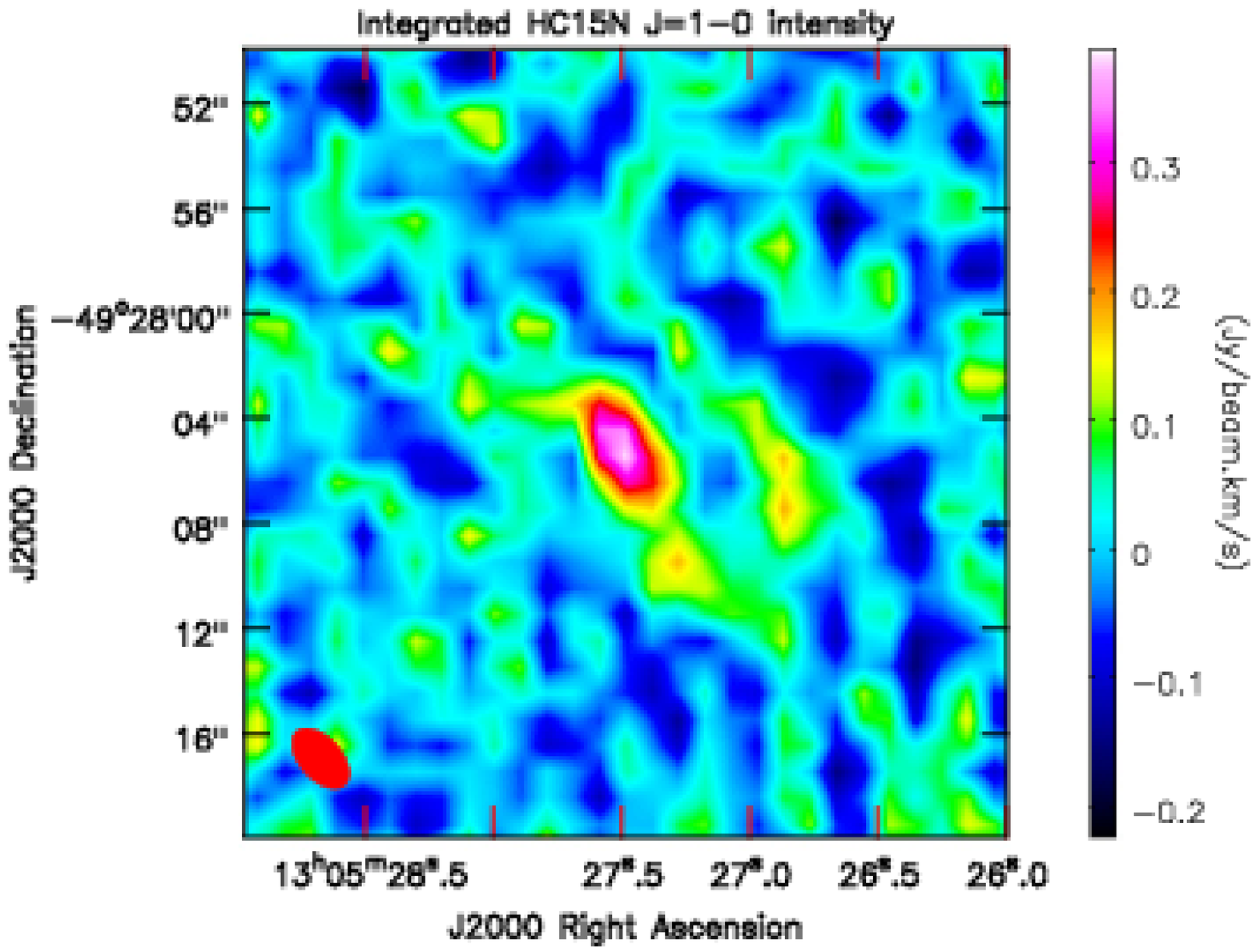}\hspace{1.5cm}\includegraphics[width=0.38\textwidth]{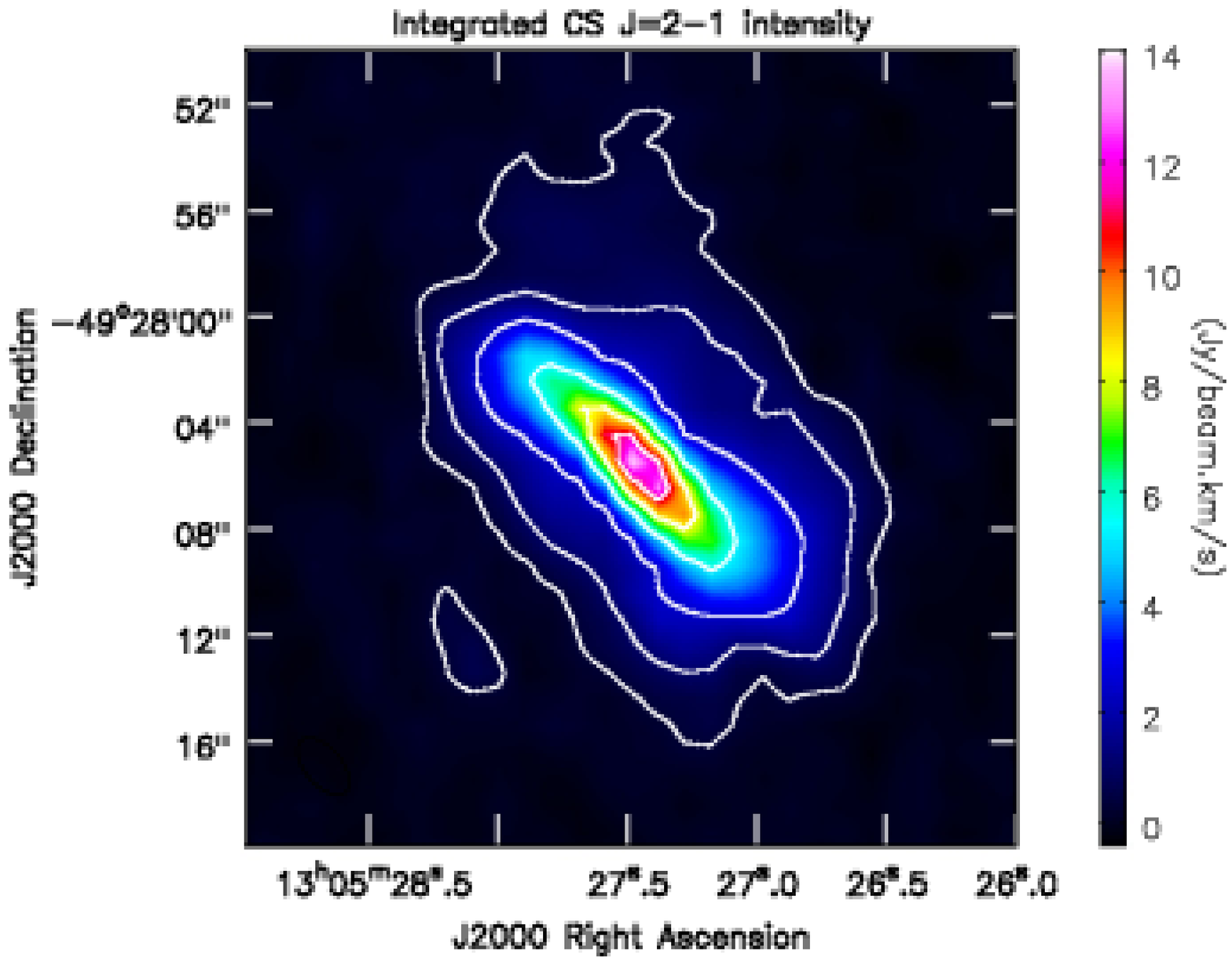}
\includegraphics[width=0.38\textwidth]{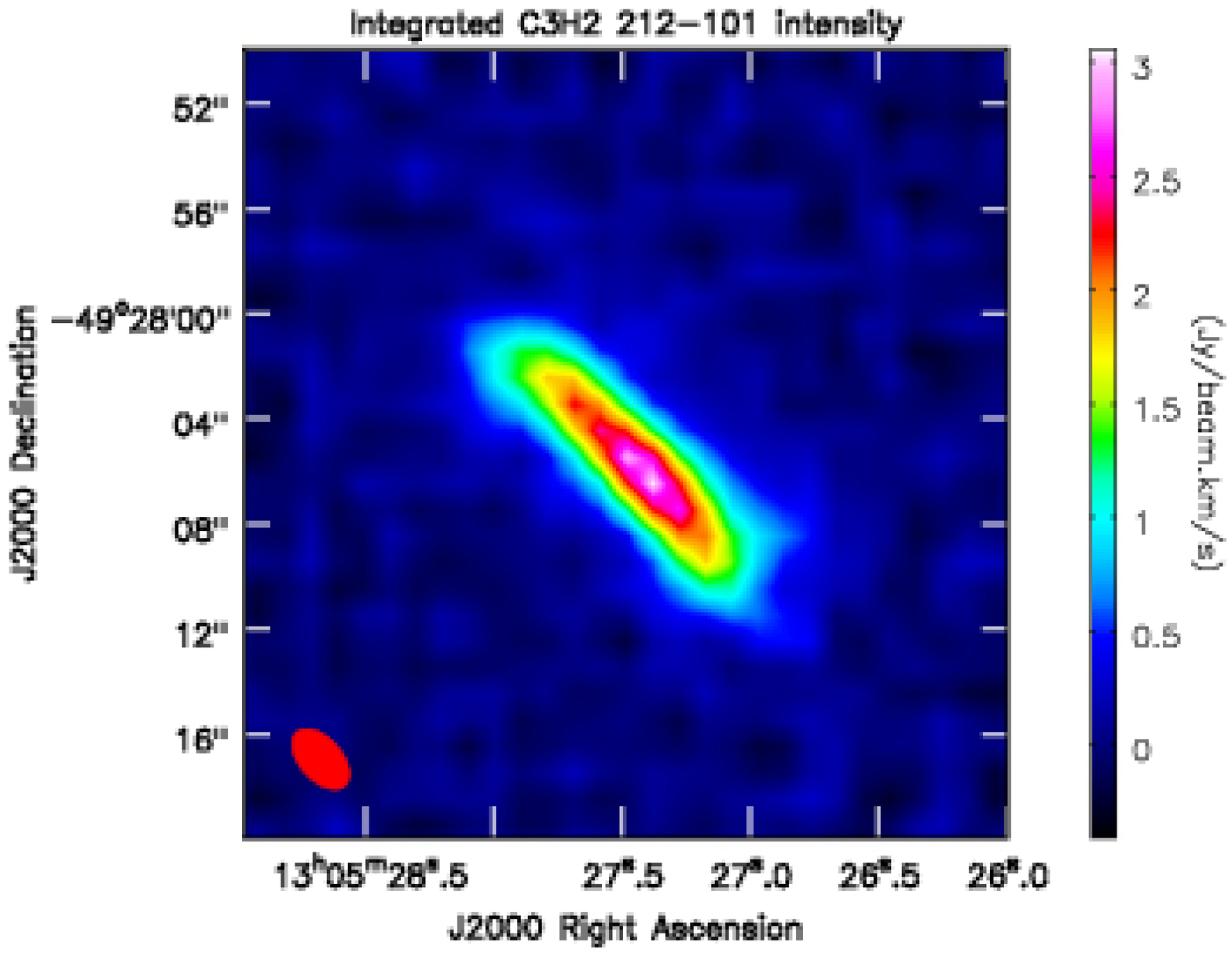}\hspace{1.5cm}\includegraphics[width=0.38\textwidth]{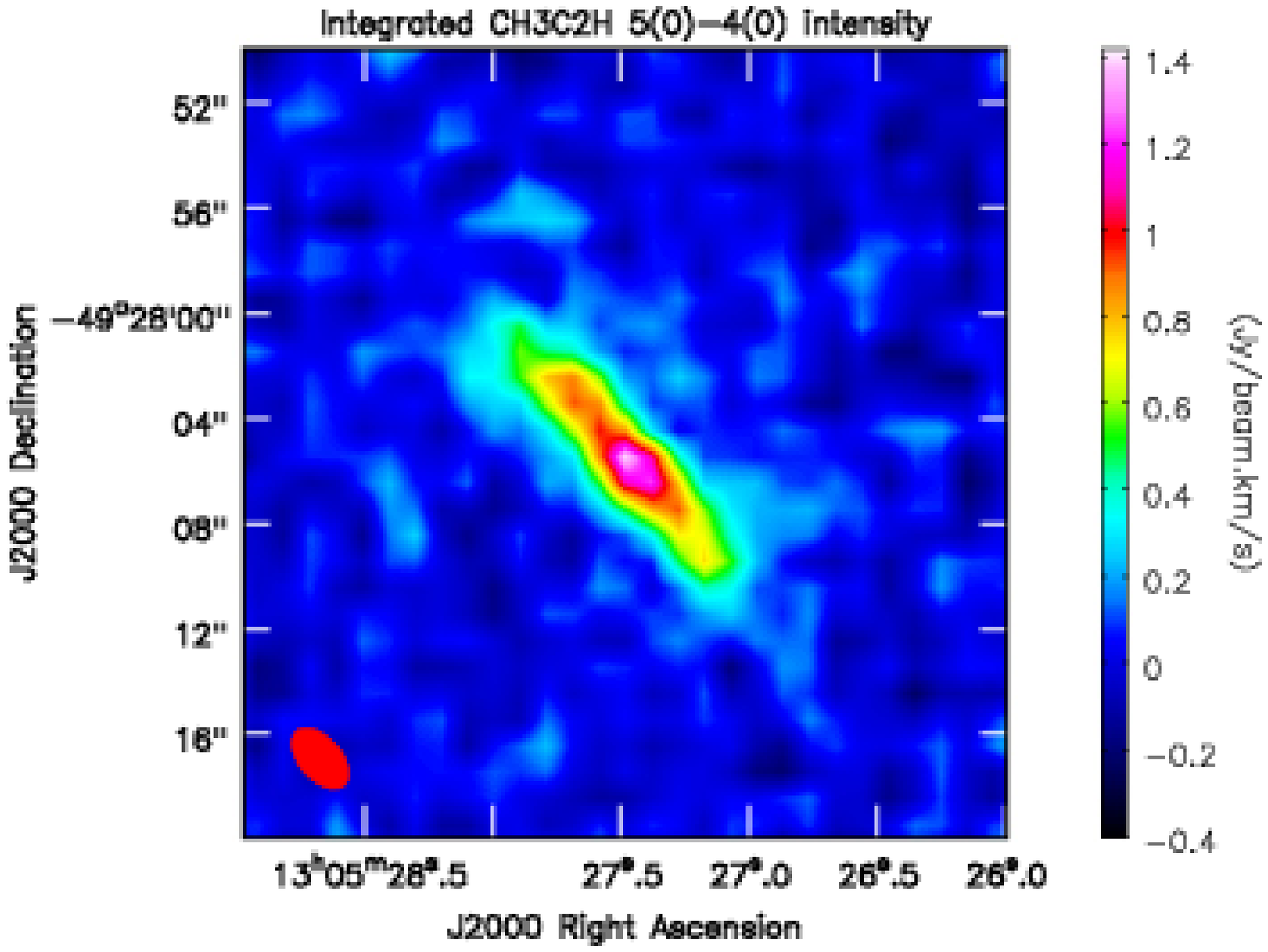}
\caption{Maps of primary beam corrected velocity integrated intensity (moment 0), obtained 
with a robustness parameter of 0.0 (Sect.\,2). Data reduced with a parameter of +2.0 
(close to natural weighting) do not reveal any additional features. 1$''$ corresponds to 
a projected linear scale of $\approx$19\,pc. {\it Upper left}: HCN $J$ = 1$\rightarrow$0 
emission with a restored beam of 2\ffas71 $\times$ 1\ffas56 and position angle PA = 43$^{\circ}$ 
(see the red spot in the lower left corner of the image). Contour levels: 5, 10, 20, 40, 60, 
and 80\% of the integrated peak flux density of 24.0\,Jy\,km\,s$^{-1}$\,beam$^{-1}$. {\it Upper 
right}: H$^{13}$CN $J$=1$\rightarrow$0 with a restored beam of 2\ffas72 $\times$ 1\ffas57, 
PA = 42$^{\circ}$. {\it Center, left}: HC$^{15}$N 1$\rightarrow$0 (only $V$ $>$ 400\,km\,s$^{-1}$) 
with 2\ffas72 $\times$ 1\ffas57, PA = 42$^{\circ}$. {\it Center right}: CS $J$ = 2$\rightarrow$1. 
Contours: 5, 10, 20, 40, 60, and 80\% of the integrated peak flux density of 14.1\,Jy\,km\,s$^{-1}$\,beam$^{-1}$. 
Restored beam: 2\ffas57 $\times$ 1\ffas38, PA = 41$^{\circ}$. {\it Lower left}: C$_3$H$_2$ 
2$_{12}$$\rightarrow$1$_{01}$ with 2\ffas72 $\times$ 1\ffas57, PA = 42.$^{\circ}$. {\it Lower 
right}: C$_3$HC$_2$H 5$_0$$\rightarrow$4$_0$. Restored beam: 2\ffas72 $\times$ 1\ffas57, PA = 
42$^{\circ}$. To convert Jy\,km\,s$^{-1}$\,beam$^{-1}$ into K\,km\,s$^{-1}$\,beam$^{-1}$, 
multiply by $\approx$37 (HCN), 39 (H$^{13}$CN, HC$^{15}$N, C$_3$H$_2$, and CH$_3$C$_2$H), or 36 (CS).} 
\label{maps-moment0}
\end{figure*}

\subsubsection{Recovered flux density}

A basic problem, when analyzing interferometric data is the amount of missing flux. 
Since our largest accessible angular scale is with $\approx$20$''$ at a 50\% intensity 
level (Sect.\,2) rather large, the amount of missing flux from the nuclear region 
should be either negligible or small (see, e.g., the single-dish maps of Dahlem et al. 
1993, Henkel et al. 1994, and Mauersberger et al. 1996). Table~\ref{tab-lines} (Cols.\,3 
and 4) provides peak flux densities derived from the 15-m SEST and those obtained by 
our measurements, for the blue- and red-shifted line peaks. It is apparent that at 
the peak velocities most of the flux has been collected by the ALMA observations. The 
only discrepancy, which may be significant, is that of CS, where the ALMA data suggest 
peak flux densities $\approx$30\% lower than those obtained with the SEST. However, even 
here the discrepancy does not reach the 3$\sigma$ level. 

Cols.\,5 and 6 of Table~\ref{tab-lines} display our integrated intensities from 2-dimensional
Gaussian fits. A comparison with SEST data obtained with a 1\,arcmin beam size indicates
that we have collected $\approx$30, 36, 44, and 50\% of the total SEST emission for CS, 
C$_3$H$_2$, H$^{13}$CN, and HCN. Uncertainties in these percentages range from $\approx$15\% 
to 35\% of the given values, depending on the strength of the respective line.

While our data apparently represent most of the flux at velocities near the line peaks,
only 30--50\% of the total velocity integrated flux density has been recovered. This can
be explained by the fact that the area covered by our ALMA data encompasses only $\approx$10\% 
of the SEST beam. Emission, mainly arising at near systemic velocities from the outer 
parts of the galaxy, is not covered by our data. As a consequence, we will assume in the 
following, that our calibration is correct within the previously estimated 10\% limits of 
uncertainty and that missing flux is not relevant within the central 10$^{''}$ (190\,pc) 
discussed below. Line ratios, discussed in Sect.\,4.3, may be even less affected as long 
as spatial distributions are similar.

\begin{figure}[t]
\vspace{0.0cm}
%\centering
\hspace{-0.0cm}
\resizebox{8.0cm}{!}{\rotatebox[origin=br]{0.0}{\includegraphics{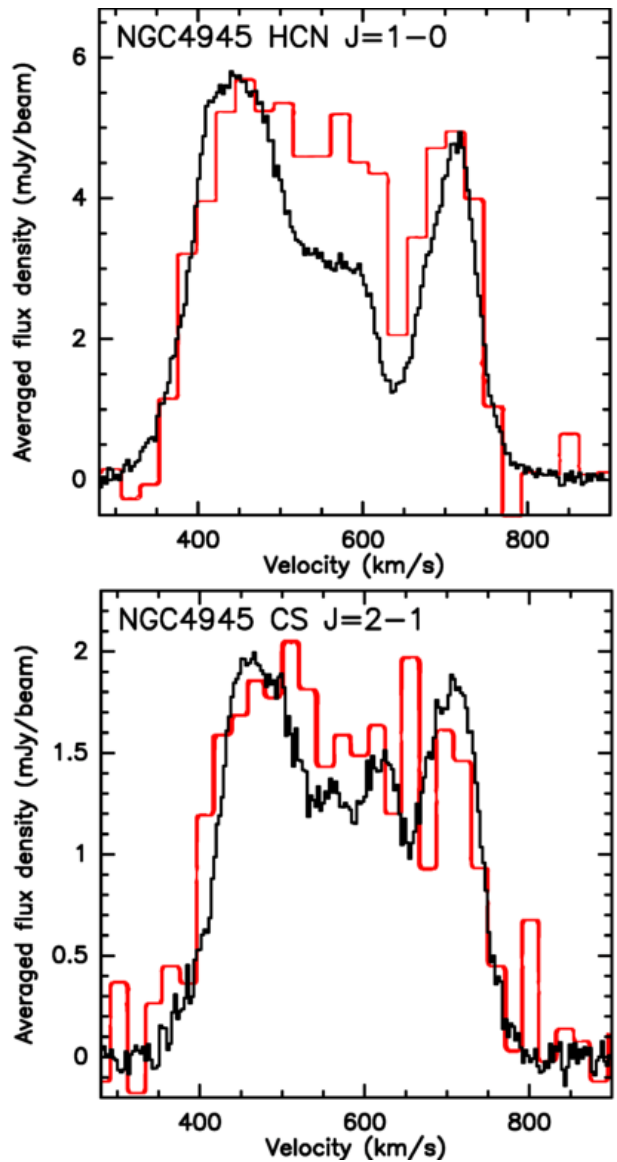}}}
\vspace{-0.0cm}
\caption{{\it Upper panel, black}: HCN $J$=1$\rightarrow$0 spectrum with a channel spacing of 
3.3\,km\,s$^{-1}$ from the central 0.25\,arcmin$^2$ of NGC~4945. The covered area is limited by 
$\alpha_{J2000}$ = 13$^{\rm h}$ 05$^{\rm m}$ 26$^{\rm s}$ and 13$^{\rm h}$ 05$^{\rm m}$ 
29$^{\rm s}$ and by $\delta_{\rm J2000}$ = --49$^{\circ}$ 28$'$ 50$''$ and --49$^{\circ}$ 
28$'$ 20$''$. The ordinate displays unweighted average values from pixels of the entire area 
(moment --1) as a function of barycentric velocity in units of mJy\,beam$^{-1}$. For spatially 
integrated flux densities, multiply by 200$\pm$20.  {\it Lower panel, black}: A CS 
$J$=2$\rightarrow$1 spectrum with a channel spacing of 3.0\,km\,s$^{_1}$ from the same region. 
Ordinate and abscissa: same units as for HCN. For spatially integrated flux densities, multiply 
by 240$\pm$24. Applied pixel size for both spectra: 1$''$$\times$1$''$. Overlayed spectra (red) 
with lower velocity resolution and larger beam size ($\approx$1$'$), matched to the amplitude 
of our spectra, are taken from Wang et al. (2004).}
\label{hcn+cs-profile}
\end{figure}

\subsubsection{The HCN $J$=1$\rightarrow$0 line}

As already mentioned, HCN $J$=1$\rightarrow$0 is the strongest spectral feature 
observed by us. Therefore it is used here as a reference for all other molecular 
lines. Table~\ref{tab-hcn} and Fig.~\ref{hcn-hotspots} display some basic information 
on detected spectral components. Fig.~\ref{hcn-channel} provides channel maps with a 
spacing of 25\,km\,s$^{-1}$, Fig.~\ref{maps-moment0} (upper left) shows a map of 
integrated HCN $J$=1$\rightarrow$0 emission, and Fig.~\ref{hcn+cs-profile} displays 
line profiles. 

The spectrum (Fig.~\ref{hcn+cs-profile}) has been obtained from an area of size 
30$''$ $\times$ 30$''$ ($\approx$570\,pc $\times$ 570\,pc). In comparison with the
SEST spectra (Henkel et al. 1990, 1994; Curran et al. 2001; Wang et al. 2004), the 
spectral intensity dip near $V$ = 630\,km\,s$^{-1}$ is more pronounced. This is 
readily explained by the large SEST beam, $\approx$1$'$, and the confinement of the 
HCN absorption (see below) to a small region and does not indicate substantial
amounts of missing flux near the systemic velocity in our data. The absorption component, 
discussed below, was already identified by Cunningham \& Whiteoak (2005) and by Green et al. 
(2016), using the Australia Telescope Compact Array (ATCA) with beam sizes of 5\ffas6 
$\times$ 3\ffas5 and 7$''$ $\times$ 7$''$, respectively. 

Most of the emission follows a narrow ridge, hereafter termed the nuclear disk, 
near position angle PA $\approx$ +45$^{\circ}$ east of north, with line peaks 
detected at low velocities slightly to the southwest and at high velocities 
slightly to the northeast of the center. At the outermost velocities, $V$ $\approx$ 
350\,km\,s$^{-1}$ (spatial Peak 2 in Fig.~\ref{hcn-hotspots} and Table~\ref{tab-hcn}) 
and 770\,km\,s$^{-1}$ (spatial Peak 11 in Fig.~\ref{hcn-hotspots} and Table~\ref{tab-hcn}), 
these peaks show an offset of 3\ffas5 in right ascension and 3\ffas5 in declination. 
The offset corresponds to $\approx$100\,pc. 

At more intermediate blue- and redshifted velocities with respect to the systemic
one, the nearly edge-on viewed (Sect\,4.5 and 4.6.2) nuclear disk becomes visible, 
directed from the center toward the southwest with its approaching and toward the 
northeast with its receding part. Combined, the morphology of the ridge emission 
resembles to some degree the mm-wave continuum distribution displayed in Fig.~\ref{continuum}, 
but the line emission is with $\approx$10$''$ (190\,pc) more extended. 

\begin{figure}[t]
\vspace{0.0cm}
%\centering
\hspace{0.3cm}
\resizebox{16.0cm}{!}{\rotatebox[origin=br]{-90.0}{\includegraphics{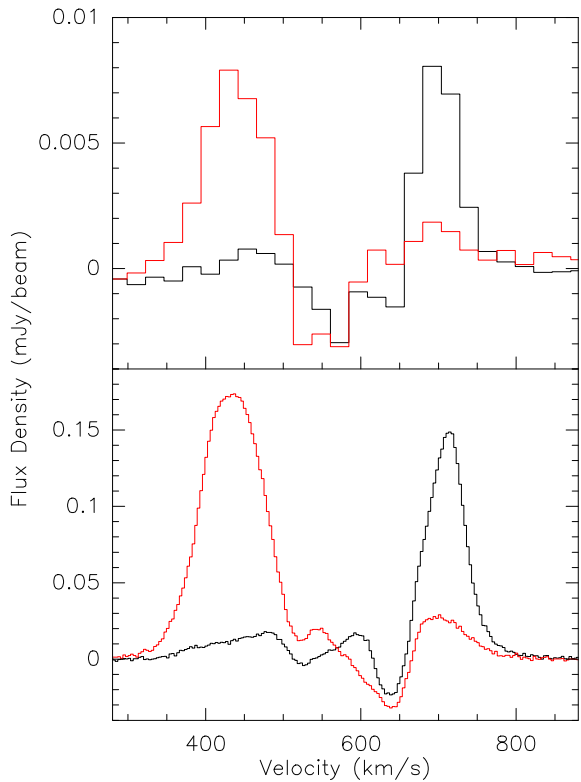}}}
\vspace{-0.8cm}
\caption{{\it Lower panel}: HCN $J$ = 1$\rightarrow$0 spectra with a channel spacing of 
3.3\,km\,s$^{-1}$. Black: spectrum from the northeastern region (peak near $V$ = 
710\,km\,s$^{-1}$) showing absorption. Red: spectrum from the southwestern region  
(peak near $V$ = 430\,km\,s$^{-1}$) showing absorption. The separation between
the two regions is the line perpendicular to the nuclear disk crossing the peak 
of the radio continuum emission (see Figs.~\ref{hcn-hotspots} and \ref{hcn-channel}).
{\it Upper panel}: The same for H$^{13}$CN $J$ = 1$\rightarrow$0, but with a channel 
width of 24\,km\,s$^{-1}$. The ordinate provides average flux densities per beam
over the covered area. The galaxy's systemic velocity is close to $V$ = 571\,km\,s$^{-1}$}.
\label{hcn-absorption}
\end{figure}

As already mentioned, the trough in the spectrum around $V$ = 630\,km\,s$^{-1}$ 
(Figs.~\ref{hcn-channel} and \ref{hcn+cs-profile}) is caused by absorption. Absorption 
is seen toward the center and slightly northeast over the widest velocity range, $V$ 
$\approx$ 510 -- 565\,km\,s$^{-1}$ and 575 -- 660\,km\,s$^{-1}$ (Fig.~\ref{hcn-channel}). 
Toward the southwest, absorption is also seen, but only between 600 and 660\,km\,s$^{-1}$. 
Between 620 and 660\,km\,s$^{-1}$, both absorption components are spatially connected and 
reveal together a morphology that closely resembles that of the mm-wave continuum 
(Fig.~\ref{continuum}). Overall, most of the absorption is seen at velocities redshifted 
with respect to the systemic one (see Figs.~\ref{hcn-channel} and \ref{hcn-absorption} and the 
discussion in Sect.\,4.7). Absolute HCN $J$ = 1$\rightarrow$0 absorption line intensities 
reach 50\% of the peak continuum flux density (Sects.~3.1 and 3.3). 

No absorption is seen beyond the extent of the central continuum source. However, 
emission is also observed well outside the ridge (Table~\ref{tab-hcn} and 
Figs.~\ref{hcn-hotspots} and \ref{hcn-channel}). To these spectral features belongs 
Peak 1 at $V$ $<$ 400\,km\,s$^{-1}$, while Peak 3, also seen at low velocities, delineates 
the southwestern edge of the nuclear disk well displaced from the very center (for details, 
see Table~\ref{tab-hcn}). Already slightly below $V$ = 500\,km\,s$^{-1}$, weakly emitting 
clouds appear to form a secondary central ridge, approximately perpendicular to the position 
angle of the main body of the galaxy. At $V$ $\sim$500 -- 700\,km\,s$^{-1}$, arms can be seen at 
the ends of this secondary ridge, $\approx$10$''$ (projected 190\,pc) from the center, 
directed toward the northeast from its northwestern edge, and toward the southwest from its 
southeastern edge (Fig.~\ref{hcn-channel} and, for further discussion, Sect.\,4.7).

\subsubsection{Rare HCN isotopologues}

\begin{figure*}[t]
\vspace{0.0cm}
\centering
\hspace{0.3cm}
\resizebox{18.4cm}{!}{\rotatebox[origin=br]{0.0}{\includegraphics{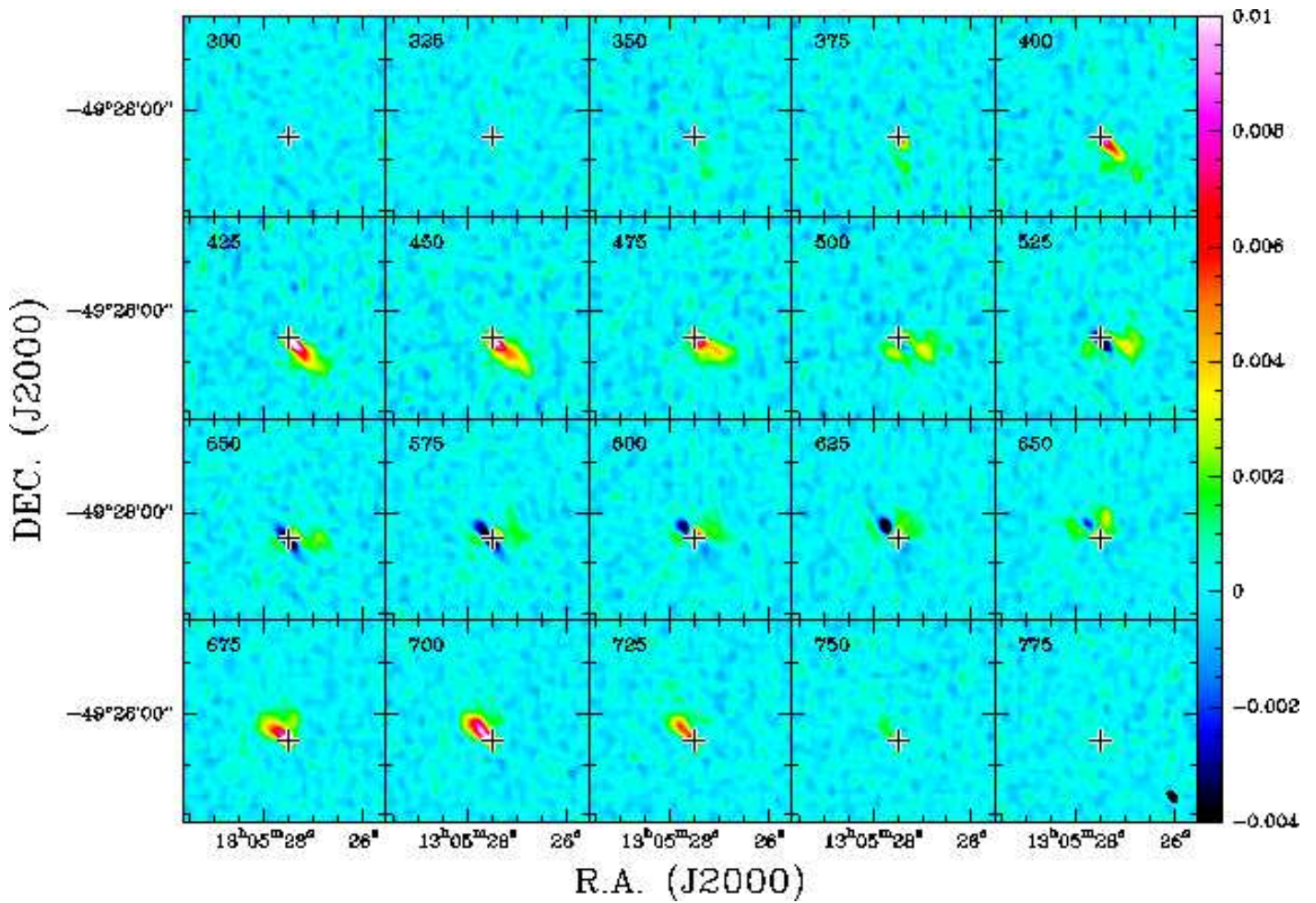}}}
\vspace{0.0cm}
\caption{H$^{13}$C$^{14}$N (H$^{13}$CN) $J$ = 1$\rightarrow$0 channel maps. For the units
of velocity, the ordinate, and the wedge as well as for the position of the crosses, see 
Fig.~\ref{hcn-channel}. As in case of the main isotopic species, absorption is present near 
the systemic velocities between 525 and 650\,km\,s$^{-1}$.  0.01\,Jy correspond to $\approx$0.42\,K.
The beam size is shown in the lower right corner of the figure.}
\label{h13cn-channel}
\end{figure*}

The H$^{13}$CN $J$ = 1$\rightarrow$0 line (see Fig.~\ref{maps-moment0}, upper right, 
and Fig.~\ref{h13cn-channel}) shows, with smaller amplitudes and lower signal-to-noise ratios, 
a similar behavior as HCN $J$ = 1$\rightarrow$0. Similar to HCN, absorption is apparent 
in the channels between $V$ = 525 and 650\,km\,s$^{-1}$. Again, we find more absorption 
at velocities slightly redshifted relative to the systemic one, but the clear asymmetry 
apparent in the HCN $J$ = 1$\rightarrow$0 line is here only seen in case of the redshifted
northeastern part of the nuclear disk (Fig.~\ref{hcn-absorption} and Sect.\,4.7 for
further discussion). As for the main isotopologue, the absorption is confined to the 
extent of the nuclear continuum source, while the surrounding gas exhibits emission. 

\begin{figure*}[t]
\vspace{-1.2cm}
%\centering
\hspace{-0.0cm}
\resizebox{19.3cm}{!}{\rotatebox[origin=br]{-90.0}{\includegraphics{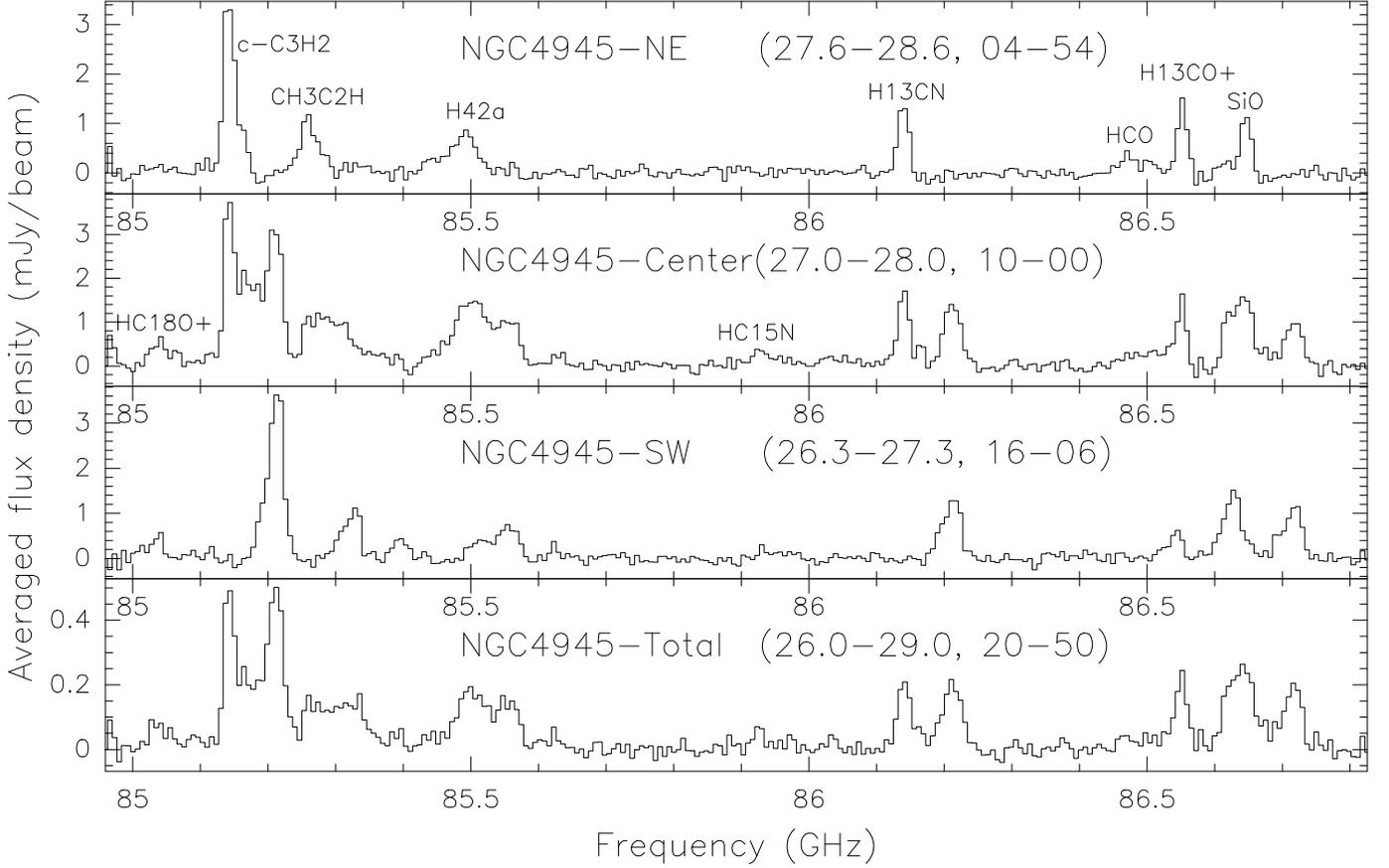}}}
\vspace{-0.3cm}
\caption{Spectra from the broad HC$^{15}$N window with seven contiguous channels smoothed, 
yielding a channel spacing of $\approx$23\,km\,s$^{-1}$. The molecular species related to detected 
spectral features are given in the two top panels and are described in more detail in 
Table~\ref{tab-lines}. For H42$\alpha$, see Bendo et al. (2016). Because almost 2\,GHz are 
covered, leading to a complex correlation between velocity and frequency, amplitudes are 
presented as a function of observed frequency. The lowest panel presents a spectrum from the 
central 0.25\,arcmin$^2$, similar to the spectra shown in Fig.~\ref{hcn+cs-profile}.
The three upper panels show spectra from smaller areas, visualized by dashed white lines 
in Fig.~\ref{hcn-hotspots}, and include (from top to bottom) regions to the northeast of the 
center, close to the nucleus, and to the southwest of the center. The four numbers in parenthesis 
on the right hand side of each of these designations provide the eastern and western edge (in 
time seconds of Right Ascension) and the southern and northern edge (in arcseconds of Declination) 
of the chosen area (compare with the moment 0 maps of Figs.~\ref{hcn-hotspots} and \ref{maps-moment0}). 
The ordinate displays unweighted average flux densities per beam size for all pixels in the given areas, 
so that the values for the largest region (lowest panel), encompassing most areas with weak 
(or even absent) emission, are smallest. Multiply by 40$\pm$4 (upper three panels) and 250$\pm$25 
to obtain spatially integrated flux densities. Applied pixel size: 1$''$$\times$1$''$.} 
\label{hc15n-profile}
\end{figure*}

Fig.~\ref{maps-moment0} and, in the Appendix, Fig.~\ref{hc15n-channel} show that we have 
also detected the $^{15}$N bearing isotopologue. While signal-to-noise ratios are low, 
HC$^{15}$N is detected in the low velocity channels and then again at $V$ $\approx$ 710\,km\,s$^{-1}$, 
displaced by a few arcseconds to the northeast, following the overall kinematics of the central 
region (Sect.\,3.2.1). Therefore, HC$^{15}$N is seen in those parts of the spectrum where the 
more abundant isotopologues show relatively strong emission. In the velocity range, where HCN 
and H$^{13}$CN are affected by weak absorption, no HC$^{15}$N signal is seen, consistent with the 
overall weakness of HCN and H$^{13}$CN in this velocity interval. Fig.~\ref{hc15n-profile} 
also shows this effect. While the line is seen in the ``total'' and central spectrum 
and tentatively also in the southwestern one, the northeastern spectrum from the 
receding part of the galaxy does not show any significant peak. The spectra of the 
more abundant HCN isotopologues are contaminated by absorption mostly in this part of the 
spectrum and it is very likely that this also holds in the case of HC$^{15}$N.

\begin{figure*}[t]
\vspace{0.0cm}
\centering
\hspace{0.3cm}
\resizebox{18.4cm}{!}{\rotatebox[origin=br]{0.0}{\includegraphics{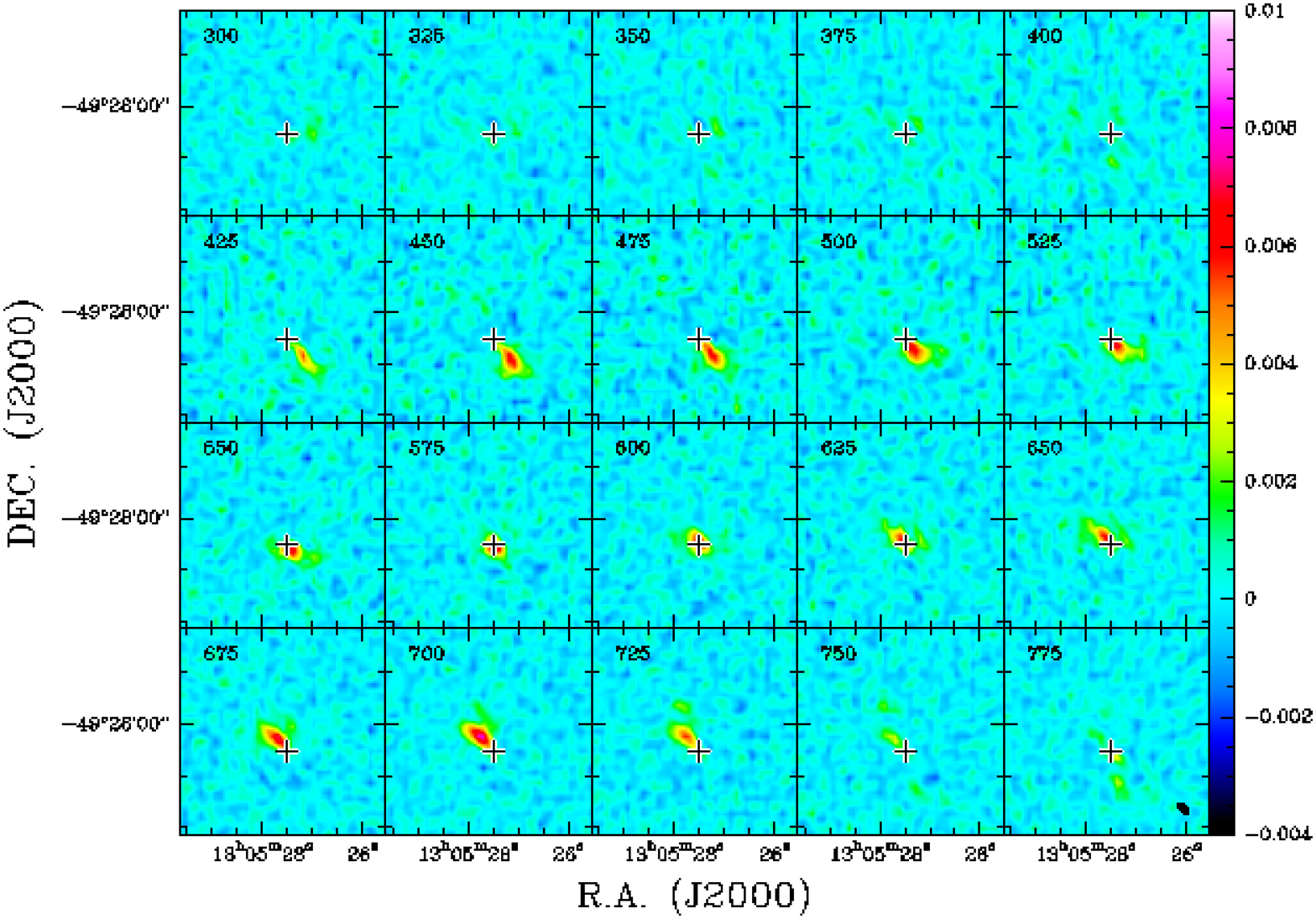}}}
\vspace{0.0cm}
\caption{CH$_3$C$_2$H 5$_0$ $\rightarrow$ 4$_0$ channel maps. For the units of velocity, ordinate, 
and the wedge as well as the position of the crosses, see Fig.~\ref{hcn-channel}. 
0.01\,Jy correspond to $\approx$0.42\,K. The beam size is shown in the lower right 
corner of the figure.}
\label{ch3c2h-channel}
\end{figure*}

\subsubsection{The CS $J$ = 2$\rightarrow$1 line}

After HCN $J$ = 1$\rightarrow$0, CS $J$ = 2$\rightarrow$1 is the strongest molecular transition 
in our spectral windows. With respect to the velocity integrated HCN line, the integrated CS 
emission (Figs.~\ref{maps-moment0}, \ref{hcn+cs-profile} and, for the channel maps, 
Fig.~\ref{cs-channel} in the Appendix) shows overall a similar spatial distribution 
(Fig.~\ref{maps-moment0}). However, being weaker than the HCN line and with all lines 
being measured with approximately the same sensitivity (Sect.\,2), the emission outside the nuclear 
ridge is near the detection threshold. Therefore, the observed ridge appears to be narrower. A 
2-dimensional Gaussian fit reveals for CS $J$ = 2$\rightarrow$1 a beam deconvolved FWHP size 
of 8\ffas8\,$\pm$\,0\ffas7 $\times$ 1\ffas9\,$\pm$\,0\ffas1 at a PA = +43\ffcirc2\,$\pm$\,1\ffcirc1. 
Obviously, the emission from the nuclear ridge or disk is, as in the case of HCN $J$ = 1$\rightarrow$0, 
more extended than that of the continuum (Sect.\,3.1). 

Because of the morphological similarity between the $\lambda$ $\approx$ 3\,mm radio continuum 
and the integrated CS $J$ = 2$\rightarrow$1 line according to Figs.~\ref{continuum} and 
\ref{maps-moment0}, we might expect that CS is even more affected by the continuum radiation 
than HCN and its weaker isotopologues. However, we find the opposite. While CS line intensities 
near the systemic velocity are also weaker than in the line wings near $V$ = 430 and 
710\,km\,s$^{-1}$ (Figs.~\ref{hcn+cs-profile} and \ref{cs-channel}), the dip near the systemic 
velocity only amounts to about one third of the peak flux density near 430\,km\,s$^{-1}$ (and 
not to more than two thirds), indicating a much smaller influence of the underlying nuclear 
3\,mm radio continuum emission. Inspection of individual channels confirms this view.

\subsection{Other lines}

Cyclic C$_3$H$_2$ (c-C$_3$H$_2$) 2$_{12}\rightarrow$1$_{01}$ is our third strongest 
line. Its emission is displayed in Figs.~\ref{maps-moment0}, \ref{hc15n-profile}, and 
\ref{c3h2-channel}. No strong emission is seen outside the nuclear disk in the moment 
0 map. However, in the channel maps of the Appendix, the outer arms are still weakly 
present.  With respect to absorption, the line shows properties, which are intermediate 
between those of HCN $J$=1$\rightarrow$0 and CS $J$=2$\rightarrow$1. c-C$_3$H$_2$ 
2$_{12}$$\rightarrow$1$_{01}$ does not exhibit absorption over the entire extent of the 
continuum source. We find instead weak absorption northeast of the center at those 
velocities, where HCN shows more widespread absorption. A comparison of the overall spectra 
(lower panels of Figs.~\ref{hcn+cs-profile} and \ref{hc15n-profile}) and individual 
channels also show that the deficit in emission (with respect to the line wings) at near 
systemic velocities ($\approx$520 -- 620\,km\,s$^{-1}$) is larger than that of CS but 
smaller than that of HCN. 

There are still a few weaker molecular spectral features to be mentioned (see 
Fig.~\ref{hc15n-profile}, where the top three panels (from the areas delineated by
dashed white lines in Fig.~\ref{hcn-hotspots}) demonstrate the shift in radial 
velocity, when moving along the plane of the galaxy from the southwest to the northeast 
with rising velocities). As is apparent from the channel maps (Fig.~\ref{ch3c2h-channel}), 
the CH$_3$C$_2$H 5$_0$$\rightarrow$4$_0$ transition shows no indication for absorption. 
Unlike CS $J$ = 2$\rightarrow$1, which still shows a dip in the overall emission spectrum 
near the systemic velocity (Fig.~\ref{hcn+cs-profile}), the CH$_3$C$_2$H profile is flat-topped 
and a detailed inspection of individual channels shows no pixel with negative flux densities. 

While HCO is too weak for a detailed analysis, both SiO $J$=2$\rightarrow$1 and 
H$^{13}$C$^{16}$O$^+$ $J$=1$\rightarrow$0 (hereafter H$^{13}$CO$^+$) show strong absorption.
Because of heavy line blending and because this will not play a major role in the discussion
(Sect.\,4), the following is not displayed in the form of figures but is only briefly noted:
Selecting the most extreme pixels, for SiO and HCO$^+$ isotopologues absolute flux density 
levels for absorption in 23\,km\,s$^{-1}$ channels reach values similar to the emission peaks. 
For SiO, these levels are $\pm$9\,mJy\,beam$^{-1}$; for H$^{13}$CO$^+$ we find $\pm$14\,mJy\,beam$^{-1}$. 
The latter is corroborated, although with lower signal-to-noise ratios, by the H$^{12}$C$^{18}$O$^+$ 
$J$ = 1$\rightarrow$0 line near 85\,GHz (hereafter HC$^{18}$O$^+$; see Fig.~\ref{hc15n-profile}), 
where the values become about $\pm$5\,mJy\,beam$^{-1}$. This is more extreme than in the cases 
of HCN and H$^{13}$CN $J$ = 1$\rightarrow$0, where we find (again for the most extreme pixels) 
absorption levels down to --48 and --9\,mJy\,beam$^{-1}$, but peak emission at 235 and 
14\,mJy\,beam$^{-1}$ in 3.3 and 23\,km\,s$^{-1}$ wide channels, respectively. For completeness, 
the corresponding values for CS $J$ = 2$\rightarrow$1 (3.0\,km\,s$^{-1}$ channels), c-C$_3$H$_2$ 
2$_{12}$$\rightarrow$1$_{01}$ and CH$_3$C$_2$H 5$_0$$\rightarrow$4$_0$ ($\approx$23\,km\,s$^{-1}$ 
channels) are 95 versus --2\,mJy\,beam$^{-1}$, 30 versus --2\,mJy\,beam$^{-1}$, and 9\,mJy\,beam$^{-1}$ 
versus no pixel with significant negative flux density.

\section{Discussion}

\subsection{Spatial morphology}

\subsubsection{The nuclear core}

The molecular emission observed by us near the center originates primarily from a highly 
inclined rotating nuclear disk (Sects.\,3.2 and 3.3). If it were a ring as suggested by 
Bergmann et al. (1992) or Wang et al. (2004), we should be able to define an inner radius. 
Like Bendo et al. (2016) for the H42$\alpha$ recombination line, however, we see no evidence 
for a central hole in our data, suggesting that any such hole in the disk has a size smaller 
than the synthesized beam, which measures 2.6" ($\approx$50\,pc) along the major axis of 
the disk.  This is consistent with Chou et al. (2007), who could also not identify such 
a central void in their CO and $^{13}$CO $J$ = 2$\rightarrow$1 maps. Note, however, that 
their data were obtained with a synthesized beam size of 5\ffas1 $\times$ 2\ffas8. Our 
measurements have higher resolution and may trace higher density gas. 

Instead of a void we find a central region with strongly enhanced integrated emission. 
This region is not larger than our beam size and may still be unresolved. It encompasses 
only $\la$40\,pc, which agrees with the less tight upper size limit obtained by Chou et al. 
(2007), $\approx$55\,pc, and is seen in all our maps of integrated intensity (Fig.~\ref{maps-moment0}). 
The cold X-ray reflector (200\,pc $\times$100\,pc, with 50\% of the X-ray emission arising 
from the innermost 30\,pc) reported by Marinucci et al. (2012) agrees well with the sizes 
of our dense and dusty nuclear disk and the unresolved central core. The feature with a steep 
velocity gradient identified in CO $J$ = 2$\rightarrow$1 by Lin et al. (2011; their fig.\,4) 
may also arise from this core. Only the c-C$_3$H$_2$ 2$_{12}$$\rightarrow$1$_{01}$ 
line (Fig.~\ref{maps-moment0}) shows a peak which is more extended than our beam size 
along the minor axis of the galaxy, possibly because it is tracing more diffuse 
gas (e.g., Thaddeus et al. 1985) in spite of its rather high critical density (see Sect.\,4.2 
for details) or because it is enhanced by photon dominated regions (for the chemistry,
see, e.g., Aladro et al. 2011) associated with the starburst ring proposed by Marconi et 
al. (2000; see below). The central $\la$40\,pc appear to host dense gas with particularly 
high column densities. While there is clearly molecular gas associated with the starburst 
ring of radius $\approx$2\ffas5 ($\approx$50\,pc), proposed by Marconi et al. (2000) on
the basis of near infrared Pa$\alpha$ data, this putative ring is not associated with peaks 
of molecular emission (with c-C$_3$H$_2$ as the possible exception; see Fig.~\ref{maps-moment0}). 
In this context we should note that the proposed Pa$\alpha$ ring is actually a sequence of line 
peaks along the main axis of the galaxy with similar extent as the nuclear disk observed by us
(see Marconi et al.'s 2000 fig.~2, upper and middle left panels).

For a better understanding of the central region, the radius of the sphere of influence of 
the nuclear engine is also of interest. With 
$$
 r_{\rm G}  = 0.11 \left(\frac{M_{\rm SMBH}}{10^8\,{\rm M_{\odot}}}\right)  
              \left(\frac{200\,{\rm km\,s^{-1}}}{\sigma_*}\right)^2 
              \left(\frac{20\,{\rm Mpc}}{D}\right) \,\, {\rm arcsec}
$$
(e.g., Barth 2003), the mass of the nuclear engine $M_{\rm SMBH}$ = (1--2) $\times$ 10$^6$\,M$_{\odot}$ 
(Greenhill et al. 1997), distance $D$ = 3.8\,Mpc (e.g., Karachentsev et al. 2007; Mould \& Sakai
2008), and adopting $\sigma_*$ = 120\,km\,s$^{-1}$ as the stellar velocity dispersion (Oliva et al. 1995), 
we obtain $r_{\rm G}$ $\approx$ 0\ffas016--0\ffas032 or 0.3--0.6\,pc. Obviously, this scale is far smaller than 
our resolution and resembles that of the size of the H$_2$O megamaser disk (Greenhill et al. 1997).

\begin{figure}[t]
\vspace{-0.0cm}
%\centering
\hspace{-0.0cm}
\resizebox{12.0cm}{!}{\rotatebox[origin=br]{-90.0}{\includegraphics{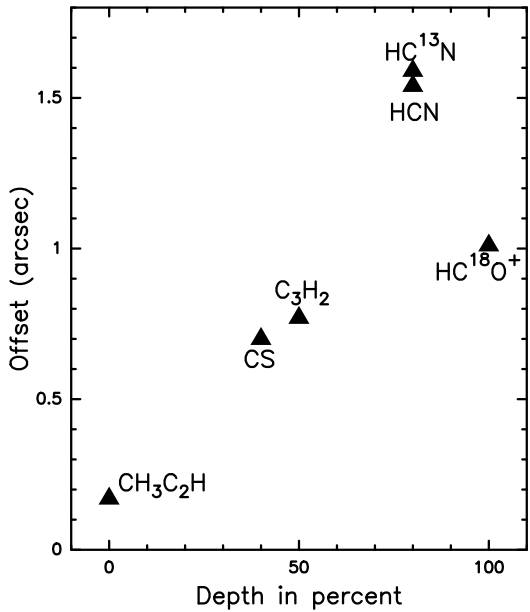}}}
\vspace{0.0cm}
\caption{Angular offsets (along the main axis with position angle 225$^{\circ}$ toward the southwest) 
of molecular versus continuum centroid position (obtained from two-dimensional Gaussian fits) as a 
function of depth of the spectral dip near the systemic velocity in percent of the line peaks at 
$V$ $\approx$ 430\,km\,s$^{-1}$. The percentages given refer to a covered area of 0.25\,arcmin$^2$ 
as described in Fig.~\ref{hcn+cs-profile}. Molecular species are indicated. For details, also 
including line assignments, formal uncertainties and rest frequencies, see Sect.\,4.1.2 
and Table~\ref{tab-lines}). }
\label{centerposition}
\end{figure}

As already mentioned, the centrally peaked molecular line intensity distributions 
(Fig.~\ref{maps-moment0}) contradict the previously proposed molecular ring structure 
originally discussed by Bergman et al. (1992; see also fig.~6 in Wang et al. 2004) on the 
basis of single-dish $^{12}$CO/$^{13}$CO line intensity ratios as a function of velocity. 
The same also holds for the molecular ring with an inner radius of 200\,pc, reported on 
the basis of single-dish CO $J$=1$\rightarrow$0 and 2$\rightarrow$1 observations by 
Dahlem et al. (1993). What we clearly see in our denser gas tracers are peaks in molecular 
line emission near $V$ = 430 and 710\,km\,s$^{-1}$, potentially forming a circumnuclear 
ring at radius $\approx$5$''$ (95\,pc) which encloses the nuclear disk (see also  
Sect.\,4.7). The nuclear disk itself extending from this radius toward the unresolved 
nuclear core provides a rich molecular environment, giving rise to strong emission 
from a multitude of molecular high density tracers.

\subsubsection{Peak positions}

Table~\ref{tab-positions} provides the coordinates of some relevant peak positions,
i.e. those from the continuum and from the redshifted, systemic, and blueshifted
emission of the CH$_3$C$_2$H line, which appears not to be affected by absorption. 
Table~\ref{tab-integral} presents the central positions of our unblended lines, obtained 
by two-dimensional Gaussian fits using the CASA-viewer. The scatter in these positions 
encompasses $\pm$0\ffs07 in right ascension and $\pm$0\ffas4 in declination and 
the weakest line, HC$^{15}$N, is not an outlier. We do not see large differences 
between the peak positions of the different lines. However, there appears to be 
a small offset relative to the peak of the continuum emission. For the seven transitions 
displayed in Table~\ref{tab-integral}, we obtain an average offset of ($\Delta 
\alpha, \Delta \delta$) = (--0\ffas60 $\pm$ 0\ffas49, --0\ffas60 $\pm$ 0\ffas33). 
Excluding HC$^{15}$N, average and sample standard deviations change 
insignificantly to (--0\ffas68 $\pm$ 0\ffas49, --0\ffas68 $\pm$ --0\ffas32). 
While the standard deviations of the mean seem to imply that the offset is not 
significant, we note that {\it all} seven molecular peaks lie slightly to the 
south-west of the maximum of the continuum emission, apparently following the 
main axis of the projected disk. The molecular peak closest to the position of 
most intense continuum emission (0\ffas2, toward the south; for formal position 
uncertainties see Table~\ref{tab-integral}) is that of CH$_3$C$_2$H, showing 
no absorption (Sect.\,3.2.5). Bendo et al. (2016) obtain 0\ffas2 for their 
H42$\alpha$ line, which is also not showing absorption.

\begin{table}
\caption[]{Relevant CH$_3$C$_2$H and continuum coordinates}
\begin{flushleft}
\begin{tabular}{lcc}
\hline
Feature                        &  $\alpha$, $\delta$      & Accuracy \\
			       &       ($J$2000)          &          \\
\hline 
                               &                          &          \\
$V$$\approx$665\,km\,s$^{-1}$  & 13 05 27.60/--49 28 03.8 &  0\ffas2 \\
3\,mm continuum                & 13 05 27.49/--49 28 05.3 &  0\ffas1 \\
$V$$\approx$570\,km\,s$^{-1}$  & 13 05 27.47/--49 28 06.2 &  0\ffas2 \\
$V$$\approx$475\,km\,s$^{-1}$  & 13 05 27.18/--49 28 08.4 &  0\ffas2 \\
			       &                          &          \\
\hline
\end{tabular}
\end{flushleft}
Approximate peak positions of the 3\,mm continuum emission (see Sect.\,3.1)
and the three locations of maximal line width of the CH$_3$C$_2$H 5$_0$$\rightarrow$4$_0$
transition, where the first column provides the associated barycentric radial 
velocity. The positions are close to the midline of the inclined nuclear disk of NGC~4945.
\label{tab-positions}
\end{table}

Keeping in mind that all data were obtained simultaneously so that relative positions 
should be very accurate (possibly more precise than the 0\ffas1 mentioned in Sect.\,3.1 
for absolute positions), we may thus ask, whether there is a connection between the 
offset and the degree of absorption a line is showing. This correlation is displayed in 
Fig.~\ref{centerposition}. Since we do not know the depth of the absorption trough 
in the HC$^{15}$N line (footnote d of Table 1 and Sect.\,3.2.3), this spectral 
feature has not been included. Errors in the depth of the dip near the systemic velocity 
may amount to $\pm$8\% while systematic position errors are difficult to quantify. We 
therefore refrain from showing a fit to the data but conclude that there appears to be 
a change in central position with increasing absorption depth. As already mentioned, 
this drift points to the southwest.  In view of this trend, shown by Fig.~\ref{centerposition}, 
the effect may be attributed to absorption predominantly occurring northeast of the dynamical 
center, shifting the fitted center of the molecular emission to positions southwest of the 
3\,mm continuum peak. This supports, in addition to the discussion of coordinates in Sect.\,3.1 
(and the moment 2 map of CH$_3$C$_2$H 5$_0$$\rightarrow$4$_0$ provided and discussed in 
Sect.\,4.6.1), our notion that the continuum peak is slightly displaced from the dynamical 
center of the galaxy.

\begin{table*}
\caption[]{Parameters of velocity integrated molecular lines$^a$}
\begin{flushleft}
\begin{tabular}{lccccccccc}
\hline
Line                                  &       Beam       &   PA  & \multicolumn{2}{c}{$\alpha,\delta_{J2000}$}& Source size    &  PA        & Inclination &  $I_{\rm mol}$    &         Integrated              \\
                                      &                  &       &                   &                        & undeconvolved  &            &             &                   &         peak flux               \\
				      & ($''\times''$)   & (deg) &                   &                        &($''\times''$)  &\multicolumn{2}{c}{(deg)} &(Jy\,km\,s$^{-1}$) &     (Jy\,km\,s$^{-1}$           \\
                                      &                  &       &                   &                        &                &            &             &                   &         beam$^{-1}$)            \\
\hline 
                                      &                  &       &                   &                        &                &            &             &                   &                                 \\
HC$^{18}$O$^+$                        & 2.72$\times$1.57 &  42   &    13 05 27.44    &     --49 28 06.1       & 7.0$\times$1.4 &  44        &    78       &        2.7        &     0.73                        \\
                                      &                  &       &                   &                        &(1.3$\times$0.3)& (03)       &   (03)      &       (0.4)       &    (0.09)                       \\
c-C$_3$H$_2$                          & 2.72$\times$1.57 &  42   &    13 05 27.44    &     --49 28 05.8       &10.5$\times$2.7 &  45        &    75       &       21.4        &     2.73                        \\
                                      &                  &       &                   &                        &(0.4$\times$0.1)& (01)       &   (01)      &       (0.7)       &    (0.08)                       \\
CH$_3$C$_2$H                          & 2.72$\times$1.57 &  42   &    13 05 27.49    &     --49 28 05.4       &10.0$\times$2.1 &  43        &    78       &        7.5        &     1.17                        \\
                                      &                  &       &                   &                        &(0.8$\times$0.1)& (01)       &   (02)      &       (0.5)       &    (0.07)                       \\
HC$^{15}$N                            & 2.72$\times$1.57 &  42   &    13 05 27.48    &     --49 28 05.5       & 6.0$\times$1.3 &  31        &    77       &        1.1        &     0.36                        \\
                                      &                  &       &                   &                        &(1.5$\times$0.5)& (05)       &   (04)      &       (0.2)       &    (0.05)                       \\
H$^{13}$CN                            & 2.72$\times$1.57 &  42   &    13 05 27.35    &     --49 28 06.0       & 9.9$\times$5.4 &  49        &    57       &        9.4        &     0.70                        \\
                                      &                  &       &                   &                        &(0.8$\times$0.4)& (05)       &   (02)      &       (0.8)       &    (0.05)                       \\
HCN                                   & 2.71$\times$1.56 &  43   &    13 05 27.38    &     --49 28 06.3       &11.4$\times$5.9 &  47        &    59       &      249.3        &    14.78                        \\
                                      &                  &       &                   &                        &(0.2$\times$0.1)& (01)       &   (01)      &       (3.0)       &    (0.17)                       \\
CS                                    & 2.57$\times$1.38 &  41   &    13 05 27.44    &     --49 28 05.7       &10.3$\times$3.3 &  47        &    71       &      112.4        &    10.47                        \\
                                      &                  &       &                   &                        &(0.3$\times$0.1)& (01)       &   (01)      &       (1.2)       &    (0.10)                       \\
HNC                                   & 5.59$\times$3.54 &  15   &    13 05 27.48    &     --49 28 05.5       & 8.5$\times$4.2 &            &             &                   &    36.80                        \\
                                      &                  &       &                   &                        &(0.5$\times$0.5)&            &             &                   &                                 \\
HCO$^+$                               & 5.59$\times$3.54 &  15   &    13 05 27.39    &     --49 28 06.9       &14.4$\times$4.8 &            &             &                   &    33.00                        \\
                                      &                  &       &                   &                        &(0.5$\times$0.5)&            &             &                   &                                 \\
3\,mm                                 & 2.55$\times$1.34 &  44   &    13 05 27.49    &     --49 28 05.3       & 5.8$\times$1.3 &  41        &   76.9      &                   &                                 \\
                                      &                  &       &                   &                        &  $<$0\ffas1    & (0.1)      &   (0.1)     &                   &                                 \\
H$_2$O                                & VLBA             &       &    13 05 27.48    &     --49 28 05.6       &                &$\approx$45 &             &                   &                                 \\
                                      &                  &       &                   &                        &                &            &             &                   &                                 \\
				      &                  &       &                   &                        &                &            &             &                   &                                 \\
\hline
\end{tabular}
\end{flushleft}
a) HCO, H$^{13}$CO, and SiO (see Table~\ref{tab-lines} and Fig.~\ref{hc15n-profile}) are severely blended 
and therefore excluded from the list. Our remaining seven molecular lines are presented on top, 
following increasing frequency. Note that for most transitions, integrated intensities 
($I_{\rm mol}$; Col.\,9) provide the difference between emission and absorption. Only CH$_3$C$_2$H 
appears to be entirely free of absorption. Formal accuracies of the peak positions in Cols.\,4 and 
5 are $\pm$0\ffas4, $\pm$0\ffas1, $\pm$0\ffas2, $\pm$0\ffas4, $\pm$0\ffas3, $\pm$0\ffas5, and 
$\pm$0\ffas5 from top to bottom for our ALMA data with $\approx$2$''$ resolution (see Col.\,2). Numbers 
in parentheses indicate standard deviations of the 2-dimensional Gaussian fits to the original (not
beam deconvolved data) and do not include the 10\% calibration uncertainty given in Sect.\,2. For 
comparison, we have also included parameters from the HNC and HCO$^+$ data of Cunningham \& Whiteoak 
(2005), from our $\lambda$ $\approx$ 3\,mm continuum data (see Sect.\,3.1), and from the 22\,GHz 
($\lambda$ $\approx$ 1.3\,cm) H$_2$O megamaser disk discovered by Greenhill et al. (1997). The uncertainty 
in the Cunningham \& Whiteoak (2005) positions is $\approx$0\ffas4; that of the megamaser disk is 0\ffas1. 
The uv-coverage of the Cunningham \& Whiteoak data is limited, leading for example to a position angle 
of 29$^{\circ}$ for the continuum. Therefore their PA values are not included and we also do not derive 
rotating disk inclinations from their data.
\label{tab-integral}
\end{table*}

It is remarkable that HCN and H$^{13}$CN, which should show (aside from potential 
saturation effects in the main isotopologue) similar spatial distributions, are 
characterized by similar offsets. HC$^{18}$O$^+$, which is even more affected by 
absorption, has its peak at a lower offset with respect to the continuum. Whether 
this is due to a lower signal-to-noise ratio (the line is weak, see Table~\ref{tab-lines} 
and Fig.~\ref{hc15n-profile}) or the consequence of a different spatial distribution 
remains an open question.

Table~\ref{tab-integral} also provides results from the $\lambda$ $\approx$ 3\,mm
lines of HNC and HCO$^+$ $J$=1$\rightarrow$0, taken from Cunningham \& Whiteoak 
(2005). While the sensitivity and uv-coverage of these data do not match those 
of our ALMA images, we nevertheless note that their HCO$^+$ profile, showing
strong absorption, appears to peak southwest of the continuum peak, while HNC,
less affected by absorption, appears to peak closer to the continuum position.

\subsubsection{Beyond the nuclear disk}

Off the central ridge (i.e. the nuclear disk), our map of integrated HCN $J$ = 1$\rightarrow$0 
emission (Figs.~\ref{hcn-hotspots} and \ref{maps-moment0}) shows a distinct secondary peak at 
$\alpha_{J2000}$ = 13$^{\rm h}$ 05$^{\rm m}$ 28\ffs1, $\delta_{\rm J2000}$ = --49$^{\circ}$ 28$'$ 
13$''$, ($\Delta\alpha$,$\Delta\delta$) = (+6$''$,--8$''$) toward the southeast with respect to 
the continuum peak. The ratio of flux densities between the secondary and primary peaks is 
0.15$\pm$0.03. Near this offset position, at (+3$''$,--6$''$), lies Knot B, following the 
nomenclature of Marconi et al. (2000, their fig.~2). Knot B is a region of enhanced near 
infrared Pa$\alpha$ emission, related to star formation, and is located above NGC~4945's 
galactic plane, toward the observer. Here we speculate that both objects are connected; Knot 
B represents merely the star forming front side of a giant molecular complex, while our data 
(not affected by obscuration) are revealing the entire cloud. In this context it is important 
to note that the region encompassing the Pa$\alpha$ and the secondary molecular peak is 
not seen in H42$\alpha$ (Bendo et al. 2016). There the 3$\sigma$ H42$\alpha$ limit is 0.15 
times its flux at the center of the galaxy. This may suggest that massive star formation is 
limited to the front side also accessible by near infrared spectroscopy. 

The huge cavity, $\approx$5$''$ ($\approx$100\,pc) in diameter, discovered by Marconi et al. 
(2000) a few arcseconds northwest of the center (see again their fig.~2 as well 
as P{\'e}rez-Beaupuits et al. 2011) might also be apparent in some of our channel maps. 
The northwestern part of the $\Psi$-shaped structure, encountered between barycentric 
$V$$\approx$530 and 570\,km\,s$^{-1}$ (see Table~\ref{tab-hcn} and Fig.\,\ref{hcn-channel}) may 
confine the bubble on its western, southern, and eastern sides. Knot C (Marconi et al. 
2000) is not clearly seen in our molecular line maps. 

For the connection between nuclear core, nuclear disk, and the outer arms mentioned in 
Sect.\,3.2.2, see Sect.\,4.7.

\subsection{Why are some lines absorbed and others not?}

There are several possible reasons why we measure some lines with signifiant
amounts of absorption, while at least one of our molecular transitions (the 
CH$_3$C$_2$H 5$_0$$\rightarrow$4$_0$ line) and the H42$\alpha$ recombination 
line (for this, see Bendo et al. 2016) appear to be seen entirely in emission. 
This may be caused by (1) different levels of line excitation, (2) different 
frequencies leading to changes in continuum flux density and morphology, (3) 
optical depth effects, (4) drastically different, directly detectable spatial 
distributions, (5) different chemical properties, or (6) a wide range of 
associated critical densities. In the following we briefly discuss these possibilities.

(1) Excitation: All molecular lines analyzed here connect levels up to $\approx$10\,K 
above the ground state. This upper limit is likely much smaller than the kinetic 
temperature throughout the nuclear disk, so that no effect due to excitation is expected.

(2) Frequencies: The frequency range of our lines is too small to allow for a 
strong influence of variations in continuum morphology and intensity on the 
strength of the absorption features. Strongest absorption is seen in the HCO$^+$ 
isotopologues and SiO. HC$^{13}$O$^+$ $J$=1$\rightarrow$0 and SiO $J$=2$\rightarrow$1 
are characterized by frequencies above that of C$_3$H$_2$ 2$_{12}$$\rightarrow$1$_{01}$
and below HCN $J$=1$\rightarrow$0 and CS $J$=2$\rightarrow$1, which are less 
affected by absorption. Considering HCO$^+$ and HCN isotopologues, the amount of 
absorption appears to be molecule dependent and does not change significantly when 
considering different isotopologues at slightly different frequencies 
(Fig.\ref{centerposition}). 

(3) Line saturation: Optical depth effects capable of reducing the effective 
critical density by a factor of $\tau$ in the optically thick case do 
not make a notable difference. This is most apparent in our data, when 
considering HCN, H$^{13}$CN, and HC$^{15}$N (Sect.\,3.2.4). The same holds
for H$^{13}$CO$^+$ and HC$^{18}$O (our data), when a spectrum from the main 
species (HCO$^+$; fig.~6 in Cunningham \& Whiteoak 2005) is added. 
 
(4) and (5) Chemistry and morphology: Aspects related to spatial distribution 
and molecular chemistry have to be considered together, because the chemistry 
may strongly affect observed morphologies in different molecular lines, and 
similarly, different environments may exhibit different chemical compositions. 
According to Table~\ref{tab-integral}, all tracers show a similar, elongated 
distribution with the main axis being consistent with the position angle of 
the nuclear disk and with a tendency for slightly smaller sizes for weaker 
lines. The latter is likely due to the noise level, which is (in absolute 
terms) similar for all lines (see Sect.\,2). There is, however, one notable 
exception.  While CS $J$=2$\rightarrow$1 is clearly stronger than c-C$_3$H$_2$ 
2$_{12}$$\rightarrow$1$_{01}$ (compare Figs.~\ref{hcn+cs-profile} and 
\ref{hc15n-profile}), c-C$_3$H$_2$ appears to be equally extended. This 
is consistent with the fact that c-C$_3$H$_2$ is quite abundant in diffuse 
low density clouds (e.g., Thaddeus et al. 1985), while CS is more 
concentrated near sites of massive star formation (e.g., Mauersberger et 
al. 1989; Shirley et al. 2003). Possibly representing more diffuse gas and 
being more widespread than CS, c-C$_3$H$_2$ should be less affected by 
the compact nuclear region with strongest continuum emission. Nevertheless, 
absorption in the c-C$_3$H$_2$ 2$_{12}$$\rightarrow$1$_{01}$ line is more 
pronounced than in CS $J$ = 2$\rightarrow$1 (cf.  Figs.~\ref{hcn+cs-profile} 
and \ref{hc15n-profile}). 

(6) Critical densities: Using the Leiden Atomic and Molecular database (LAMBDA; 
Sch{\"o}ier et al. 2005; van der Tak et al. 2007) and for CH$_3$C$_2$H Askne et 
al. (1984), we can obtain critical densities (where rates of spontaneous radiative 
decay are matched by collisional de-excitation) for all observed molecular transitions. 
Accounting only for the Einstein-$A$ coefficient and the collision rate (downwards) 
related to the measured transition in the LAMBDA database, we obtain in order 
of increasing relevance of absorption for CH$_3$C$_2$H, CS, C$_3$H$_2$, HCN, SiO, 
and HCO$^+$ density values of $n_{\rm crit}$ $\approx$ 1$\times$10$^4$, 3$\times$10$^5$, 
1$\times$10$^6$, 2$\times$10$^6$, 2$\times$10$^6$, and 2$\times$10$^5$\,cm$^{-3}$, 
respectively. CS, for example, shows a critical density similar to that of HCO$^{+}$, 
but it is only HCO$^+$, which shows deep absorption, comparable to that of SiO 
(Sect.\,3.3). Thus we do not see a dependence between critical density and absorption 
depth in most molecular lines.  The inclusion of the HNC $J$ = 1$\rightarrow$0 data 
by Cunningham \& Whiteoak (2005) with $n_{\rm crit}$ $\approx$3$\times$10$^5$\,cm$^{-3}$ 
and a systemic velocity emission level of order 25\% of the line peak does not make 
the problem more transparent. We note, however, that CH$_3$C$_2$H has a significantly 
lower critical density than all the other lines observed by us, like the CO $J$ = 
2$\rightarrow$1 line measured by Chou et al. (2007); in both cases, no absorption 
is apparent. 

To summarize, notable absorption appears to be only found
when critical densities are well in excess of $n_{\rm crit}$ $\approx$
10$^{4}$\,cm$^{-3}$. This is likely due to different spatial distributions,
which however cannot be assessed with the angular resolutions attained 
by Chou et al. (2007) and our study. With 84\%$\pm$10\% of the continuum 
caused by bremsstrahlung (Bendo et al. 2016), the higher density tracers 
are likely concentrated close to the many individual continuum sources 
related to massive star formation, while those with low critical densities 
should emit from more extended regions. While we think that this is the 
most plausible explanation, there still may be a caveat. CH$_3$C$_2$H 
5$_0$$\rightarrow$4$_0$ line emission is weaker than that of HCN/HCO$^+$
$J$ = 1$\rightarrow$0 and CS $J$ = 2$\rightarrow$1 emission and therefore its
optical depth may be lower. More severe photon trapping in the stronger lines
may reduce their effective critical densities below those given here (e.g., 
Shirley 2015). However, this is not supported by our low estimate of the HCN 
$J$ = 1$\rightarrow$0 optical depth in Sect. 4.4. In any case, subarcsecond 
resolution is required to reveal the differences in the small scale distributions 
of the molecular tracers and the mm-wave continuum (for comparison, see, e.g., 
the high resolution 2.3\,GHz continuum data presented by Lenc \& Tingay 2009).

\begin{table}
\caption[]{Peak flux densities in single 1$''$ sized pixels$^{a)}$}
\begin{flushleft}
\begin{tabular}{lcc}
\hline
Species                        &    \multicolumn{2}{c}{Peak flux}       \\
			       & 430\,km\,s$^{-1}$ & 710\,km\,s$^{-1}$  \\
                               &    \multicolumn{2}{c}{mJy/beam}        \\
\hline 
                               &                   &                    \\
HCN                            &        235.0      &      200.0         \\
CS                             &         95.0      &       80.0         \\
HC$^{18}$O                     &          5.5      &        5.5         \\
c-C$_3$H$_2$                   &         27.0      &       30.0         \\
CH$_3$C$_2$H                   &          6.5      &        9.0         \\
H$^{13}$CN                     &         14.0      &       13.5         \\
H$^{13}$CO$^+$                 &          7.0      &       14.0         \\
SiO                            &          9.0      &        9.0         \\
			       &                   &                    \\
\hline
\end{tabular}
\end{flushleft}
a) For the specific transitions and frequencies, see Table~\ref{tab-lines}.
Given flux densities may be accurate to $\pm$10\% (see Sect.\,1 and, for given
values, also the end of Sect.\,3.3). Flux density ratios should not be affected by 
calibration errors (estimated to be of order 10\%; see Sect.\,2) but only by noise. 
In Sect.\,4.4 we therefore conservatively adopt for such ratios also a $\pm$10\% 
uncertainty.
\label{tab-fluxes}
\end{table}

\subsection{Line intensity ratios}

Single dish observations have already addressed a number of chemical fingerprints of 
the central region of NGC~4945 (e.g., Henkel et al. 1990, 1994; Curran et al. 2001; Wang et 
al. 2004; Monje et al. 2014). The second-to-last of these studies contains a particularly 
large collection of molecular tracers between 81 and 355\,GHz. However, due to a lack 
of interferometric data these studies could not account for the influence of absorption 
onto overall line profiles and integrated intensities and, in addition, detailed 
comparisons with spectral properties of other galaxies were not yet possible more than 
a decade ago. Our study only covers a small frequency interval, but with superior 
sensitivity and angular resolution. And we can also refer to spectral line surveys of 
other galaxies. Here we focus on the line wings of NGC~4945 near barycentric $V$ = 430 
and 710\,km\,s$^{-1}$, where absorption does not play a role. 

With respect to the molecular species we have measured, the CS/HCN ratio has been claimed 
to be a possible starburst/AGN tracer (Meijerink et al. 2007; Izumi et al. 2013, 2016), 
while CH$_3$C$_2$H, c-C$_3$H$_2$, and HCO have been identified as tracers of UV fields 
in galaxies (e.g., Garc\'{\i}a-Burillo et al. 2002; Mart\'{\i}n et al. 2009; Aladro et al. 
2015). SiO can highlight the presence of shocks (e.g., Gusdorf et al. 2008; Guillet et al. 
2009; Duarte-Cabral et al. 2014). 

In the following we compare line intensities $I$ (see Table~\ref{tab-fluxes}) at the positions 
of strongest emission of the southwestern blue- and northeastern redshifted line peaks (note, 
that the values differ from those given in Table~\ref{tab-lines}, where area integrated 
intensities are derived; taking flux density or temperature ratios is not critical in this
context). Since a detailed discussion of chemical issues is well beyond the 
scope of this article, the following offers merely some highly superficial first insights into 
the chemical complexity of the gas in the nuclear disk of NGC~4945. Following fig.~17 of
Meijerink et al. (2007), our $I$(HCN)/$I$(CS) line peak intensity ratios of $\approx$2.5  
with an estimated uncertainty of 10\% (see Table~\ref{tab-ratios}) are consistent with a photon 
dominated region (PDR) of density $n$(H$_2$) $\approx$ 10$^6$\,cm$^{-3}$ and $\approx$10$^3$ times 
the local interstellar radiation field as well as with an XDR of density 10$^{4...5}$\,cm$^{-3}$, 
but a more detailed discussion on physical and chemical boundary conditions is clearly still 
required.

\begin{table}
\caption[]{Integrated line intensity ratios$^{a)}$}
\begin{flushleft}
\begin{tabular}{lrrr}
\hline
Species                                &    \multicolumn{2}{c}{NGC~4945}        &   NGC~253      \\
			               & 430\,km\,s$^{-1}$ & 710\,km\,s$^{-1}$  &                \\
\hline 
                                       &                   &                    &                \\
$I$(HCN)/$I$(CS)                       &   2.5$\pm$0.3     &   2.5$\pm$0.3      &   2.0$\pm$0.2  \\
$I$(HCN)/$I$(c-C$_3$H$_2$)             &   8.7$\pm$0.9     &   6.7$\pm$0.7      &  11.4$\pm$0.2  \\
$I$(HCN)/$I$(H$^{13}$CN)               &  16.7$\pm$1.7     &  14.8$\pm$1.5      &  14.5$\pm$0.4  \\
$I$(c-C$_3$H$_2$/$I$(HC$^{18}$O$^+$)   &   4.9$\pm$0.5     &   5.5$\pm$0.6      &   7.1$\pm$0.5  \\
$I$(HCN)/$I$(SiO)                      &  26.1$\pm$2.6     &  22.2$\pm$2.2      &  14.4$\pm$0.2  \\
$I$(CS)/$I$(SiO)                       &  10.6$\pm$1.1     &   8.9$\pm$0.9      &   7.1$\pm$0.1  \\
$I$(c-C$_3$H$_2$)/$I$(CH$_3$C$_2$H)    &   4.2$\pm$0.4     &   3.3$\pm$0.3      &   2.3$\pm$0.1  \\
$I$(H$^{13}$CN)/$I$(H$^{13}$CO$^+$)    &   2.0$\pm$0.2     &   1.0$\pm$0.1      &   1.2$\pm$0.1  \\
                                       &                   &                    &                \\
\hline
\end{tabular}
\end{flushleft}
a) For the specific $\lambda$ $\approx$ 3\,mm transitions and frequencies, see Table~\ref{tab-lines}.
For the NGC~4945 ratios, derived from Table~\ref{tab-fluxes}, an uncertainty of 10\% has been conservatively 
adopted. For NGC~253, see Table~A.2 of Aladro et al. (2015). 
\label{tab-ratios}
\end{table}

A comparison with the eight galaxies covered by the $\lambda$ $\approx$ 3\,mm line survey of 
Aladro et al. (2015) provides additional insights. Our $\lambda$ $\approx$ 3\,mm $I$(HCN)/$I$(CS) 
line intensity ratios are in line with seven of the eight targets studied by Aladro et al.
(2015). Among these seven, NGC~253 is outstanding. It is the strongest molecular line emitter 
outside the Local Group and has a distance, an inclination and an infrared luminosity similar to 
that of NGC~4945 (see also Sect.\,4.8). It is therefore selected as the representative one of 
the seven for Table~\ref{tab-ratios}. Only in M~51, the ratio is, at 5.2$\pm$0.1, significantly 
higher.  M~51 is a Seyfert galaxy where the inner spiral arms were also part of the measurement. 
This adds complexity but the high ratio is likely related to the presence of a nuclear jet leading 
to shock enhanced HCN (Matsushita et al. 2015). Our $I$(HCN)/$I$(H$^{13}$CN) peak flux density 
ratios are, at $\approx$15, smaller than for most of the galaxies of the Aladro sample, indicating 
a relatively high optical depth. This makes sense because the data from Aladro et al. (2015) 
were taken with $\approx$25$''$ resolution, while in NGC~4945 we consider exclusively 
those two compact $\approx$2$''$ ($\approx$40\,pc) sized regions with highest unabsorbed intensity. 
Nevertheless, the ratios for the two ULIRGs in their study, Arp~220 and Mrk~231, are even smaller 
(4--7) in spite of particularly large $^{12}$C/$^{13}$C isotope ratios (Henkel et al. 2014), 
indicating that there the bulk of the molecular gas is assembled in a compact volume of very 
high opacity (e.g., Rangwala et al. 2011; Aalto et al. 2015a, b; Mart\'{\i}n et al. 2016; Scoville 
et al. 2017). 

The galaxy with a similar $I$(HCN)/$I$(H$^{13}$CN) ratio in the Aladro et al. (2015) 
sample is NGC~253, which is also supposed to have a similar $^{12}$C/$^{13}$C ratio
(see Wang et al. 2004; Henkel et al. 2014) of $\approx$40 versus 40--50. Therefore
opacity effects should not be significant when comparing the NGC~253 and NGC~4945 data 
sets. From Table~\ref{tab-ratios} we find that ratios are qualitatively similar, but 
nevertheless there are some notable differences. Following Mart\'{\i}n et al. (2006) 
the chemistry in the nuclear region of NGC~253 is dominated by low-velocity shocks, 
which can be traced by SiO. Our line ratios from NGC~4945 indicate that SiO is less 
prominent with respect to other tracers, so shocks may still play an important but 
slightly less dominant role. Surprisingly, the ``ubiquitous'' (Thaddeus et al. 1985) 
c-C$_3$H$_2$ is less prominent in NGC~253 than in our NGC~4945 molecular hotspots. 
In view of the large region studied in NGC~253, this result appears to be counterintuitive. 
It can possibly be explained by massive star formation and a more pronounced dominance 
of PDRs in NGC~4945. Another interesting feature is the underabundance of CH$_3$C$_2$H 
with respect to c-C$_3$H$_2$ in NGC~4945 (Table~\ref{tab-ratios}). If high temperatures 
support the formation of CH$_3$C$_2$H as suggested by Aladro et al. (2015), this 
may hint at higher temperatures in the central region of NGC~253 than in our two 
selected locations of NGC~4945. We conclude that the active galactic nucleus in
NGC~4945 (NGC~253 is devoid of an AGN) has not a great influence. The greater dominance
of PDRs in NGC~4945, a possibly higher $T_{\rm kin}$ in the nuclear region of NGC~253,
and the small radius of the sphere of influence of the nuclear engine in NGC~4945
(Sect.\,4.1.1) all suggest that the presence or absence of an AGN is not dominating
the differences outlined above.

The $I$(HCN)/$I$(H$^{13}$CN) ratios at barycentric $V$ = 430 and 710\,km\,s$^{-1}$ are, 
near 15, identical within uncertainties. If the gas is well mixed, this indicates 
similar opacities, in apparent disagreement with the strong gradient in $I$(CO 
2$\rightarrow$1)/$I$(CO 1$\rightarrow$0) along the major axis, from 0.9 to 1.6, 
obtained by Dahlem et al. (1993, their fig.~6). CO 2$\rightarrow$1/1$\rightarrow$0 
line ratios much larger than unity, as seen near 700\,km\,s$^{-1}$, are indicative 
of (1) highly excited gas with low CO opacity, (2) partially displaced 
CO $J$ = 1$\rightarrow$0 and 2$\rightarrow$1 emitting regions with different levels 
of excitation or (3) relative pointing errors between the CO $J$=1$\rightarrow$0 and 
2$\rightarrow$1 line data. Very high excitation is indicated by the detection of 
the 321\,GHz line of H$_2$O near $v$ = 700\,km\,s$^{-1}$ (but not at other
velocities), about 1850\,K above the ground state (Pesce et al. 2016). Similar 
$I$(HCN)/$I$(H$^{13}$CN) ratios and drastically different CO 2$\rightarrow$1/1$\rightarrow$0 
ratios near barycentric $V$ = 400 and 710\,km\,s$^{-1}$ (0.9 versus 1.6) would suggest 
in the absence of CO-pointing errors that unlike our data from high column density 
positions the single-dish CO data of Dahlem et al. (1993) include large amounts 
of more diffuse gas.

\subsection{The nitrogen isotope ratio}

As already mentioned in the introduction, Chin et al. (1999) reported a 
tentative detection of HC$^{15}$N $J$=1$\rightarrow$0 emission, suggesting 
an H$^{13}$CN/HC$^{15}$N ratio of order two. This has been interpreted as a
hint for efficient stellar $^{15}$N ejection into the ISM of starbursts, due 
to rotationally induced mixing of protons into the helium-burning shells of 
young massive stars (e.g., Woosley \& Weaver 1995; Timmes et al. 1995). 
NGC~4945, with its nuclear starburst, is an ideal target to test this 
scenario. 

In view of the weakness of the HC$^{15}$N line, single pixel peak
flux densities as displayed in Table~\ref{tab-fluxes} cannot be used. Covering
instead the entire measured region, we find from Table~\ref{tab-lines} (Col.\,6)
and Table~\ref{tab-integral} (Col.\,9) for the $J$=1$\rightarrow$0 lines 
velocity integrated intensities of $I$(HCN)/$I$(H$^{13}$CN) = 26.5$\pm$2.3, 
$I$(HCN)/$I$(HC$^{15}$N) = 227$\pm$41, and $I$(H$^{13}$CN)/$I$(HC$^{15}$N) 
= 8.5$\pm$1.7 (the errors do not include calibration uncertainties; for these, see 
Sect.\,2). For $^{12}$C/$^{13}$C $\approx$ 40--50 (Wang et al. 2004; Hitschfeld et al. 2008; 
for NGC~253, see Henkel et al. 2014), this would imply a plausible average HCN opacity 
of two and a $^{14}$N/$^{15}$N ratio of $\approx$400. The latter matches the value
encountered in the local Galactic disk (Dahmen et al. 1995) and is much higher
than that tentatively deduced by Chin et al. (1999).

\begin{figure}[t]
%\vspace{-0.6cm}
%\centering
%\hspace{-0.8cm}
\resizebox{8.8cm}{!}{\rotatebox[origin=br]{0.0}{\includegraphics{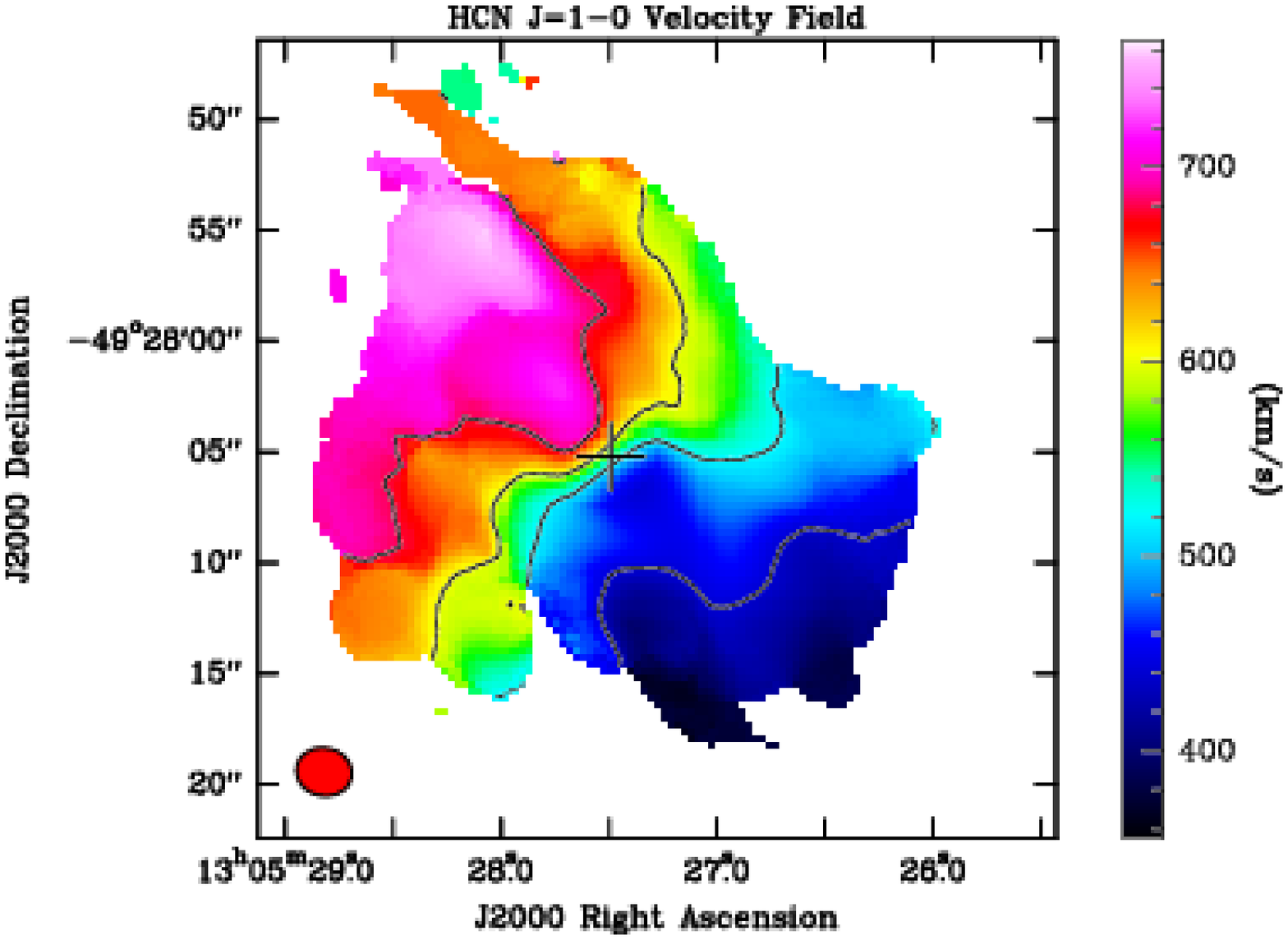}}}
%\vspace{-9.5cm}
\caption{The HCN $J$ = 1$\rightarrow$0 moment~1 velocity field, clipped at the 5$\sigma$ level. 
Contours are 440, 520, 600, and 680\,km\,s$^{-1}$ from lower right to upper left. Primary 
beam correction and Briggs weighting (robustness parameter +2.0) has been applied. 
The restored beam (2\ffas52 $\times$ 2\ffas17, PA = 78$^{\circ}$) is shown in the lower 
left corner of the image. The cross indicates the $\lambda$ $\approx$ 3\,mm radio continuum 
peak. }
\label{hcn-moment1}
\end{figure}

\begin{figure}[t]
%\vspace{-0.6cm}
%\centering
%\hspace{-0.8cm}
\resizebox{8.8cm}{!}{\rotatebox[origin=br]{0.0}{\includegraphics{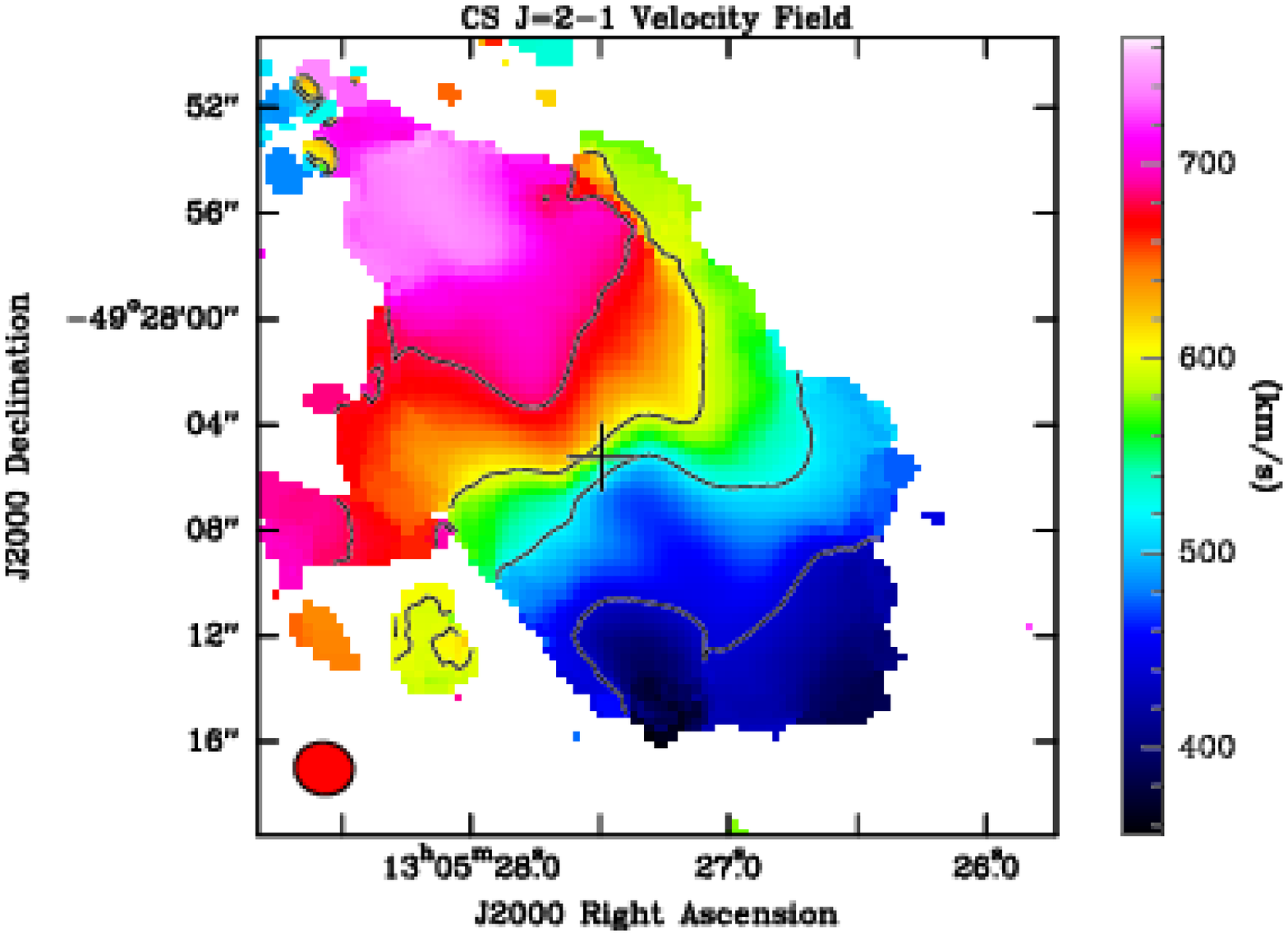}}}
%\vspace{-9.5cm}
\caption{The CS $J$ = 2$\rightarrow$1 moment~1 velocity field, clipped at the 5$\sigma$ level. 
Contours are 440, 520, 600, and 680\,km\,s$^{-1}$ from lower right to upper left. Primary 
beam correction and Briggs weighting (robustness parameter +2.0) has been applied. 
The restored beam (2\ffas22 $\times$ 1\ffas99, PA = 78$^{\circ}$) is shown in the lower 
left corner of the image. The cross indicates the $\lambda$ $\approx$ 3\,mm radio continuum 
peak. }
\label{cs-moment1}
\end{figure}

Nevertheless, in view of the possible influence of absorption and differences 
between the northeastern and southwestern edge of the inner disk (see, e.g., the 
huge difference in the CO $J$ = 2$\rightarrow$1/1$\rightarrow$0 ratio as a 
function of velocity (Dahlem et al. 1993, their fig.~6)), a closer inspection is 
warranted. Smoothing seven contiguous channels like in Fig.~\ref{hc15n-profile},
we obtain peak flux densities of 13.9 (H$^{13}$CN-southwest), 13.5 
(H$^{13}$CN-northeast), 3.5 (HC$^{15}$N-southwest), and 2.2\,mJy\,beam$^{-1}$
(HC$^{15}$N-northeast). The latter is barely above the noise level and
does not necessarily imply different abundance ratios in the southwest and
northeast. Adding and dividing the respective flux densities, we obtain
with this method an $I$(H$^{13}$CN)/$I$(HC$^{15}$N) ratio of order five. Again,
this is well above the ratio deduced from the tentative detection of Chin
et al. (1999) and would imply a $^{14}$N/$^{15}$N value similar or slightly 
below that of the solar system, 270 (e.g., Dahmen et al. 1995), where some 
enrichment due to massive star ejecta may have played a role (Chin et al. 1999). 
To summarize, the measured range of ratios, $^{14}$N/$^{15}$N $\approx$ 200 
-- 500, which we find in NGC~4945, is not in contradiction with the traditional 
view of $^{15}$N being mainly synthesized in low mass stars. More recently,
this has been also  proposed by Adande \& Ziurys (2012), based on Galactic data 
(see also Ritchey et al. 2015 and Furuya 2017). 

While our observations demonstrate that the ratio is $\gg$100, it remains to be 
seen whether it is $\approx$200 or $\approx$450. This will await ALMA measurements 
with higher sensitivity than those presented here with 30\,minutes of on-source 
integration time. Furthermore, in subsequent studies $^{13}$C/$^{15}$N ratios 
should be quantified in more than one molecular species to account for possible 
fractionation effects as modeled by Roueff et al. (2015; see their fig.~3),
even though they may not be significant because of the high kinetic temperatures
likely characterizing the molecular gas in the nuclear disk of NGC~4945.

\subsection{Inclinations}

Assuming that the observed emission arises from a flat disk with azimuthal symmetry, 
the ratio of the lengths of the minor to the major axis of our 2-dimensional Gaussian 
fits provides a direct measure of the inclination. Such an estimate is not biased (as 
at optical or NIR wavelengths) by extinction due to dust. The resulting inclination 
angles are given in Col.\,8 of Table~\ref{tab-integral}. For the outer disk, $i$ = 
78$^{\circ}$$\pm$3$^{\circ}$ (Ott et al. 2001), while for the nuclear region
(hereafter ``inner disk'') Chou et al. (2007) proposed $i$ = 62$^{\circ}$$\pm$2$^{\circ}$ 
on the basis of CO $J$ = 2$\rightarrow$1 data. We also get for some lines inclinations
below or even far below those of the large scale disk, i.e. 71$^{\circ}$ from CS 
$J$=2$\rightarrow$1 and values close to that of Chou et al. (2007) from HCN and 
H$^{13}$CN $J$ = 1$\rightarrow$0. However, the lines also collect emission from 
outside the nuclear disk (Sects.\,3.2.2, 4.1.3 and in particular 4.7) and thus 
broaden the extent of the emission along the minor axis, yielding unrealistically 
low inclinations. Our other four lines, HC$^{18}$O$^+$ $J$=1$\rightarrow$0, 
c-C$_3$H$_2$ 2$_{12}$$\rightarrow$1$_{01}$, CH$_3$C$_2$H 5$_{0}$$\rightarrow$4$_0$
and HC$^{15}$N $J$=1$\rightarrow$0, being too weak to trace much of the less intense 
emission outside the nuclear disk, yield instead $i$ = 75$^{\circ}$ -- 78$^{\circ}$, 
i.e. values perfectly consistent with the inclination on larger scales. We conclude 
that the inner disk still retains the inclination of the outer one, while some 
molecular lines are contaminated by emission from regions outside the inner 
$\approx$10$''$$\times$2$''$ sized region representing the nuclear disk. While this 
is an obvious explanation for CO (Chou et al. 2007), HCN, and CS, it is 
remarkable that the optically thinner H$^{13}$CN (but {\it not} HC$^{15}$N) 
is following HCN. We conclude that four of our seven analyzed transitions 
show the same inclination as the outer disk, which would not be the case if
the inclinations were different. Furthermore, the radio continuum arising 
predominantly from bremsstrahlung related to star formation with associated 
molecular clouds (Bendo et al. 2016), provides with its extent given in Sect.\,3.1
an inclination of $i$ = 76.9$^{\circ}$$\pm$0.1$^{\circ}$ (see also Sect.\,4.6.2
for a modeled inclination). In this sense NGC~4945 is remarkable, because a 
correlation between nuclear and large scale disk inclination is absent in 
statistically relevant samples of active spirals (e.g., Ulvestad \& Wilson 1984 
and Sect.\,4.7 below).

\begin{figure}[t]
%\vspace{-0.6cm}
%\centering
%\hspace{-0.8cm}
\resizebox{8.8cm}{!}{\rotatebox[origin=br]{0.0}{\includegraphics{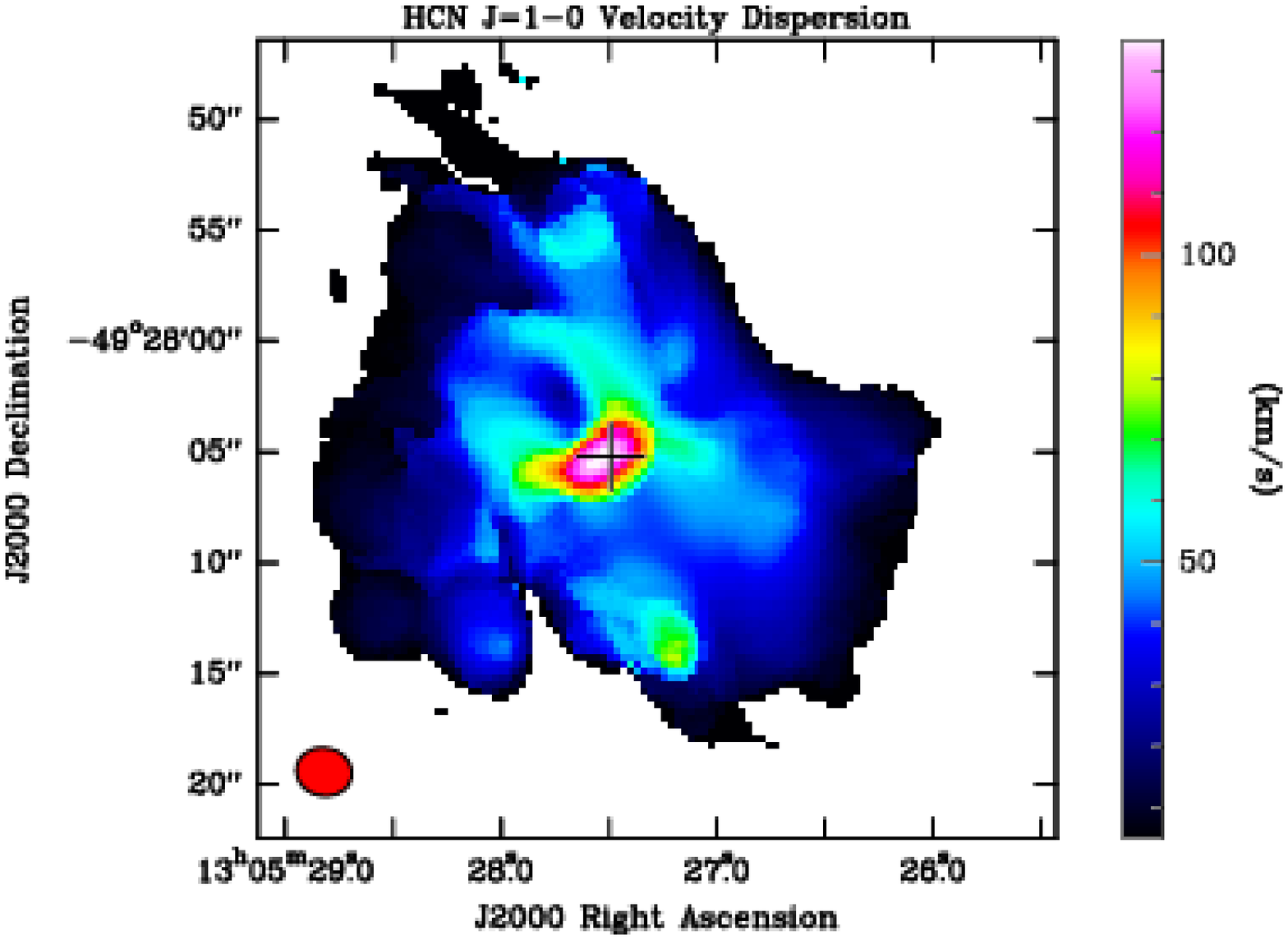}}}
%\vspace{-9.5cm}
\caption{A map of the HCN $J$ = 1$\rightarrow$0 moment~2 velocity dispersion, clipped at the 
5$\sigma$ level. Primary beam correction and Briggs weighting (robustness parameter +2.0) 
has been applied. The restored beam (2\ffas52 $\times$ 2\ffas17, PA = 78$^{\circ}$) is shown 
in the lower left corner of the image. The cross indicates the $\lambda$ $\approx$ 3\,mm radio 
continuum peak, while the position of the secondary maximum in the south (also seen in 
Fig.~\ref{cs-moment2}) is inconspicuous in all other aspects.} 
\label{hcn-moment2}
\end{figure}

\begin{figure}[t]
%\vspace{-0.6cm}
%\centering
%\hspace{-0.8cm}
\resizebox{8.8cm}{!}{\rotatebox[origin=br]{0.0}{\includegraphics{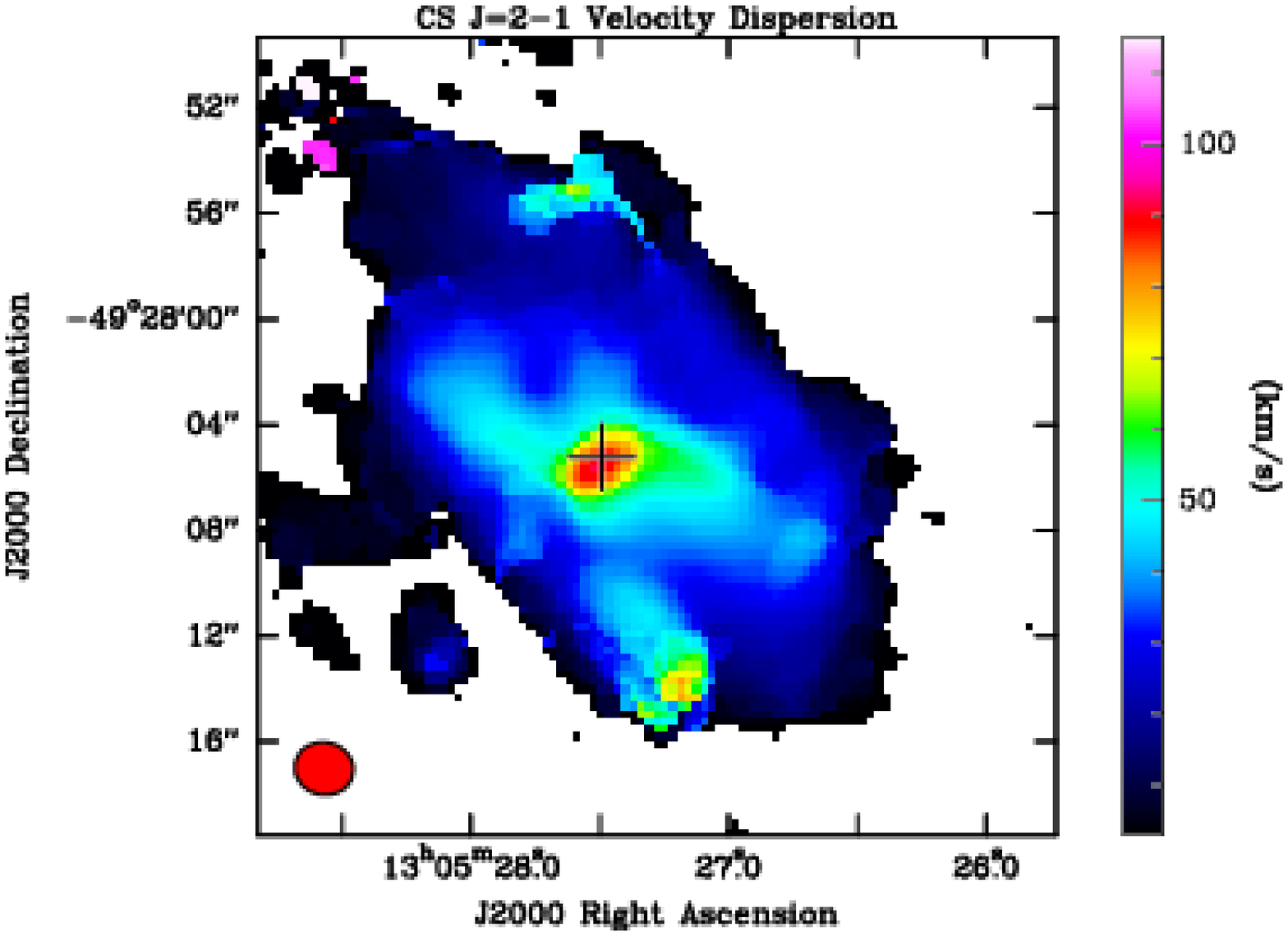}}}
%\vspace{-9.5cm}
\caption{A map of the CS $J$ = 2$\rightarrow$1 moment~2 velocity dispersion, clipped at the 5$\sigma$ 
level. Primary beam correction and Briggs weighting (robustness parameter +2.0) has been 
applied. The restored beam (2\ffas22 $\times$ 1\ffas99, PA = 79$^{\circ}$) is shown in the lower left 
corner of the image. The cross indicates the $\lambda$ $\approx$ 3\,mm radio continuum peak, while
the secondary peak in the south (also seen in Fig.\ref{hcn-moment2}) is inconspicuous 
in all other aspects. }
\label{cs-moment2}
\end{figure}

\subsection{Kinematics}

\subsubsection{Observational constraints}

Figs.~\ref{hcn-moment1} -- \ref{cs-moment2} provide images of the intensity 
($I_{\rm i}$) weighted line-of-sight velocity distributions ($I$ = $\Sigma I_{\rm i}$)
$$
      V = \frac{\Sigma (I_{\rm i} V_{\rm i})}{I} 
$$
(moment 1) and the intensity weighted velocity dispersions 
$$
     \Delta V = \sqrt{\frac{\Sigma I_{\rm i} (V_{\rm i} - V)^2}{I}}
$$ 
(moment 2) of the two strongest lines, those of HCN and CS. Emphasizing here not basic 
morphological features but trying to also account for as much large scale structure 
as possible, these figures present Briggs weighted images (weighting robustness 
parameter +2.0) to come close to natural weighting. While being affected by different 
degrees of absorption, the velocity fields of HCN and CS look quite similar and are at 
first order consistent with rotation of a disk around a central source. Furthermore, 
the systemic velocity contour passes, in spite of this absorption, within the limits 
of accuracy ($\approx$1$"$) through the center of the galaxy (see also Cunningham \& 
Whiteoak 2005). However, our isovelocity contours are not quite as regular as those 
of the H42$\alpha$ recombination line. There, isophotes are approximately 
perpendicular to the main axis of the galaxy (PA $\approx$ --45$^{\circ}$; see 
Bendo et al. 2016), giving rise to the impression that the innermost gas is rotating 
like a rigid body. A comparison with the H91$\alpha$+H92$\alpha$ data of Roy 
et al. (2010, their fig.~2), obtained with an even smaller beam size (1\ffas4 
$\times$ 1\ffas2 instead of the 2\ffas0 to 2\ffas5 used here) also reveals 
this difference. While in the innermost $\approx$4$''$ ($\approx$75\,pc) their isovelocity 
contours are roughly parallel to each other and perpendicular to the position angle of 
the galaxy, their contours in the northeast (particularly the 680\,km\,s$^{-1}$ 
contour) are displaced northwards from the main axis. This also holds in 
the case of CO $J$ = 2$\rightarrow$1 (Chou et al. 2007). However, our HCN data 
(Fig.~\ref{hcn-moment1}) show instead a twist to the south, while CS 
(Fig.~\ref{cs-moment1}) is neutral in this respect. While it might be that 
absorption is significantly affecting the observed velocity fields, the main 
cause of the differences may be emission from outside the nuclear disk (Sect.\,4.7).

\begin{figure}[t]
%\vspace{-0.6cm}
%\centering
%\hspace{-0.8cm}
\resizebox{8.8cm}{!}{\rotatebox[origin=br]{0.0}{\includegraphics{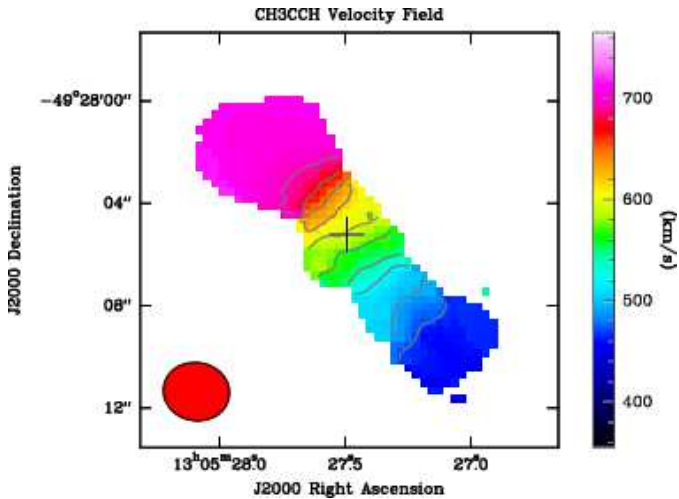}}}
%\vspace{-9.5cm}
\caption{The CH$_3$C$_2$H 5$_0$$\rightarrow$4$_0$ velocity field, clipped at the 5$\sigma$ level. 
Unlike in Figs.~\ref{hcn-moment1} and \ref{cs-moment1}, here the emission is confined to the nuclear 
disk. Primary beam correction and Briggs weighting (robustness parameter +2.0) has been applied. 
The contours start at 470\,km\,s$^{-1}$ and reach 695\,km\,s$^{-1}$; their spacing is 25\,km\,s$^{-1}$. 
The restored beam (2\ffas65 $\times$ 2\ffas28, PA = 80$^{\circ}$) is shown in the lower left corner 
of the image. The cross indicates the $\lambda$ $\approx$ 3\,mm radio continuum peak. }
\label{ch3c2h-moment1}
\end{figure}

\begin{figure}[t]
%\vspace{-0.6cm}
%\centering
%\hspace{-0.8cm}
\resizebox{8.8cm}{!}{\rotatebox[origin=br]{0.0}{\includegraphics{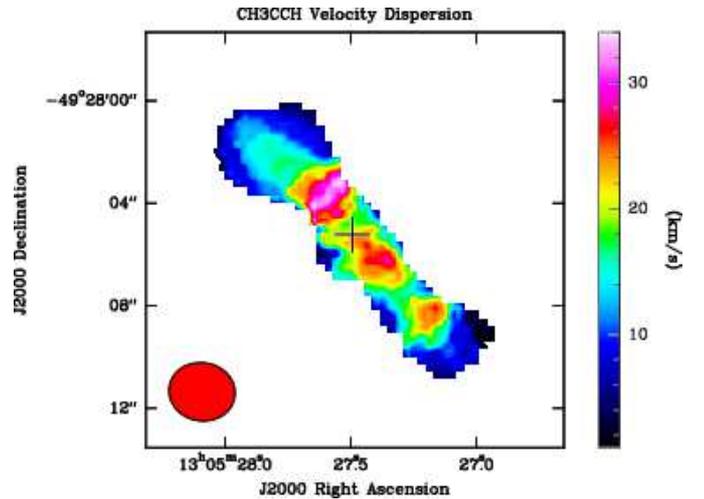}}}
%\vspace{-9.5cm}
\caption{A map of the CH$_3$C$_2$H 5$_0$$\rightarrow$4$_0$ velocity dispersion, clipped at the 
5$\sigma$ level. Primary beam correction and Briggs weighting (robustness parameter +2.0)
has been applied. The restored beam (2\ffas65 $\times$ 2\ffas28, PA = 80$^{\circ}$) is shown 
in the lower left corner of the image. The cross indicates the $\lambda$ $\approx$ 3\,mm radio 
continuum peak. }
\label{ch3c2h-moment2}
\end{figure}

When considering the distribution of line widths, there is another notable 
feature: The region with largest dispersion ($\ga$80\,km\,s$^{-1}$ in 
Figs.~\ref{hcn-moment2} and \ref{cs-moment2}) is, as expected, located at 
the center of the galaxy. Perhaps a part of this could be related to the 
non-thermal source with jet-like morphology detected by Lenc \& Tingay (2009) 
at an offset of $\approx$1$''$ from the dynamical center. However, a detailed 
inspection of the spectra indicates that the fits of the central region are 
questionable because of absorption near systemic velocities (see, e.g., 
Sect.\,3.2.2 and Table~\ref{tab-hcn}). Toward the northwest and southeast 
slightly off the region exhibiting absorption (e.g., Fig.~\ref{hcn-channel}), 
wide line emission is seen, because here parts of the approaching and 
receding disk are seen simultaneously. Toward the absorbing region itself, 
velocities near the systemic ones are blocked, inhibiting meaningful 
fits of the line widths. 

In view of the line contamination by absorption and, likely even more important, by gas 
arising from outside the nuclear disk (Sects.\,3.2.2, 4.1.3, and 4.7), we thus prefer 
to choose the CH$_3$C$_2$H 5$_0$$\rightarrow$4$_0$ line. While its emission is $\approx$35 
times weaker than that of HCN $J$=1$\rightarrow$0, it is not significantly affected by 
absorption and traces solely the nuclear disk. Figs.~\ref{maps-moment0} and 
\ref{ch3c2h-moment1}--\ref{ch3c2h-moment2} show the distributions of velocity integrated 
intensity, velocity, and velocity dispersion exclusively for the nuclear disk. The 
velocity field turns out to be highly regular, similar to that derived for the H42$\alpha$ 
recombination line (Bendo et al. 2016). Twists to the north or south (e.g., Chou et al. 2007) 
are absent and the most blueshifted and redshifted velocities are found in the 
outermost regions of the nuclear disk along its major axis. 

While the velocity field is quite regular, the velocity dispersion looks more complex. 
Instead of one peak extended along a southeast-northwest axis (Figs.~\ref{hcn-moment2}, 
\ref{cs-moment2}), we find three (Fig.~\ref{ch3c2h-moment2}). The central one, elongated 
along the main axis of the galaxy and not perpendicular to it, is accompanied by two 
additional regions with wider lines to the northeast and southwest, separated by a 
total of 5\ffas9$\pm$0\ffas2. At these locations we observe the transition between 
rigid body rotation and steep rotation curve to a flat one (see also Chou et al. 2007). 
At a radius of 2\ffas45 ($\approx$45\,pc) and with a rotation velocity of 140\,km\,s$^{-1}$ we 
obtain with equation\,(1) of Mauersberger et al. (1996; $\eta$ = 1) an enclosed mass of 
$M_{2.45}$ = 2.1 $\times$ 10$^8$\,M$_{\odot}$ with an estimated error of 10\%, in good
agreement with Cunningham \& Whiteoak (2005). For comparison, Roy et al. (2010) obtain 
with the H92$\alpha$ line 3$\times$10$^7$\,M$_{\odot}$ for the mass inside a radius of 1$''$ 
($\approx$19\,pc). Furthermore, we note that the nuclear region is slightly lopsided: The center 
of the line connecting the two outer peaks of line width is located $\approx$1\ffas4 
southwest of the continuum peak (see Table~\ref{tab-positions}).  Systemic velocities 
($V_{\rm barycentric}$ $\approx$571\,km\,s$^{-1}$) are found about 0\ffas9 ($\approx$15--20\,pc) 
southwest of the continuum peak and may be closer to the position of the maser disk (Greenhill 
et al.  1997) given in Sect.\,3.1. While uncertainties with respect to the value of the systemic 
velocity (Sect.\,2) and the relative positions of the continuum peak and the maser disk 
are significant, most of the star formation represented by the continuum emission (see 
Bendo et al 2016) appears to originate slightly northeast of the dynamical center. The 
lopsidedness of the central region of NGC~4945 also explains why molecular lines with 
significant absorption show integrated intensity peaks slightly shifted to the southwest: 
With the bulk of the continuum arising from the northeast, line emission is likely more 
quenched by absorption at this side of the center than in the southwestern part of the 
inclined nuclear disk (see Sect.\,4.2 and Fig.~\ref{centerposition}).

\begin{figure}[t]
\vspace{-0.5cm}
%\centering
\hspace{-0.8cm}
\resizebox{8.4cm}{!}{\rotatebox[origin=br]{0.0}{\includegraphics{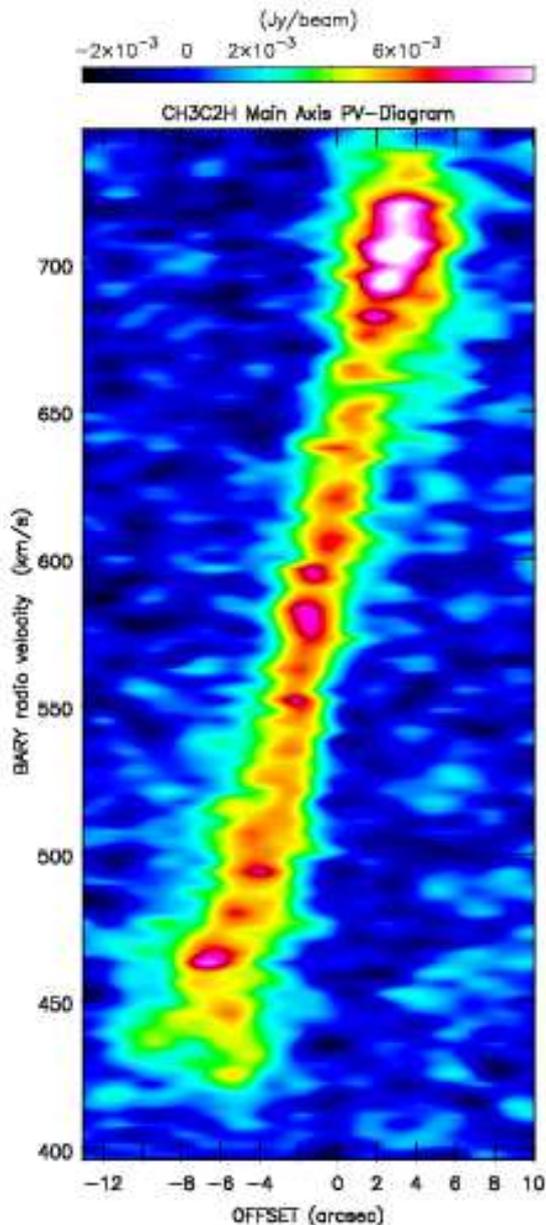}}}
\vspace{-1.0cm}
\caption{Position-velocity plot of CH$_3$C$_2$H 5$_0$$\rightarrow$4$_0$ along the major axis
with the abscissa scaled relative to the continuum peak. Contour: 7\,mJy\,beam$^{-1}$.
The plot emphasizes (a) the clumpiness of the gas, (b) the steepening of the rotation 
curve between barycentric $V$ $\approx$ 500\,km\,s$^{-1}$ and the systemic velocity of 
$V$ $\approx$ 571\,km\,s$^{-1}$ and (c) its flattening between 470 and 500\,km\,s$^{-1}$.}
\label{ch3c2h-pv}
\end{figure}

Fig.~\ref{ch3c2h-pv} shows a position velocity (PV) diagram along the major axis, 
which may be compared with the CO based diagrams of Chou et al. (2007) and Lin et al. 
(2011). At a 3\,mJy\,beam$^{-1}$ level, no structure can be recognized except particularly 
prominent emission near 710\,km\,s$^{-1}$ and a less prominent enhancement near 430\,km\,s$^{-1}$. 
At levels $>$4.5\,mJy\,beam$^{-1}$ the clumpiness of the gas becomes apparent. 
The size of these condensations at a given velocity is of order 2$^{\prime\prime}$, 
roughly corresponding to 40\,pc, which is not only the linear size of our beam but is 
also a characteristic linear scale of giant molecular clouds (see, e.g., Rosolowski 
et al. 2003 for M\,33). Furthermore, the high intensity backbone of the rotation 
curve is steeper between barycentric $V$ $\approx$ 500 and 570\,km\,s$^{-1}$ than over 
the entire velocity range, and flatter between 470 and 500\,km\,s$^{-1}$. Overall, 
the PV diagram from CH$_3$C$_2$H shows much less complexity than that of CO $J$ = 
2$\rightarrow$1 (compare with figs.~1 and 4 of Lin et al. 2011).

\subsubsection{A kinematical model of the nuclear disk involving tilted rings}

From the data cube (e.g., Fig~\ref{ch3c2h-moment1}) it is evident that NGC~4945 contains a 
rather well-behaved, regularly rotating nuclear disk that is best seen in our CH$_3$C$_2$H 
5$_0$$\rightarrow$4$_0$ line, (1) because of an absence of significant absorption and 
(2) because weaker spatial components outside the disk are below our sensitivity limit. 
It is in principle straightforward to perform a parametrisation of this (and other) line(s)
in terms of a (tilted-) ring analysis (Rogstad et al. 1974). As, however, the disk is 
observed to be at high inclination and the point spread function's half power width is 
large compared to the disk's diameter along the minor axis, a kinematical analysis on a 
velocity field might fail: line-of-sight and smearing effects will render a velocity field 
unreliable. In addition we are interested in the radial profiles of the line strenghts of 
the detected lines and we want to test the feasibility of simultaneous fits of several lines. 
This is only possible with a 3-d modeling approach, in which an observation is simulated, 
based on input parameters, to then adjust the parameters based on a comparison of observed 
and modeled data. Here we hence attempt a (tilted-)ring modeling with the software {\sc 
TiRiFiC} (J{\'o}zsa et al. 2007; J{\'o}zsa 2016). 

The naturally weighted data cube was primary-beam corrected and converted to radio 
velocity, using $85.45727\,{\rm GHz}$ (see Table~\ref{tab-lines}) as reference frequency. 
Accordingly, the CH$_3$C$_2$H line is tracked at the correct velocity, while in 
addition the c-C$_3$H$_2$ 2$_{12}$--1$_{01}$ line and H42$\alpha$ (for the latter,
see Bendo et al. 2016) are also part of this analysis. Furthermore, the analyzed 
position-velocity plots show weak emission from lines likely belonging to 
HOCO$^+$ 4$_{04}$ $\rightarrow$ 3$_{03}$ ($\nu_{\rm rest}$ = 85.53148\,GHz) and 
c-C$_3$H$_2$ 4$_{32}$ $\rightarrow$ 4$_{23}$ ($\nu_{\rm rest}$ = 85.65642\,GHz), 
which are too inconspicuous to be apparent in the spectra shown in Fig.~\ref{hc15n-profile}. 
To accelerate the analysis the data cube was binned along the spatial axes (still 
providing a pixel size fully sampling the synthesised beam).

The fitting strategy follows an established path (see, e.g. Zschaechner et al. 2012;
Gentile et al. 2013; Kamphuis et al. 2013; Schmidt et al. 2014; de Blok et al. 2014).
Including a large set of parameters in the model (e.g. inward or outward motion, vertical 
motion, several modes of surface brightness inhomogeneities, etc.) will necessarily lead to 
a good fit at the expense of a more and more arbitrary model parametrisation. Too 
many parameters can be degenerate to some degree. We hence intend to find the simplest 
model which compares well to the data. We point out that both the definition of simple 
as well as the notion of a good fit is subjective. We also cannot guarantee to have 
found a unique solution. On the other hand we tested a wide range (see below) of 
parameter combinations, such that the solution presented here is rather well 
established.

\begin{figure}[h]
\resizebox{8.8cm}{!}{\rotatebox[origin=br]{0.0}{\includegraphics{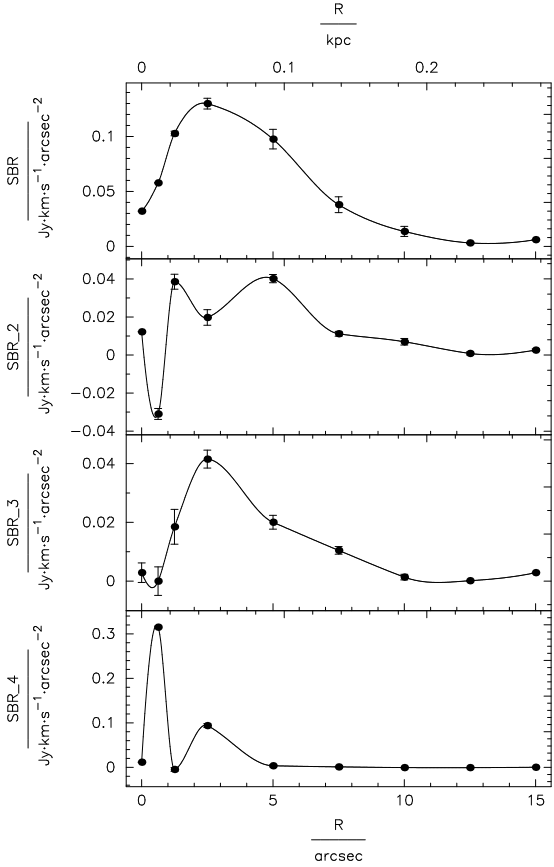}}}
\caption{Final parametrisation: surface brightness. Negative values might reflect statistical 
errors, but might also be due to absorption. From top to bottom: c-C$_3$H$_2$ $J$ =
2$_{12}\rightarrow1_{10}$ (SBR), CH$_3$C$_2$H 5$_0$$\rightarrow$4$_0$ (SBR-2) and, 
overlapping, c-C$_3$H$_2$ $J$=4$_{32}\rightarrow4_{23}$ (SBR-3) and H\,42$\alpha$ (SBR-4; 
see Table~\ref{tab-lines} for details to the transitions). The abscissa provides galactocentric 
radii and the ordinate displays emission in units of Jy\,km\,s$^{-1}$\,arcsec$^{-2}$. Further 
information is given in Sect.\,4.6.2. }
\label{parameters-sbr}
\end{figure}

\begin{figure}[h]
\resizebox{8.8cm}{!}{\rotatebox[origin=br]{0.0}{\includegraphics{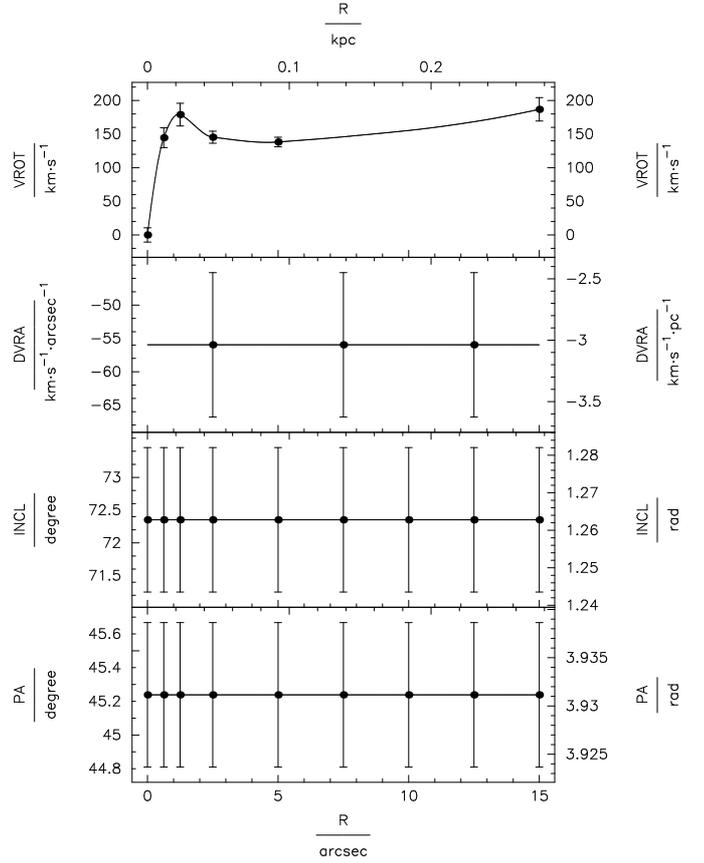}}}
\caption{Final parametrisation (from top to bottom): Rotation velocity, vertical velocity
gradient $DVRA$ (see Sect.\,4.6.2), inclination, and position angle as a function of galactocentric 
radius. These are the parameters, which are assumed to be shared by all the lines included in 
Figs.~\ref{parameters-sbr}, \ref{parameters-scaleheight}, \ref{parameters-dispersion} and 
\ref{slices-model-26}.}
\label{parameters-vrot}
\end{figure}

Fitting started with a flat-disk model, assuming a flat rotation curve. The number 
of nodes between which the model is interpolated using an Akima interpolation is 
varied from parameter to parameter. Beyond a radius of 15$''$, there is no meaningful 
observational contribution to any model. Near 2$''$, we are limited by our angular 
resolution. The rotation curve rises steeply at the center, which is why we choose 
a denser sampling there.

A simple rotating disk model already gives a reasonable fit to the data. In the 
following we fit four disks simultaneously to the data, also with the aim to check whether 
CH$_3$C$_2$H and H42$\alpha$ model parameters are consistent with those of C$_3$H$_2$ 
2$_{12}$$\rightarrow$1$_{10}$ (hereafter c-C$_3$H$_2$--2) and the line assigned to 
c-C$_3$H$_2$ $J$ = 4$_{32}$$\rightarrow$4$_{23}$ (hereafter c-C$_3$H$_2$--4). It is 
assumed that all lines share rotation velocity, inclination, position angle, and central 
position. Notably, the c-C$_3$H$_2$--4 and H42$\alpha$ lines have some channels in common, 
which is why the weak C$_3$H$_2$ $J$ = 4 feature had to be included in the model. 
Otherwise {\sc TiRiFiC}'s $\chi^2$-minimisation would be pushed into a wrong minimum. 
The parameterisation of the c-C$_3$H$_2$--4 disk is however unreliable. c-C$_3$H$_2$--2 
shows strong absorption, which is why the model surface brightness (Fig.~\ref{parameters-sbr}, 
upper panel) is unreliable near the centre.

Both a radially varying scale height as well as a radially varying 
velocity dispersion can improve fits to the data. These parameters were found to provide 
best fits if varied independently from each other for each line.

\begin{figure}[h]
\resizebox{8.8cm}{!}{\rotatebox[origin=br]{0.0}{\includegraphics{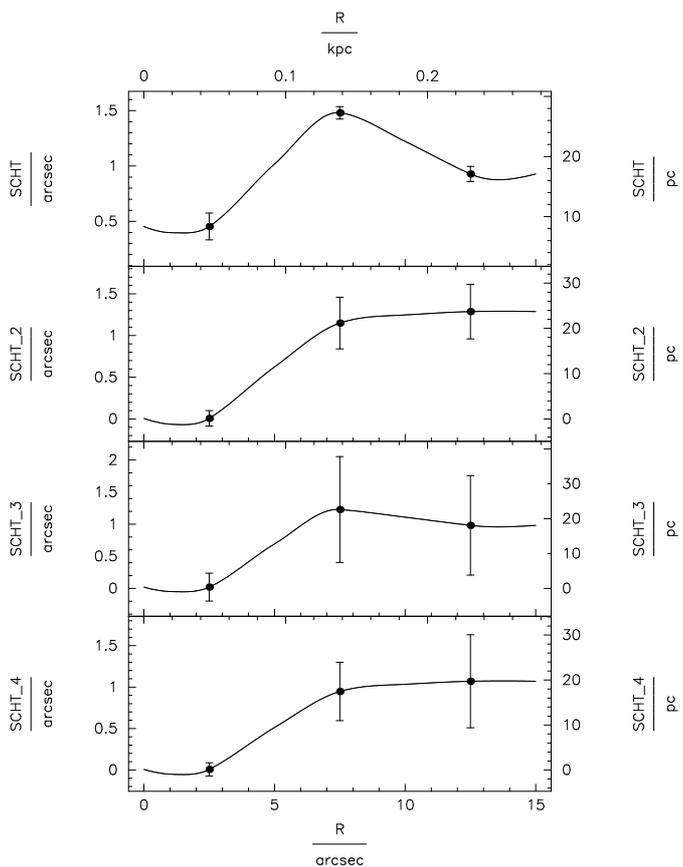}}}
\caption{Final parametrisation: full scale height. From top to bottom in the same order as in 
Fig.~\ref{parameters-sbr}: c-C$_3$H$_2$ (SCHT), CH$_3$C$_2$H (SCHT-2) and, overlapping, 
c-C$_3$H$_2$ (SCHT-3) and H\,42$\alpha$ (SCHT-4). Ordinate on the left hand side: scale 
height in arcseconds; ordinate on the right hand side: scale height in pc, both as a 
function of galactocentric radius.}
\label{parameters-scaleheight}
\end{figure}

\begin{figure}[h]
\resizebox{8.8cm}{!}{\rotatebox[origin=br]{0.0}{\includegraphics{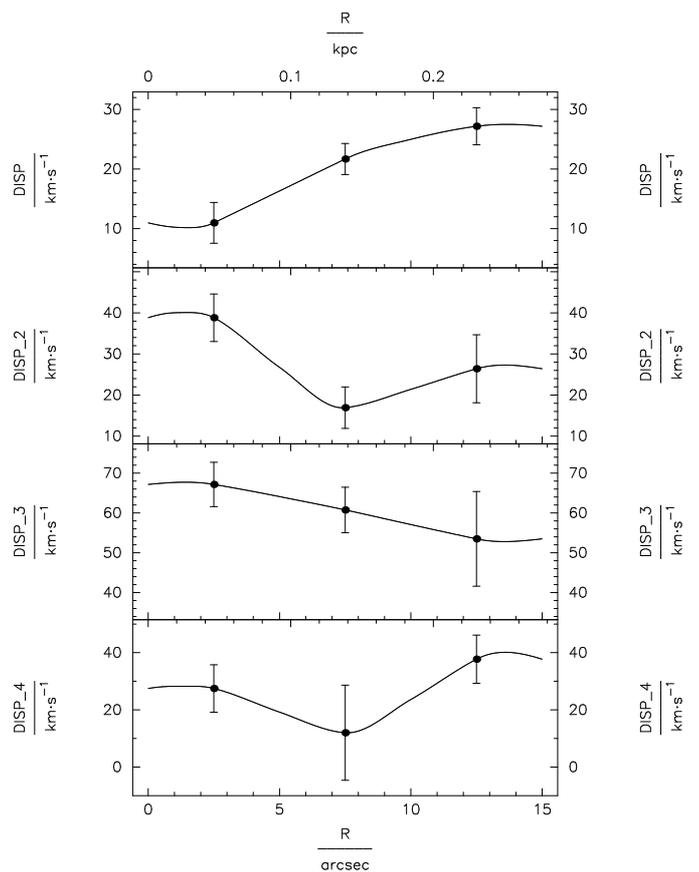}}}
\caption{Final parametrisation: Intrinsic dispersion (see Sect.\,4.6.1) as a function of 
galactocentric radius (this may be compared with Fig.~\ref{ch3c2h-moment2}). From top to bottom 
as in Fig.~\ref{parameters-sbr}: c-C$_3$H$_2$ (DSP), CH$_3$C$_2$H (DSP-2) and, overlapping, 
c-C$_3$H$_2$ (DSP-3) and H\,42$\alpha$ (DSP-4; see Table~\ref{tab-lines}).}
\label{parameters-dispersion}
\end{figure}

{\rm TiRiFiC} allows us to test the following kinematic behavior: a central radial 
(in- or outward) motion (parameter $VRAD$) can vary linearly with height $|z|$ above 
or below the plane (parameter $DVRA$, see also Schmidt et al. 2014), i.e. 
$$
        V_{\rm exp} = VRAD  +  |z| \times\ DVRA.
$$
This motion is performed parallel to the main disk. Tested were the following scenarios, 
keeping parameters identical for all lines: (1) radially varying radial motion in 
addition to radially varying vertical gradients of the radial motion, (2) constant radial 
motion at all radii, and (3) constant radial motion at a given vertical height at all
radii and a constant vertical gradient of the radial motion at all radii. 

As the latter provided a comparable fit to the other solutions and was slightly favourable 
compared to a model with a constant radial motion throughout the disk, we used this model 
to reduce the numbers of nodes for the rotation curve to obtain the final solution 
(Figs.~\ref{parameters-sbr} -- \ref{parameters-dispersion}). Fitting started with 
a flat-disk model, assuming a flat rotation curve. Initial parameters (constant surface 
brightness, inclination, position angle, rotation speed) were based on a rough inspection 
of the cube. While TiRiFiC in the chosen nested-interval mode depends on an initial
guess of parameters, it generally converges quickly starting from a rough guess if 
the number of initial parameters is low. Consecutively, the number of parameters were 
increased to allow for example for an implementation of a radial surface brightness profile, 
a rotation curve, and additional parameters. When it was found that the introduction of 
additional parameters did not visibly improve the model, we chose not to use those parameters.
Errors were estimated by fitting the two (in the model identical) halves of the galaxy 
independently (segments 1, 2, 3, and 4 on one and segments 5, 6, 7, and 8 on the 
other side, covering in this way the four lines) and taking the average deviation from 
the final solution averaged over this and the one or two neighbouring nodes as the error.

\begin{figure}[t]
\resizebox{8.8cm}{!}{\rotatebox[origin=br]{0.0}{\includegraphics{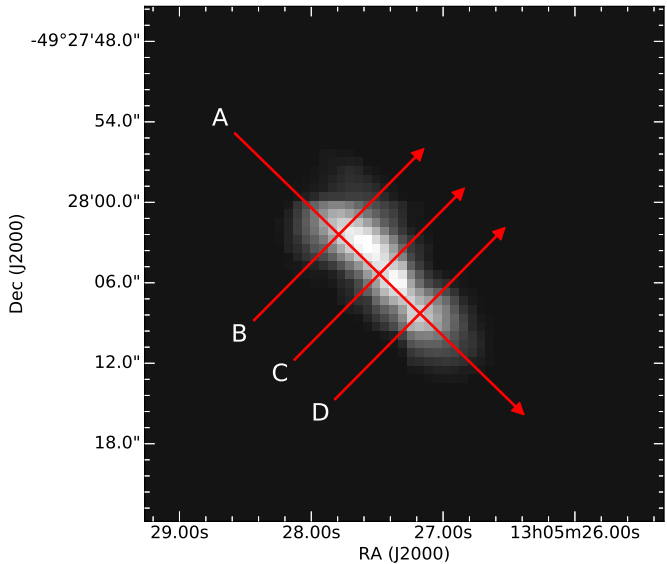}}}
\caption{Slice positions of PV-diagrams in Fig.~\ref{slices-model-26} overlaid on a 
velocity integrated intensity map of the CH$_3$C$_2$H 5$_0$$\rightarrow$4$_0$ line (see, 
e.g., Table~\ref{tab-lines} and Fig.~\ref{maps-moment0}.) The PV-diagrams in 
Fig.~\ref{slices-model-26} are all directed from left (east) to right (west).}
\label{slice-positions}
\end{figure}

Displaying the results for modeled segments 1, 2, 3, and 4, Fig.~\ref{parameters-sbr} 
shows the surface brightness of the different tracers as a function of galactocentric 
radius, indicating peak values in the inner 5$''$. Figure~\ref{parameters-vrot} provides  
the rotation curve (upper panel), rising fast and settling between 130 and 180\,km\,s$^{-1}$ 
at larger radii, consistent with, e.g., Ott et al. (2001). Position angle and inclination 
could be kept constant, the latter to a slightly lower value than the one favored in 
Sect.\,4.5, which was based solely on the ellipticity of the measured distributions and
the assumption of azimuthal symmetry. An inclination of 75$^{\circ}$ to 78$^{\circ}$ 
(see Sect.\,4.5) would correspond to deviations of order 2.5 -- 5.5$\sigma$, given 
the error bars in Fig.~\ref{parameters-vrot}. Fig.~\ref{parameters-scaleheight} provides scale 
heights of order 20--30\,pc at galactocentric radii of 5$''$ -- 15$''$, while the velocity 
dispersion (Fig.~\ref{parameters-dispersion}) is of order 20\,km\,s$^{-1}$ except for our 
C$_3$H$_2$--4 line, where signal-to-noise ratios and blending with H42$\alpha$ likely lead 
to unreliable results.  

\begin{figure*}[t]
\resizebox{8.8cm}{!}{\rotatebox[origin=br]{0.0}{\includegraphics{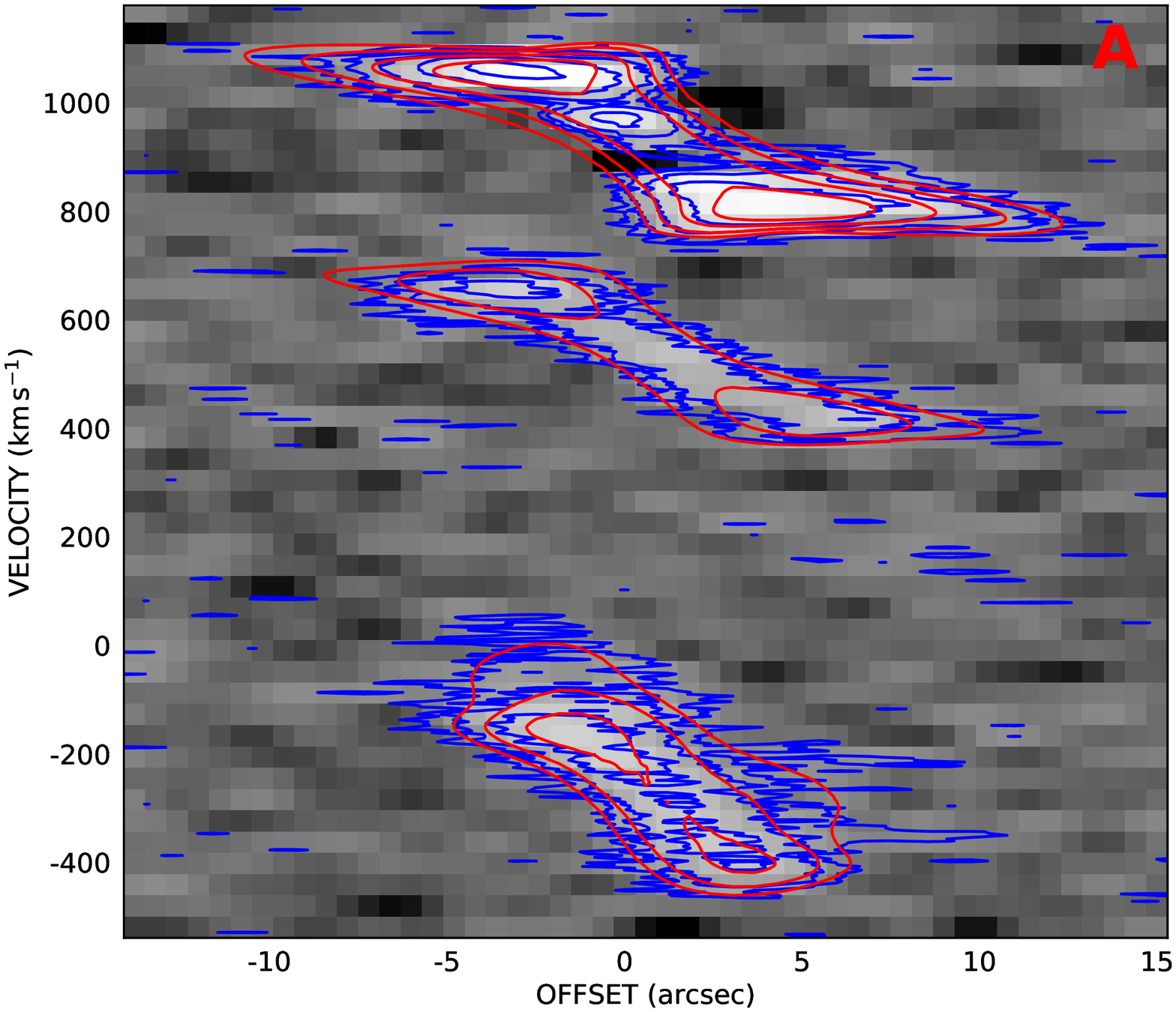}}}
\resizebox{8.8cm}{!}{\rotatebox[origin=br]{0.0}{\includegraphics{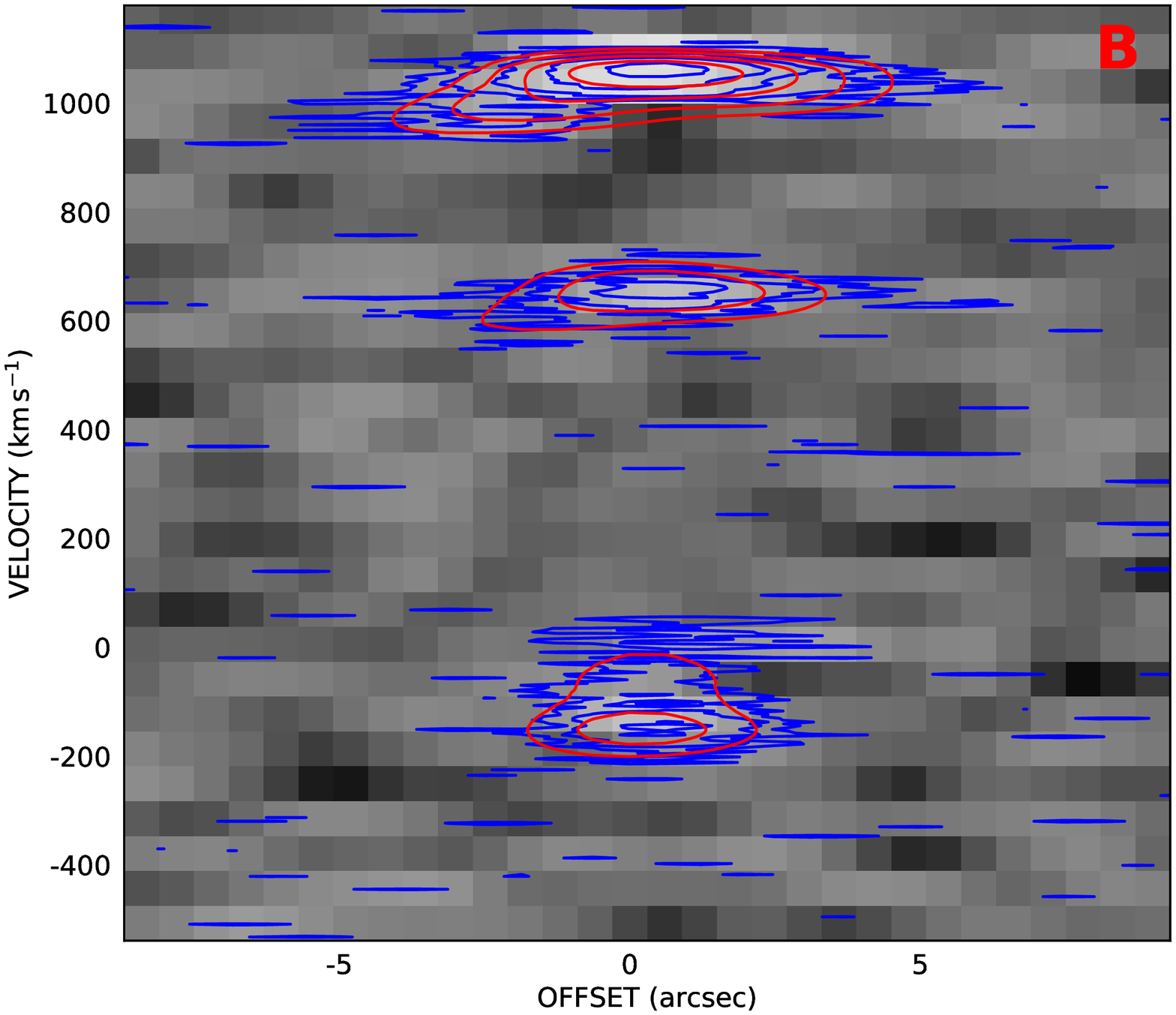}}}
\resizebox{8.8cm}{!}{\rotatebox[origin=br]{0.0}{\includegraphics{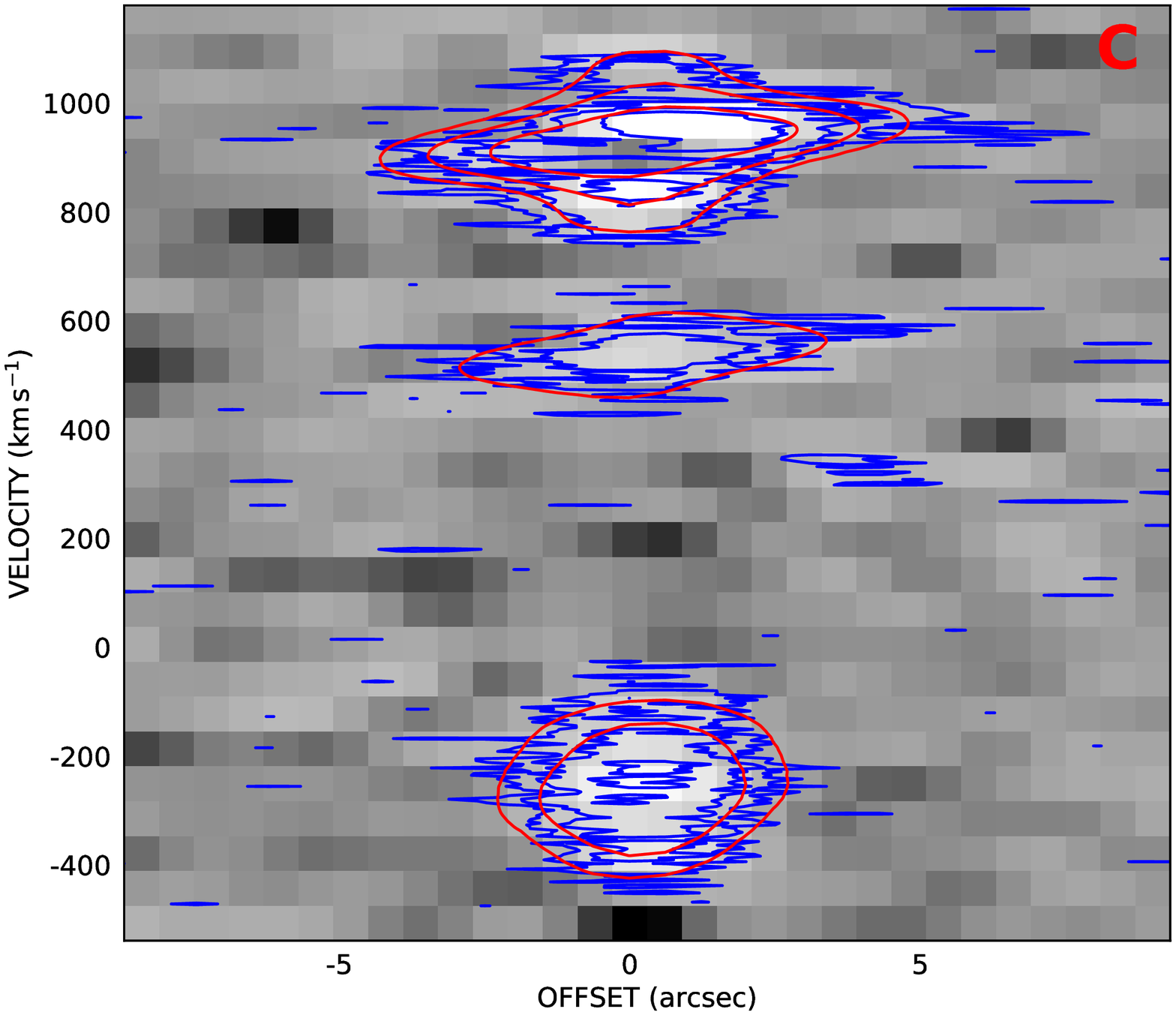}}}
\hspace{0.55cm}
\resizebox{8.8cm}{!}{\rotatebox[origin=br]{0.0}{\includegraphics{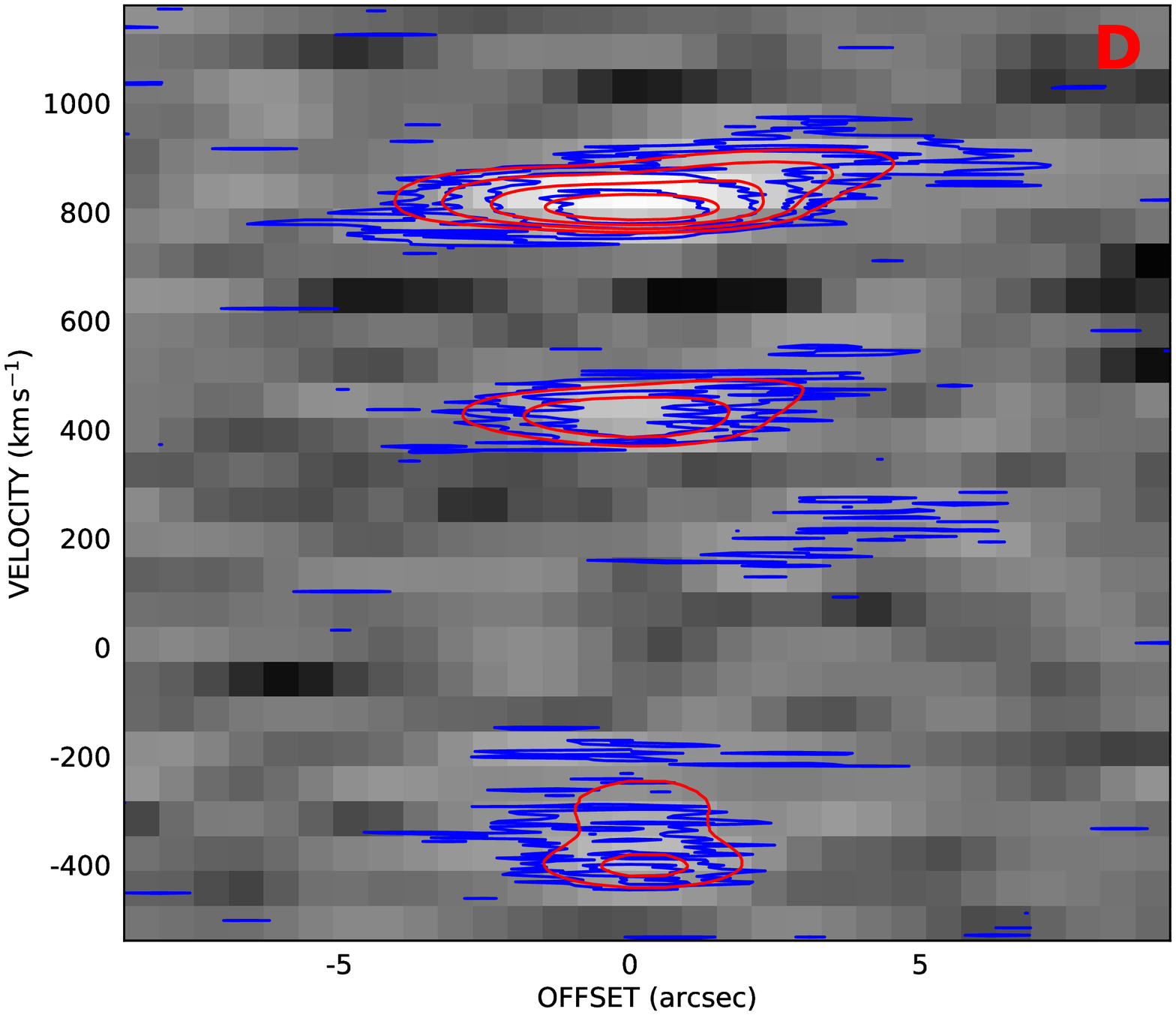}}}
\caption{PV-diagrams along the lines shown in Fig.~\ref{slice-positions}, each with (from top to 
bottom) c-C$_3$H$_2$ 2$_{12}$$\rightarrow$1$_{01}$, CH$_3$C$_2$H 5$_0$$\rightarrow$4$_{0}$ and, 
overlapping, C$_3$H$_2$ 4$_{32}$$\rightarrow$4$_{23}$ and H42$\alpha$. The velocity scale refers 
to that of the CH$_3$C$_2$H line (see Sect.\,4.6.2 and Table~\ref{tab-lines}). Dark: low intensity; 
light grey: high intensity. {\it Top left panel}: cut A (see also Fig.~\ref{slice-positions}) along 
the major axis from the northeast toward the southwest. {\it Top right panel}: cut B perpendicular 
to the major axis crossing the 710\,km\,s$^{-1}$ peak from the southeast toward the northwest. {\it 
Bottom left panel}: cut C perpendicular to the major axis, central line. {\it Bottom right panel}: 
cut D perpendicular to the major axis crossing the 430\,km\,s$^{-1}$ peak. A, B, C or D are displayed 
in the upper right corner of their respective panels. Blue contours: original data cube. Red contours: 
model data cube with constant vertical gradient in radial motion and final rotation curve (see top two 
panels of Fig.~\ref{parameters-vrot}). Contours: 0.1, 0.2, 0.4, 0.8\,mJy\,beam$^{-1}$.} 
\label{slices-model-26}
\end{figure*}

Using the obtained model parameters, we created position-velocity diagrams along the most 
important lines, one cut along the major axis and three perpendicular to it crossing either 
the dynamical center of the galaxy or the southwestern 430\,km\,s$^{-1}$ and northeastern 
710\,km\,s$^{-1}$ peaks (Fig.~\ref{slice-positions}). Figure~\ref{slices-model-26} shows both 
the observed (blue contours) and the modeled (red contours) distributions in the form of 
PV-diagrams along these four cuts from the northeast to the southwest (upper left panel) and 
from the southeast to the northwest (other panels). The overall agreement is good. A careful 
comparison of the minor axis PV-diagrams shows that particularly the c-C$_3$H$_2$--2 and 
CH$_3$C$_2$H lines (some and likely no absorption, respectively; see Sect.\,3.3)) provide 
evidence for non-circular motion. Here, from left to right, rising velocities (and not merely 
widened line widths) are encountered along the minor axis. With the southeastern (left) parts 
of the minor axis cuts representing the near and the northwestern (right) parts the far side 
of the galaxy (Sect.\,3.2), the rising velocities toward the far side imply radially 
outflowing gas.  The second panel of Fig.~\ref{parameters-vrot} provides the absolute value 
of the $DVRA$ parameter, where we set $VRAD$ = 0. A careful inspection of the velocity 
gradients in Fig.~\ref{slices-model-26} along the minor axis reveals outflow velocities 
of order 50\,km\,s$^{-1}$. This corresponds in the model to the DVRA-value at 1$''$ above or below 
the plane of the galaxy (Fig.~\ref{parameters-vrot}, second panel), which is roughly consistent 
with the modeled thickness of the nuclear disk (Fig.~\ref{parameters-scaleheight}).
Higher sensitivity and resolution measurements of various tracers would be worthwhile 
to obtain more information on the distribution, kinematics, mass, momentum and energy 
of this gas in the nuclear disk.

\subsection{The bar}

From systematic observations of spiral galaxies with high resolution it is known that the 
orientation of the outer large scale disk is not connected with that of the innermost
accretion disk (e.g., Ulvestad \& Wilson 1984). This has been established in a particularly 
convincing form for those H$_2$O megamaser galaxies, where the 22\,GHz water vapor hotspots
form a close to edge-on circumnuclear disk with Keplerian rotation and a characteristic size 
of 1\,pc, providing direct information on the innermost geometry of the galaxy. NGC~4945 
(Greenhill et al. 1997) is like NGC~1068 (see below and, e.g., Greenhill et al.  
1996; Gallimore et al. 2001) one of these objects. 

For the gas to travel from kpc-scales to the accretion structures at $\approx$1\,pc,
almost the entire angular momentum must be lost to feed the central monster. Gravitational 
torques may provide an efficient mechanism. The bars inside bars model proposed by Shlosman 
et al. (1989) postulates the presence of a series of embedded bars, gradually removing the 
angular momentum from the gas, distorting its orientation and funneling it into the central 
region of a galaxy. While the determination of disk orientations in the central parsec requires 
interferometric observations at radio wavelengths, measurements of such galaxies on larger 
scales down to $\approx$100\,pc are possible thanks to the Hubble Space Telescope (HST). 
Greene et al. (2013) and Pjanka et al. (2017) have investigated the orientation of 18 such 
disk-megamaser galaxies and find that structures on a $\approx$100\,pc scale are consistent 
with being randomly distributed with repect to both, the kpc-sized large scale and the 
pc-sized megamaser disks.

Based on H\,{\sc i} and CO data with 23$''$ resolution from NGC~4945, Ott et al. (2001) 
found systematic departures from uniform circular motion and S-shaped velocity contours 
suggesting the presence of a large scale kpc-sized bar at position angle 
$\approx$35$^{\circ}$ and approximate azimuthal angle (counterclockwise) of 40$^{\circ}$ 
extending out to $\sim$150$''$ from the nucleus. 

Lin et al. (2011) modeled the inner few 100\,pc of NGC~4945, which is closer to the 
scales discussed in this work. They demonstrated that S-shaped CO iso-velocity contours can be
reproduced by a shock along spiral density waves, which are excited by a rapidly rotating bar. 
In view of the radio recombination line analysis of Bendo et al. (2016) and the CH$_3$C$_2$H 
velocity field presented in Fig.~\ref{ch3c2h-moment1}, we cannot confirm the presence of 
S-shaped iso-velocity contours in the 10$''$ $\times$ 2$''$ sized nuclear disk. Furthermore, 
their adopted inclination for this nuclear disk (62$^{\circ}$) appears to be too low by at 
least 10$^{\circ}$ (Sect.\,4.1.1 and Fig.~\ref{parameters-vrot}). Nevertheless, making use 
of our most intense line, that of the high density tracer HCN $J$ = 1$\rightarrow$0, we can 
study the connection between the nuclear disk and its dense surrounding gas in unprecedented 
detail.

\begin{figure}[t]
\resizebox{9.1cm}{!}{\rotatebox[origin=br]{0.0}{\includegraphics{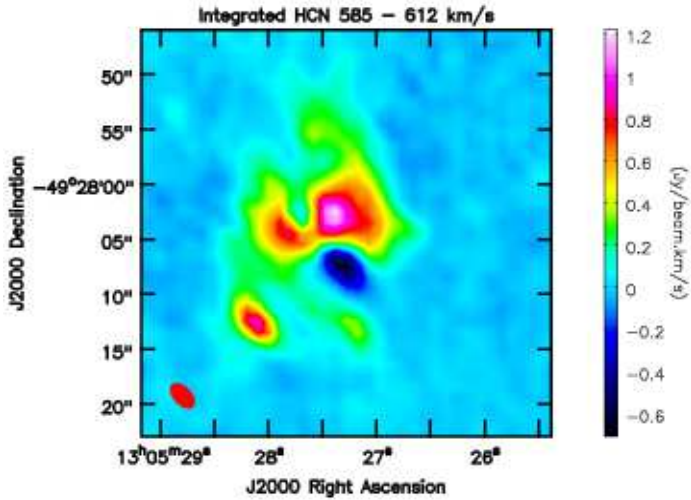}}}
\caption{Primary beam corrected HCN $J$ = 1$\rightarrow$0 emission (Briggs weighting with a
robustness parameter of 0.0) integrated over the velocity interval 585 -- 612\,km\,s$^{-1}$, the near 
systemic velocity range, where nuclear absorption is not as strong as at adjacent velocities. 
The arms (mainly green color, see Sect.\,4.7) are visible in the northwest and southeast, 
likely connected by the elongated region of strong emission along an east-west axis. See 
Fig.~\ref{cs-bar} for an even clearer image, where absorption plays a lesser role.}
\label{hcn-bar}
\end{figure}

What we find off the major axis of the nuclear disk, in a broad velocity range 
($\approx$100\,km\,s$^{-1}$) around the systemic velocity, are not outer rings or tori 
but two arms, the western one bending to the north and northeast, the eastern one 
bending toward the south and southwest (Figs.~\ref{hcn-channel} and \ref{hcn-bar}). 
Each of them extends mainly along the declination axis over at least $\approx$15$''$
(this is close to the largest accessible angular scale of our measuements, see Sect.\,2), 
corresponding to 250--300\,pc. Length, orientation and curvature are in impressive agreement 
with the two spiral arms proposed by Lin et al. (2011; their figs.~4 and 6), except near 
the center, where the arms do not wrap around the nucleus, but appear to start on
their respective side.

\begin{figure}[t]
\hspace{-0.4cm}
\resizebox{9.6cm}{!}{\rotatebox[origin=br]{0.0}{\includegraphics{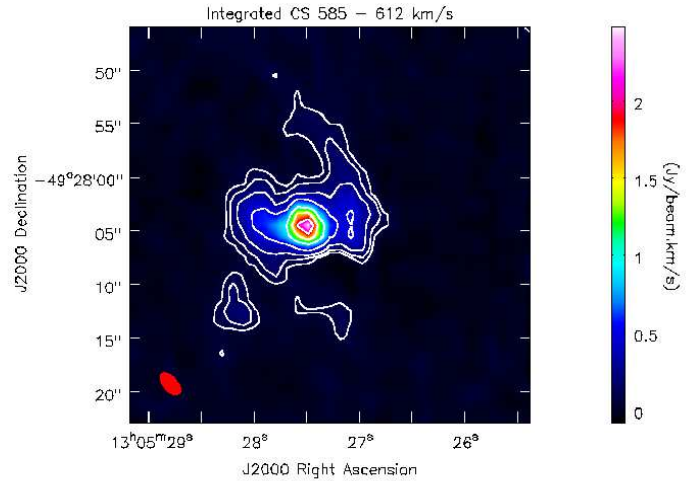}}}
\caption{Primary beam corrected CS $J$ = 2$\rightarrow$1 emission (Briggs weighting with 
a robustness parameter of 0.0) integrated over the velocity interval 585 -- 612\,km\,s$^{-1}$, 
the near systemic velocity range, where nuclear absorption is not as strong as at adjacent 
velocities. Contours: 0.025, 0.05, 0.1, 0.2, 0.4, 0.6, and 0.8 times the integrated peak flux 
density of 2.48\,Jy\,km\,s$^{-1}$\,beam$^{-1}$. Overall, line absorption plays a lesser role 
than in Fig.~\ref{hcn-bar}. While the southern region of enhanced emission is part of the
southeastern arm, the hotspot around 7$''$ farther to the east is (as in Fig.~\ref{hcn-bar}) 
a separate feature, possibly related to Marconi et al.'s (2000) Knot B (Sect.\,4.1.3).}
\label{cs-bar}
\end{figure}

\begin{figure}[ht]
\hspace{-0.4cm}
\resizebox{8.8cm}{!}{\rotatebox[origin=br]{0.0}{\includegraphics{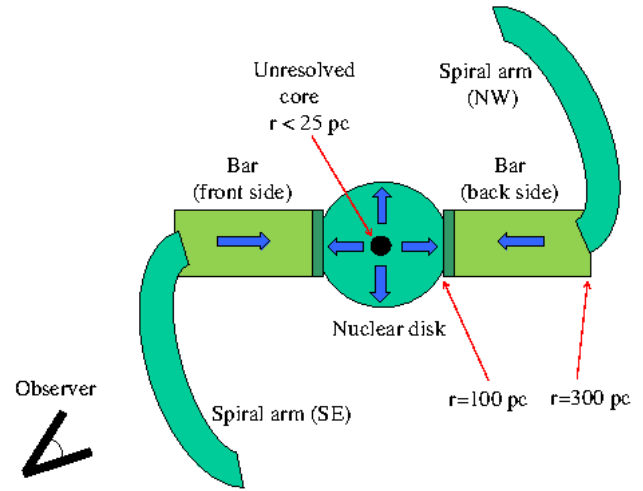}}}
\caption{A sketch of the central parts of NGC~4945 viewed from a position above the 
plane and on top of the center of the galaxy. The blue arrows indicate modeled
or observationally suggested radial motions of the gas.} 
\label{sketch}
\end{figure}

Near the systemic velocity there is a narrow velocity range, where absorption toward 
the nucleus is relatively weak and emission can be viewed partially also in between the inner 
starting points of the arms. Here we may recognize in the HCN $J$ = 1$\rightarrow$0 image 
a direct connection between the two arms (Fig.~\ref{hcn-bar} and also Fig.~\ref{hcn-channel}). While 
the CS $J$ = 2$\rightarrow$1 line is overall weaker (e.g., Tables~\ref{tab-lines} and \ref{tab-integral} 
and Fig.~\ref{hcn+cs-profile}), it is less affected by nuclear absorption (Fig.~\ref{hcn+cs-profile} 
and Sect.\,4.2) and therefore provides a clearer view onto the connection between the arms and 
the nuclear disk. What we find (Fig.~\ref{cs-bar}) is a broad east-west connection, extending 
(due to a lesser influence of absorption) more to the south than in the HCN $J$=1$\rightarrow$0 
image. This is the signature of a bar-like feature having a total length of slightly less than 
10$''$ ($\approx$150\,pc linear size, corresponding to $\approx$600\,pc at $i$ $\approx$ 75$^{\circ}$). 
This connecting bar-like feature (hereafter termed ``the bar''; note that it is smaller
than the one proposed by Ott et al. 2001) apparently touches the nuclear disk at its near and far side at 
a galactocentric radius of $\approx$100\,pc (assuming azimuthal symmetry for the nuclear disk) and 
may allow for substantial inflow of gas. In projection, the bar is not located exactly perpendicular 
to the nuclear disk, but is elongated roughly along the east-west direction and may form an 
azimuthal angle of $\approx$60$^{\circ}$, counterclockwise, to the midline of the nuclear disk
in the plane of the sky. The eastern side, which is the front side of the bar, is located 
partially in front of the northeastern side of the continuum source and might therefore provide 
more severe absorption there than in the southwest. Such an enhanced absorption is not 
clearly seen by our HCN and H$^{13}$CN $J$ = 1$\rightarrow$0 spectra (Sect.\,3.2.2 and 
Fig.~\ref{hcn-absorption}). However, most of the northeastern absorption of HCN and H$^{13}$CN $J$ 
= 1$\rightarrow$0 (Figs.~\ref{hcn-channel}, \ref{hcn-absorption} and \ref{h13cn-channel}) is 
seen  at receding velocities with respect to the systemic one and therefore suggests inflow. 
These are the kinematics expected in the case of a bar. Using the Herschel Space Observatory
and analyzing HF $J$ = 1$\rightarrow$0 absorption lines toward the nuclear region of NGC~4945, 
Monje et al.  (2014) find a velocity component centered at $\approx$640\,km\,s$^{-1}$, redshifted 
with respect to the systemic velocity of the galaxy. This is interpreted as inflow from radial 
distances $\la$200\,pc, in agreement with our much higher resolution data. The scenario resulting from 
our measurements including Sects.\,4.6.2 and 4.7 is schematically displayed in Fig.~\ref{sketch}. 

The recently detected 36\,GHz Class I methanol maser (McCarthy et al. 2017), located $\approx$10$''$
southeast of the center, is (if part of the plane of the galaxy) several 100\,pc off the 
AGN and is thus likely either part of the front side of the bar or part of the southeastern 
arm. This agrees well with the findings of Ellingsen et al. (2017) that the 36\,GHz Class I 
methanol masers in NGC~253 are located at a projected distance of order 300\,pc 
off the center.

To summarize, we propose inflow of gas through the bar from galactocentric radii of 
$\approx$300 to $\approx$100\,pc toward the nuclear disk. On the other hand, the rotation of the 
nuclear disk, reaching galactocentric radii of up to $\approx$100\,pc at azimuthal angles
$\approx$60$^{\circ}$ off the bar, is disturbed by outflowing gas. The strong emission 
from the outer edges of the nuclear disk, seen at 430 and 710\,km\,s$^{-1}$, may thus 
not merely be caused by its geometry, with tangentially viewed locations of a highly 
inclined rotating disk yielding particularly large column densities near $V_{\rm sys}$
$\pm$ $V_{\rm rot}$. Instead, the peaks in line emission may be further enhanced by the 
fact that here inflow from the bar meets outflow from inside the nuclear disk, leading
to the formation of a ring near $r$ = 100\,pc.

\subsection{The specific case of NGC~4945}

To place our analysis in a slightly broader context, here we compare our findings 
with respect to the properties of the nuclear region of NGC~4945 with those of two 
additional nearby galaxies. One of these has been selected on the basis of being 
as similar as possible with respect to NGC~4945. Furthermore, it can be considered as
{\it the} template for star forming galaxies outside the Local Group. This galaxy has 
been observed with a linear resolution similar to ours and is, like NGC~4945, also 
suitable for a comparison with the less active Central Molecular Zone (CMZ) of our Galaxy. 
The other target is a slightly more distant object that is even more active than NGC~4945, 
to check how differences in overall activity (i.e. infrared luminosity) may influence 
central regions on $\approx$100\,pc sized linear scales.

The first of these two galaxies is NGC~253. NGC~4945 and NGC~253 share many 
properties. Both are starbursting galaxies with similar infrared luminosity, distance, 
inclination and position angle (e.g., Henkel et al. 1986; Engelbracht et al. 1998; 
Walter et al. 2017) and also appear to show similar $^{12}$C/$^{13}$C ratios and 
HCN $J$ = 1$\rightarrow$0 optical depths (Sect.\,4.3). Even the position angles of 
their bars are not far apart ($\approx$90$^{\circ}$ for NGC~4945 and $\approx$65$^{\circ}$ 
for NGC~253; Canzian et al.  1988; Engelbracht et al. 1998; Iodice et al. 2014). 
However, projected sizes of their nuclear disks are with 10$''$ versus 30$''$ ($\approx$200 
versus 600\,pc) quite different. The more widespread dense gas in NGC~253, not 
containing an AGN, is not showing the kind of nuclear concentration seen in NGC~4945. 
The NH$_3$ data from Lebr{\'o}n et al. (2011) and the multiline images from Meier 
et al. (2015) and Walter et al. (2017) provide good examples for the morphology of the 
dense gas in NGC~253. Maps of these non-CO molecular tracers reveal no overwhelmingly 
strong emission from NGC~253's nucleus, but exhibit instead a chain of blobs along 
the plane of the galaxy, extending over more than half a kpc. Note, however, that 
if the bar in NGC~4945 would not be seen ends-on but from the side, we would see 
molecular emission along a similarly extended ridge as toward NGC~253. NGC~253 
also hosts a (likely starburst driven) wind with a rate of roughly 9\,M$_{\odot}$, 
limiting nuclear star formation activity (Bolatto et al. 2013), as well as a 
spectacular 1.5 kpc sized OH plume also indicating outflowing molecular material 
from the nuclear region of NGC~253 (Turner 1985). The above similarities and the 
molecular outflow in NGC~253 suggest that both galaxies, NGC~4945 (see Sect.\,4.6.2) 
and NGC~253, share outflowing molecular gas in their central parts and possibly inflowing 
molecular gas associated with their respective bars farther outside. 

NGC~1068, the other galaxy, shows a much higher degree of activity. Recent ALMA 
studies were obtained by Garc\'{\i}a-Burillo et al. (2014, 2016, 2017) and Viti et al. 
(2014). This LIRG (to our knowledge the nearest one) hosts a slightly off-centered 
molecular disk of size $\approx$350\,pc with a warped innermost disk-like structure of 
a few parsec in length.  Outside the disk, CO traces two elongated filaments parallel 
to the major axis of a large scale bar, while a pseudo-ring consisting of two arms is 
encountered at its outer end. Inflow is found inside this ring and the outer bar region. 
However, the CND kinematics and a bow shock at a galactocentric radius of $\approx$400\,pc 
reveal further inside an outward flow perturbed by rotation. The molecular outflow 
rate is much higher than the star formation rate, thus being likely AGN driven. 

Apparently, in NGC~1068 the boundary between inner outflowing and outer inflowing gas 
is located at much larger radii than in NGC~4945, which is possibly related to the 
different activity levels of their respective AGN. To summarize, all three well studied 
galaxies, NGC~253, NGC~1068, and NGC~4945 share outflowing gas in their inner regions, 
while farther out at least NGC~1068 and (likely, because of the bar-like structures) 
NGC~253 and NGC~4945 experience inflow. In this context we still have to emphasize that 
the outflow seen by us in the nuclear disk does not include the spatially unresolved 
molecular core of size $\la$2$''$, which is much smaller in linear size than the 
nuclear disk in NGC~1068. Again, a stronger AGN activity in NGC~1068 might be 
responsible for this difference.

As already mentioned, bars can change the orientation of parts of their parent 
galaxies and the presence of several of them can distort the geometry in a way 
that even on intermediate ($\approx$100\,pc) scales there is little connection to 
the orientations at largest and smallest scales (e.g., Pjanka et al. 2017). However, 
NGC~4945 does not show any such change, in spite of the above suggested presence 
of a bar (Figs.~\ref{hcn-bar} and \ref{cs-bar}) reaching the nuclear disk at a 
galactocentric radius of $\approx$100\,pc and the more extended bar at a different 
azimuthal angle farther out proposed by Ott et al. (2001). This is a puzzle which 
remains, for the moment, unexplained, leaving NGC~4945 as one of the few known 
megamaser galaxies (Pjanka et al. 2017) forming a pc-scale 22\,GHz H$_2$O disk, 
a nuclear 200\,pc scale disk and a large scale disk with similar orientation.

\section{Conclusions}

We have observed the dense molecular gas with linear resolution of $\approx$40\,pc
in the nuclear region of the prominent southern starburst galaxy NGC~4945 using 
32 ALMA antennas in compact configuration during Cycle 1. $\lambda$ $\approx$ 3\,mm 
(Band 3) spectral line imaging in the range (rest frame) 85.1 -- 86.9\,GHz and 
around 88.6 and 98.0\,GHz led to the following main results:

(1) While showing a higher degree of activity, the central part of NGC~4945 is viewed
from a similar perspective as our Galactic center region ($i$ $\approx$ 75$^{\circ}$ versus
90$^{\circ}$), even with a not highly different position angle (45$^{\circ}$ versus 
63$^{\circ}$), with the receding side also being located in the northeast, and with a 
bar only a few 10$^{\circ}$ off our line of sight. In addition, the central part of 
NGC~4945 may exhibit a similar lopsidedness as the Central Molecular Zone (CMZ) of our
Galaxy (e.g., Ginsburg et al. 2016) with star formation most vigorously occurring 
slightly northeast of the center (for NGC~4945 here defined as the location of the maser 
disk (Greenhill et al. 1997)). The radio continuum representing free-free emission 
from star forming regions appears to peak slightly (less than an arcsecond) northeast of 
the central megamaser disk of the galaxy. This is corroborated (a) by a comparison with 
the position of the megamaser disk, (b) by a comparison with molecular line peak
positions, (c) by a comparison with the moment~2 distribution of our CH$_3$C$_2$H 
5$_0$$\rightarrow$4$_0$ map, and (d) by a comparison with the line of systemic velocity 
in the corresponding moment~1 map.

(2) Most molecular lines observed by us are affected by absorption toward the 
nuclear mm-wave continuum source at velocities from about --60 to +90\,km\,s$^{-1}$ 
relative to the systemic velocity (here taken as 571\,km\,s$^{-1}$ on a barycentric 
scale). The only line not affected by significant absorption is the CH$_3$C$_2$H 5$_0$ 
$\rightarrow$ 4$_0$ transition, likely due to its low critical density with respect to 
the other molecular lines studied here. With the radio continuum apparently peaking 
slightly northeast of the maser disk, line emission being affected by absorption 
against this continuum component is shifted toward the southwest. 

(3) Inside the roughly 10$''$ $\times$ 2$''$ (200\,pc $\times$ 40\,pc) sized nuclear 
disk there is a spatially unresolved molecular core of size $\la$2$''$ ($\la$40\,pc) 
with particularly intense molecular line emission, consistent in size with the core of 
the nuclear X-ray source detected by {\it Chandra} (Marinucci et al. 2012). Outside, 
starting at deprojected galactocentric radii of $\approx$300\,pc (adopting an inclination 
$i$ = 75$^{\circ}$), we find two spiral-like arms, one in the west turning toward the 
northeast and one in the east turning toward the southwest. The front and back sides of 
the nuclear disk are connected to the arms by a thick bar-like gaseous feature extending 
(again adopting $i$ = 75$^{\circ}$) from galactocentric radii of $\approx$100\,pc out to 
$\approx$300\,pc. This bar-like structure is seen almost ends-on at a position angle of 
$\approx$90$^{\circ}$ east of north, with the eastern part being the front side. This 
configuration may be responsible for the fact that the northeastern part of the continuum 
source is absorbed over a wider range of velocities than the southwestern one. If the 
bar-like component would be seen from the side, the extent of the central molecular ridge 
would approximately triple from 10$''$ to 30$''$ and would then be similar in size to 
the nuclear region encountered in NGC~253, but containing a prominent compact core which 
is absent in NGC~253. The recently detected Class I methanol maser in NGC~4945 (McCarthy 
et al. 2017) is likely either part of the front side of the bar or part of the southeastern 
spiral-like arm and thus, as the corresponding masers in NGC~253 (Ellingsen et al. 2017), 
quite far from the dynamical center of the galaxy.

(4) Position-velocity cuts perpendicular to the plane of the galaxy, following the minor 
axis, reveal outflowing gas with a characteristic deprojected velocity of 50\,km\,s$^{-1}$ for the 
nuclear disk. Farther out, if the bridge between the outer arms and the nuclear disk 
represents a bar, gas should be inflowing. Thus the 430 and 710\,km\,s$^{-1}$ components 
at the outer edges of the nuclear disk at galactocentric radii of $\approx$100\,pc may 
not only be prominent because of geometric reasons but also because here inflow from 
outside may meet outflow from inside the nuclear disk, possibly forming a ring confining
the nuclear disk.

(5) It is remarkable that on all scales, from the extended disk encompassing $>$20\,kpc
to the H$_2$O megamaser disk highlighting the central parsec, position angle and 
inclination appear not greatly changed. While the assumption of azimuthal symmetry 
suggests $i$ = 77$^{\circ}$ for the nuclear 10$''$ $\times$ 2$''$ disk, the application 
of a ring model leads to a best value of 72$^{\circ}$ -- 73$^{\circ}$. 

(6) Initially designed to prove that the rare $^{15}$N isotope is mainly synthesized 
in massive rotating stars, our project does not provide a clearcut result in this respect. 
Instead of confirming a previous tentatively obtained very low $^{14}$N/$^{15}$N ratio of 
$\approx$100 (Chin et al. 1999) explained by large amounts of $^{15}$N being ejected 
by rotating massive stars in the starburst region, we obtain a much higher ratio in 
the more common range between those of the Solar system and the local interstellar medium
(200 -- 500). Thus, $^{15}$N being mainly formed by lower mass ($<$8\,M$_{\odot}$) 
stars as proposed by Adande \& Ziurys (2012) is a viable alternative, also in view of 
the data from NGC~4945. 

(7) A position-velocity diagram of the nuclear disk in the CH$_3$C$_2$H 
5$_0$$\rightarrow$4$_0$ line indicates slight deviations from a straight line in 
the southwestern part. Furthermore, individual molecular complexes of size $\la$40\,pc 
could be identified, possibly representing giant molecular clouds.

(8) We neither find a pronounced molecular counterpart to the star forming ring 
at $r$ $\approx$ 50\,pc proposed by Marconi et al. (2000) and based on near 
infrared Pa$\alpha$ line data nor emission from their Knot C, northwest of the 
nuclear disk. Their Knot B in the southeast, however, is likely detected delivering 
the highest HCN $J$ = 1$\rightarrow$0 and CS $J$ = 2$\rightarrow$1 emission peak 
outside the nuclear disk.  Furthermore, the northwestern part of the $\Psi$-shaped 
structure encountered in HCN $J$ = 1$\rightarrow$0 between barycentric $V$ $\approx$ 
530 and 570\,km\,s$^{-1}$ (Table~\ref{tab-hcn} and Fig.~\ref{hcn-channel}) may 
confine the supernova-driven superbubble reported by Marconi et al. (2000) in the 
east, south and west.

Now knowing which tracer can provide a realistic description of the distribution 
of dense molecular gas in the innermost regions of NGC~4945 without being strongly 
affected by absorption, more sensitive measurements of CH$_3$C$_2$H with similar 
resolution as those presented here are not only mandatory but will also be feasible 
making use of the high sensitivity of ALMA. An essential part of such studies would be 
a characterization of the outflow, e.g. AGN or starburst related, and the properties
of the gas in the bar-like structure. To achieve these goals, also measurements of the 
excitation conditions of the gas would be desirable. Most urgent, however, are observations 
with significantly higher resolution, aiming to resolve the $\la$2$''$ molecular core. 
This core, giving rise to strong quasi-thermal emission, has so far not been studied 
but has the potential to close the main remaining gap in our knowledge between the 
large scale properties and the nuclear pc-scale 22\,GHz H$_2$O maser disk in one of 
our nearest and most prominent extragalactic molecular goldmines, NGC~4945.

\acknowledgements
We wish to thank an anaonymous referee for helpful comments. This work was 
partially carried out within the Collaborative Research Council 956, subproject 
A6, funded by the Deutsche Forschungsgemeinschaft (DFG). Y. Gong was supported 
by the National key research and development program under grant 2017YFA0402702 
and the National Natural Science Foundation of China (NSFC grant nos. 11127903).
S. Garc\'{\i}a-Burillo thanks for economic support
from Spanish grants ESP2015-68964-P and AYA2016-76682-C3-2-P. R.~S. Klessen 
acknowledges support from the DFG via SFB~881 (subprojects B1, B2, and B8) and
in SPP~1573 (grant numbers KL~1358/18.1, KL~1358/19.2). RSK furthermore thanks 
the European Research Council for funding in the ERC Advanced Grant STARLIGHT
(project number 339177). The paper is based on observations carried out with 
the Atacama Large Millimeter/Submillimeter Array (ALMA). Project: 
ADS/JAO.ALMA\#2012.1.00912.S (Cycle 1). ALMA is a partnership 
of ESO (representing its member states), NSF (USA) and NINS (Japan), together with 
NRC (Canada) and NSC and ASIAA (Taiwan) and KASI (Republic of Korea), in 
cooperation with the Republic of Chile. The Joint ALMA Observatory is 
operated by ESO, AUI/NRAO and NAOJ. We wish to thank the members of the 
German ALMA Regional Center (ARC) node at the Argelander-Institut f{\"u}r 
Astronomie of the University of Bonn for their support and help with the 
data reduction. This research has made use of NASA's Astrophysical Data System.
The NASA/IPAC Extragalactic Database (NED), also used, is operated by the Jet 
Propulsion Laboratory, California Institute of Technology, 
under contract with the National Aeronautics and Space Administration. Finally, we 
made use of HyperLeda (http://leda.univ-lyon1.fr) from the University of Lyon.

\begin{appendix}

\section{Channel maps}

Complementing the HCN, H$^{13}$CN, and CH$_3$C$_2$H channel maps in the main
text (Figs.~\ref{hcn-channel}, \ref{h13cn-channel}, and \ref{ch3c2h-channel}), 
here channel maps of the most prominent remaining {\it unblended} spectral features 
are presented. These are the HC$^{15}$N $J$ = 1$\rightarrow$0, the CS $J$ = 
2$\rightarrow$1 and the c-C$_3$H$_2$ 2$_{12}$ $\rightarrow$ 1$_{01}$ transitions.

\begin{figure*}[t]
\vspace{0.0cm}
\centering
\hspace{0.3cm}
\resizebox{18.4cm}{!}{\rotatebox[origin=br]{0.0}{\includegraphics{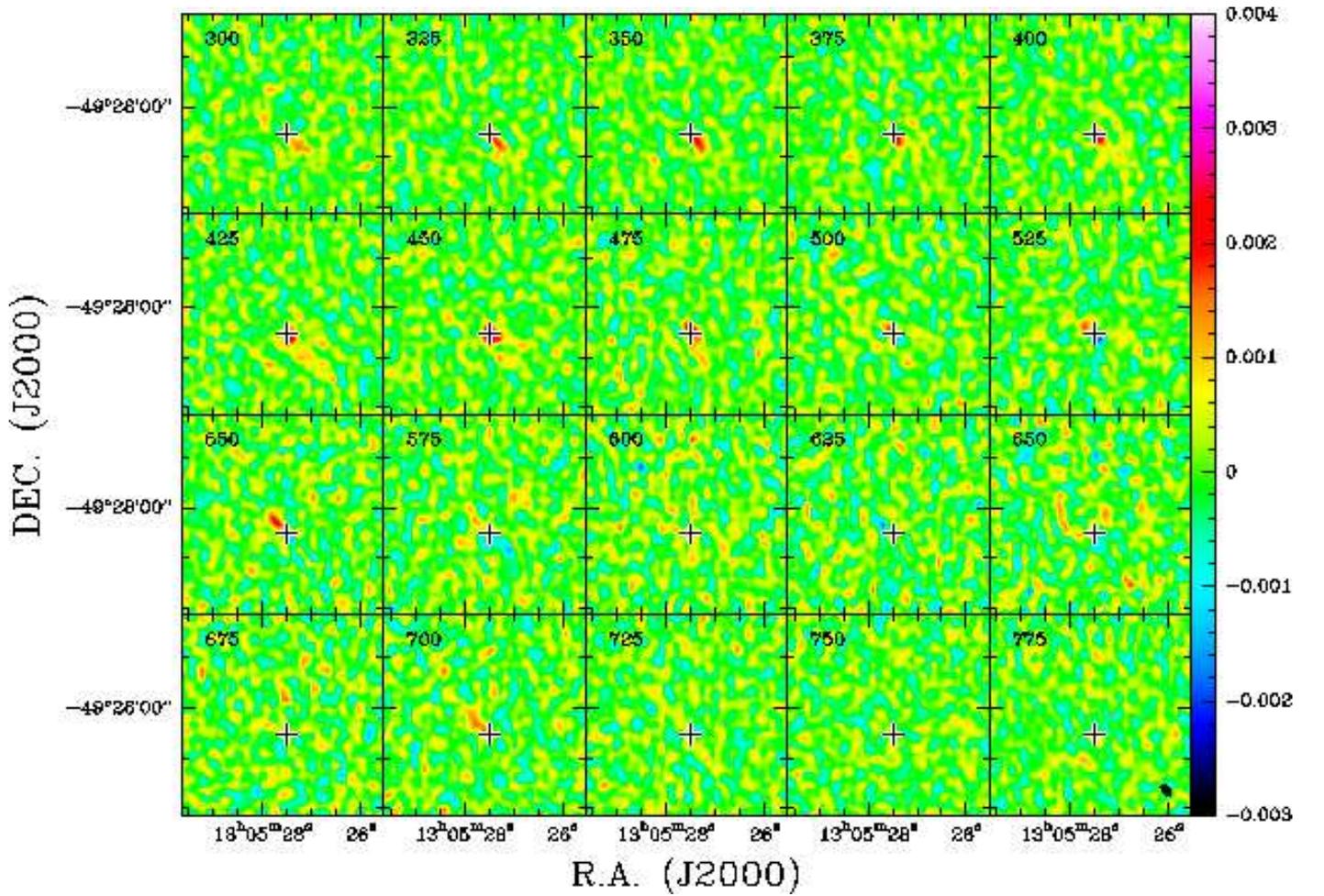}}}
\vspace{0.0cm}
\caption{H$^{12}$C$^{15}$N (HC$^{15}$N) $J$ = 1$\rightarrow$0 channel maps. For the units
of velocity, ordinate, and wedge as well as the position of the crosses, see Fig.~\ref{hcn-channel}. 
Note that the signal seen is very close to the noise level. This or a weak transition of 
SO or CH$_3$OH may explain the ``emission'' at $V$ $\approx$ 350\,km\,s$^{-1}$. Near the 
systemic velocity, where the more abundant HCN isotopologues exhibit weak absorption, no 
signal is seen, likely also a consequence of too low signal-to-noise ratios. 0.002\,Jy 
correspond to $\approx$85\,mK. The beam size is shown in the lower right corner of the figure.}
\label{hc15n-channel}
\end{figure*}

\begin{figure*}[t]
\vspace{0.0cm}
\centering
\hspace{0.3cm}
\resizebox{18.4cm}{!}{\rotatebox[origin=br]{0.0}{\includegraphics{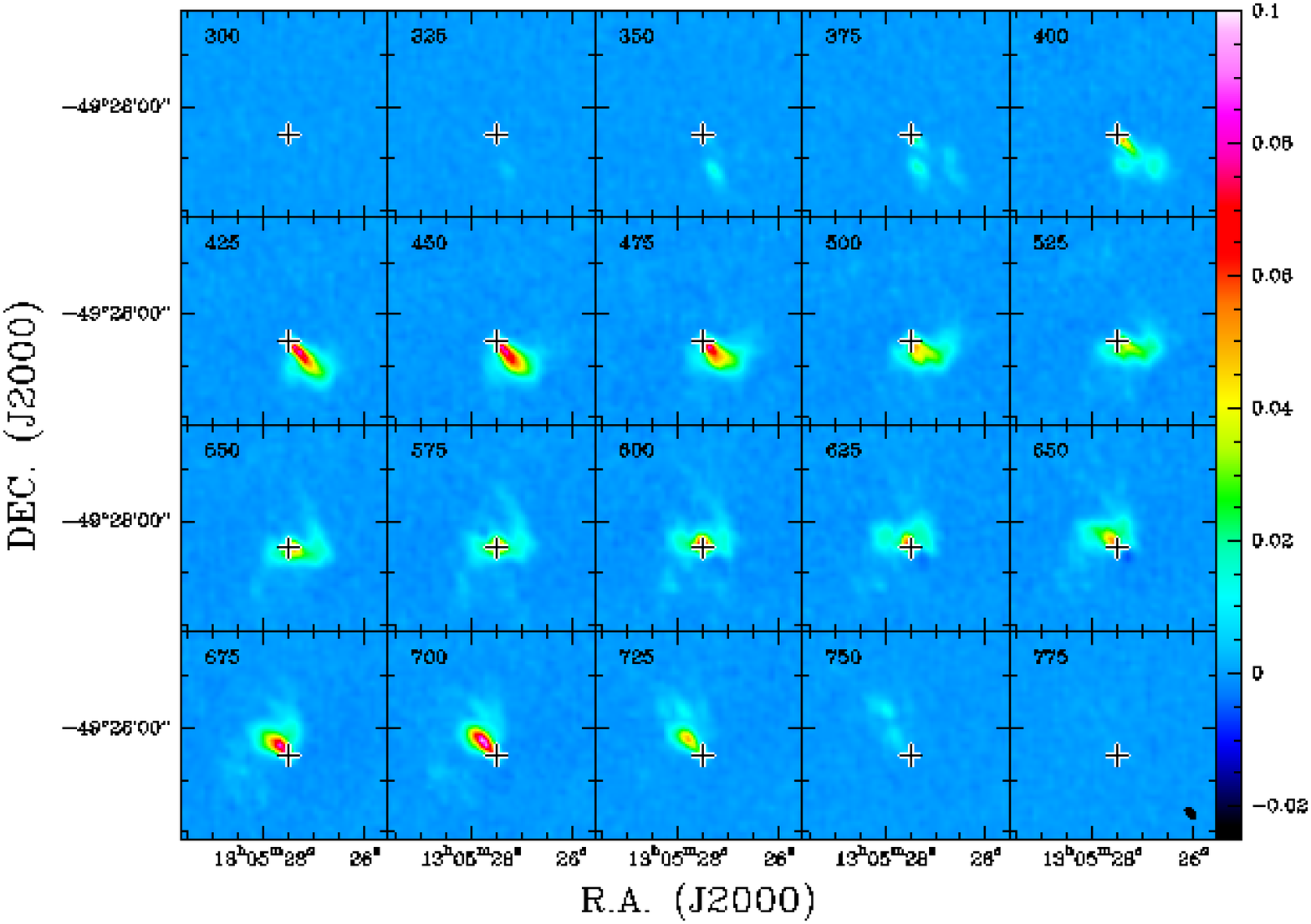}}}
\vspace{0.0cm}
\caption{$^{12}$C$^{32}$S (CS) $J$ = 2$\rightarrow$1 channel maps. For the units of velocity, 
ordinate, and the wedge as well as for the position of the crosses, see Fig.~\ref{hcn-channel}. 
0.1\,Jy correspond to $\approx$3.3\,K. The beam size is shown in the lower right corner of the 
figure.}
\label{cs-channel}
\end{figure*}

\begin{figure*}[t]
\vspace{0.0cm}
\centering
\hspace{0.3cm}
\resizebox{18.4cm}{!}{\rotatebox[origin=br]{0.0}{\includegraphics{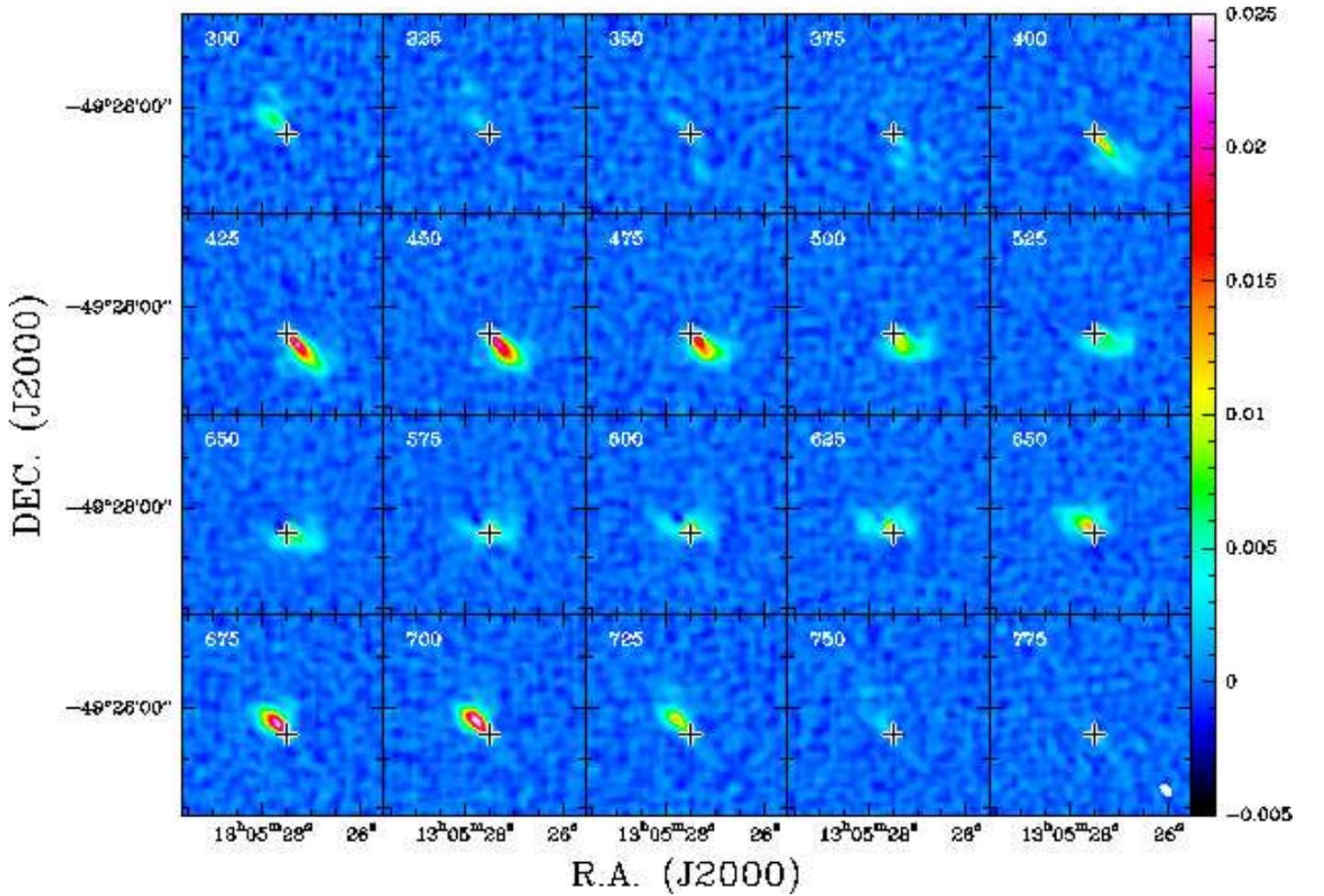}}}
\vspace{0.0cm}
\caption{c-C$_3$H$_2$ 2$_{12}$ $\rightarrow$ 1$_{01}$ channel maps. For the units of 
velocity, ordinate, and the wedge as well as for the position of the crosses, see Fig.~\ref{hcn-channel}. 
0.02\,Jy correspond to $\approx$0.85\,K. The beam size is shown in the lower right corner of the figure.
The emission seen at an apparent c-C$_3$H$_2$ velocity of $\approx$300\,km\,s$^{-1}$ is caused by the 
redshifted part of the CH$_3$C$_2$H $J$=5$_0$$\rightarrow$4$_0$ line, emitting near 710\,km\,s$^{-1}$ 
(see also Figs.~\ref{hc15n-profile} and \ref{ch3c2h-channel})).}
\label{c3h2-channel}
\end{figure*}

\end{appendix}

\end{document}